\documentclass[iop]{emulateapj}
\usepackage{graphicx}

\scshape

\newcommand{\mum}{$\mu$m}
\newcommand{\msun}{M$_{\odot}$}
\newcommand{\lsun}{L$_{\odot}$}
\newcommand{\lbol}{L$_{\rm bol}$}
\newcommand{\starf}{star-forming}
\newcommand{\msunmyr}{M$_{\odot}~{\rm Myr}^{-1}$}
\newcommand{\msunmyrpc}{M$_{\odot}~{\rm Myr}^{-1}~{\rm pc}^{-2}$}
\newcommand{\msunpc}{M$_{\odot}~{\rm pc}^{-2}$}
\newcommand{\mipslam}{24~$\mu$m}
\newcommand{\sigsfr}{$\Sigma_{\rm SFR}$}
\newcommand{\siggas}{$\Sigma_{\rm gas}$}
\newcommand{\sigsfrav}{$\overline\Sigma_{\rm SFR}$}
\newcommand{\siggasav}{$\overline\Sigma_{\rm gas}$}
\newcommand{\sigth}{$\Sigma_{\rm th}$}
\newcommand{\hii}{H{\small II}}
\newcommand{\co}{$^{13}$CO}
\newcommand{\kms}{km\,s$^{-1}$}
\newcommand{\cmsq}{cm$^{-2}$}
\newcommand{\nhtwo}{$N_{{\rm H}_2}$}

\date{}
\shorttitle{The star formation Law in Galactic molecular clouds}
\shortauthors{Retes-Romero et al.}

\begin{document}
\title{The star formation law in Galactic high-mass star-forming molecular clouds}


\author{R. Retes-Romero, Y. D. Mayya, A. Luna and L. Carrasco}
\affil{Instituto Nacional de Astrof\'isica, \'Optica y Electr\'onica,
Luis Enrique Erro 1, Tonantzintla, Puebla, C.P. 72840, Mexico.}
\email{rretes@inaoep.mx, ydm@inaoep.mx, aluna@inaoep.mx, carrasco@inaoep.mx}
To appear in ApJ

\begin{abstract}
We study the star formation (SF) law in 12 Galactic 
molecular clouds with ongoing high-mass star formation (HMSF) activity, as 
traced by the presence of a bright IRAS source and other HMSF tracers. 
We define the molecular cloud (MC) associated to each IRAS source using 
$^{13}$CO line emission, and count the young stellar objects (YSOs) within 
these clouds using GLIMPSE and MIPSGAL \mipslam\ Spitzer databases.
The masses for high luminosity YSOs (\lbol$>$10~\lsun) are determined 
individually using Pre Main Sequence evolutionary tracks and the evolutionary 
stages of the sources, whereas a mean mass of 0.5~\msun\ was adopted to
determine the masses in the low luminosity YSO population. The star formation 
rate surface density (\sigsfr) corresponding to a gas surface density (\siggas) 
in each MC
is obtained by counting the number of the YSOs within successive contours 
of $^{13}$CO line emission. We find a break in the relation between \sigsfr\ 
and \siggas, with the relation being power-law (\sigsfr$\propto$\siggas$^N$) 
with the index $N$ varying between 1.4 and 3.6 above the break. The \siggas\ 
at the break is between 150--360~\msunpc for the sample clouds, which 
compares well with the threshold gas density found in recent studies of 
Galactic \starf~regions. 
Our clouds treated as a whole lie between the
Kennicutt (1998) relation and the linear relation for Galactic and 
extra-galactic dense \starf~regions. 
We find a tendency for the high-mass YSOs to be found preferentially in dense 
regions at densities higher than 1200~\msunpc\ ($\rm \sim0.25~g~cm^{-2}$).
\end{abstract}


\section{Introduction}

The knowledge of physical processes driving the conversion of interstellar 
gas into stars is fundamental to the development of a predictive physical 
theory of star formation. A basic step in this direction is to obtain empirical 
relations between parameters that are related to star formation.
\citet{schmidt59} suggested a relation between the
star formation rate (SFR) and the density of the gas, which in 
recent decades has been reformulated by \citet{kennicutt98} as a relation 
between SFR surface density (\sigsfr) and the surface density of the 
gas (\siggas). This relation, often referred to as Kennicutt-Schmidt (KS) 
relation, has a power-law form \sigsfr$\propto$\siggas$^N$ with the index 
N=1.4. The relation was established over kiloparsec (kpc) scales, 
using the HI and CO lines to trace the gas content and holds over more than 4 
orders of magnitude in \siggas. The sampled regions encompass low density 
gas in disks of galaxies as well as high density gas in the infrared-bright 
circumnuclear regions. 

Stars form predominantly in dense clumps, of sizes of about one parsec.
These clumps themselves form part of larger molecular 
clouds (MCs). The CO line is the most commonly used tracer of the 
MC mass. Yet, it is a poor tracer of the high-density gas that resides in clumps. 
The denser regions are traced by high-density molecular lines such as 
NH$_3$ or by mapping the dust continuum submm emission 
\citep{heyer+16}. \citet{gao+04}
found a linear relation between SFR and the total mass of dense gas in luminous
infrared galaxies. The non-linear KS relation and the linear relation
between the SFR and the mass of the dense gas, jointly imply an increase
in the fraction of total gas in dense form as $\sim$\siggas$^{0.4}$
\citep{heiderman+10}. Systems with very high SFRs, such as ultraluminous
galaxies, have almost 100\% of their gas in the star-forming dense phase, whereas at lower
SFRs, this fraction can be as small as 1\% \citep{kennicutt+12}.

In recent years, the relation between the SF and gas mass has been explored 
at the scale of clumps and down to core scale in Galactic \starf~ regions 
\citep[e.g.][]{lada+10,heiderman+10} using extinction maps at infrared (IR)
wavelengths and/or dense gas tracers, to estimate the \siggas, and number counts 
of Young Stellar Objects (YSOs) along with a mean value of stellar mass per YSO 
and lifetime of Class II phase to estimate the SFR.
These studies found a linear relations, much like the relation found by 
\citet{gao+04} using dense gas tracers in luminous external galaxies.  
Additionally, \citet{wu+10} and \citet{heyer+16} have found linear relationships 
for dense clumps using dense gas tracers.
Indeed, \citet{heyer+16} found for dense clumps a strong linear correlation between 
\sigsfr~and \siggas~normalized by free-fall and clump crossing times, suggesting 
the star formation is regulated at local scales.  
\citet{heiderman+10} found that the linear relation holds above a threshold 
gas density of \sigth=129~\msunpc. For densities below the threshold \sigth, the SFR 
drops steeply with a power index $N$ as large as 4.6. \citet{heiderman+10} also 
found that the SFR at a given \siggas\ in Galactic \starf\ regions lies 
above the KS-derived SFR by factors of up to 17. 
They argued that this difference might arise due to the 
kiloparsec-size beams used to determine \siggas\ in extra-galactic studies,
which mostly contains non-star forming diffuse CO gas below \sigth.
Recent simulations by \citet{calzetti+2012} on the effects of 
sampling scale on the KS law support this idea.

Galactic studies that obtained a linear relation between the SFR and the
gas density, have restricted their analysis to clumps, that are expected
to transform almost all their mass into stars. Typical MCs contain gas at a 
variety of column densities, covering the entire range of \siggas\ found in 
extra-galactic studies, with the clumps representing the high-end of the \siggas\ 
distribution. Hence, the star-formation law can be  studied locally
within individual MCs. Indeed the original conjecture of \citet{schmidt59}
pertains to SF law within the clouds that are actively forming stars \citep{lada+13}.
\citet{gutermuth+11} studied eight nearby low-mass \starf\ clouds,
and found a power-law relation with N ranging between 1.37--3.8.
\citet{lada+13} carried out a similar study for a sample of 
four nearby clouds, finding N=2.04 for Orion A, Taurus, and California and 
N=3.3 for Orion B. More recently, \citet{willis+15} studied six massive \starf\ clouds, 
finding an average slope of $N=2.15\pm0.41$.
They also found that the dispersion of the relationship within individual
clouds is much lower than the differences in the $N$ values from one cloud to another. 

In summary, the value of $N$ within \starf\ Galactic MCs varies more than the 
range of values found for extra-galactic \starf\ regions \citep{kennicutt+12, bigiel+08}.
In other words, there is no preferred value of $N$ within MCs.
\citet{lada+13} have pointed out that even clouds having similar $N$ values could have 
vastly different levels of star formation activity, as the latter 
depends on the density structure within clouds, which is found to
vary significantly from cloud to cloud. Similar conclusions were drawn by 
\citet{burkert+hartmann13} using an analysis of the surface density structure 
within the Galactic \starf~clouds.

In the present study, we derive star formation laws for 
high-mass \starf\ MCs at sub-pc to parsec scale spatial resolutions.
Our approach differs from most previous explorations of this relationship for 
the Galactic MCs in three aspects: (1) the chosen MCs do not have a known 
optical nebula associated to them, indicating that 
SF activity has started recently in our sample of clouds.
This ensures that the physical condition of the gas has not been altered by
a previous generation of high-mass stars;
(2) the chosen MCs contain at least one high-mass YSO, an IRAS source, 
which ensures that our MCs are high-mass \starf\ regions, and
(3) we use $^{13}$CO data, instead of extinction maps, to derive the gas density. 
This procedure allows the exploration of embedded SF, even at relatively 
low surface densities. Our approach  also allows us to quantify  
the difference in SFRs at a given \siggas\ between Galactic and extra-galactic
studies, that used the same tracer, namely CO.

In this work, we carry out a search for Young Stellar Objects (YSOs) in 
the MCs associated to 12 IRAS sources at distances from 1~kpc to 5~kpc.
In \S2, we discuss the criteria for sample selection, and the method we have
followed for defining the MCs associated to the IRAS sources.  
The sample of YSOs is discussed in \S3, and in \S4, the mass function (MF)
is presented. A detailed analysis of the star formation law for the sample of \starf~
regions is reported in \S5. Our conclusions are summarized in \S6.

\section{The sample and Observational Data}

Our star-forming MCs for the study of the YSO population were selected using the 
following criteria:
(a) they contain an IRAS source with characteristics of an
ultra-compact \hii\ region \citep{wood+89},
(b) the IRAS sources are associated with the dense cores detected by CS(2-1) line emission \citet{bronfman+96},
(c) the line of sight (LOS) for each cloud is devoid of any other
foreground molecular component associated to other MCs as
inferred by the $^{13}$CO(J=1-0) spectra,
and (d) they have GLIMPSE \citep{churchwell+09} and 
\mipslam-MIPSGAL \citep{rieke+04} Spitzer public data.
The first two criteria ensure us that the selected clouds contain
high-mass star formation sites. The 3$^{\rm rd}$ criterion
is imposed to guarantee that all the YSOs associated geometrically to a
MC are physically associated to it. The last criterion
is the basis for the identification of the YSOs. Eighty clouds
of the first quadrant of the Galaxy satisfied the first 2 criteria.
However, the 3$^{\rm rd}$ criterion was satisfied by only 12 of these
clouds, thus restricting our sample size to 12 MCs. All these have GLIMPSE
and MIPSGAL public data available. Typical Spitzer RGB images 
(3.6~$\mu$m, 8.0~$\mu$m and \mipslam) of the resulting sample of clouds 
are shown in Figure~\ref{figure1}, where the position of IRAS source is identified 
by cross symbol. Table~1 lists the properties of the IRAS sources compiled 
from the literature such as IRAS name, galactic coordinates, bolometric 
luminosities, distance to the object etc.

\begin{figure*}
\includegraphics[width=8.5cm]{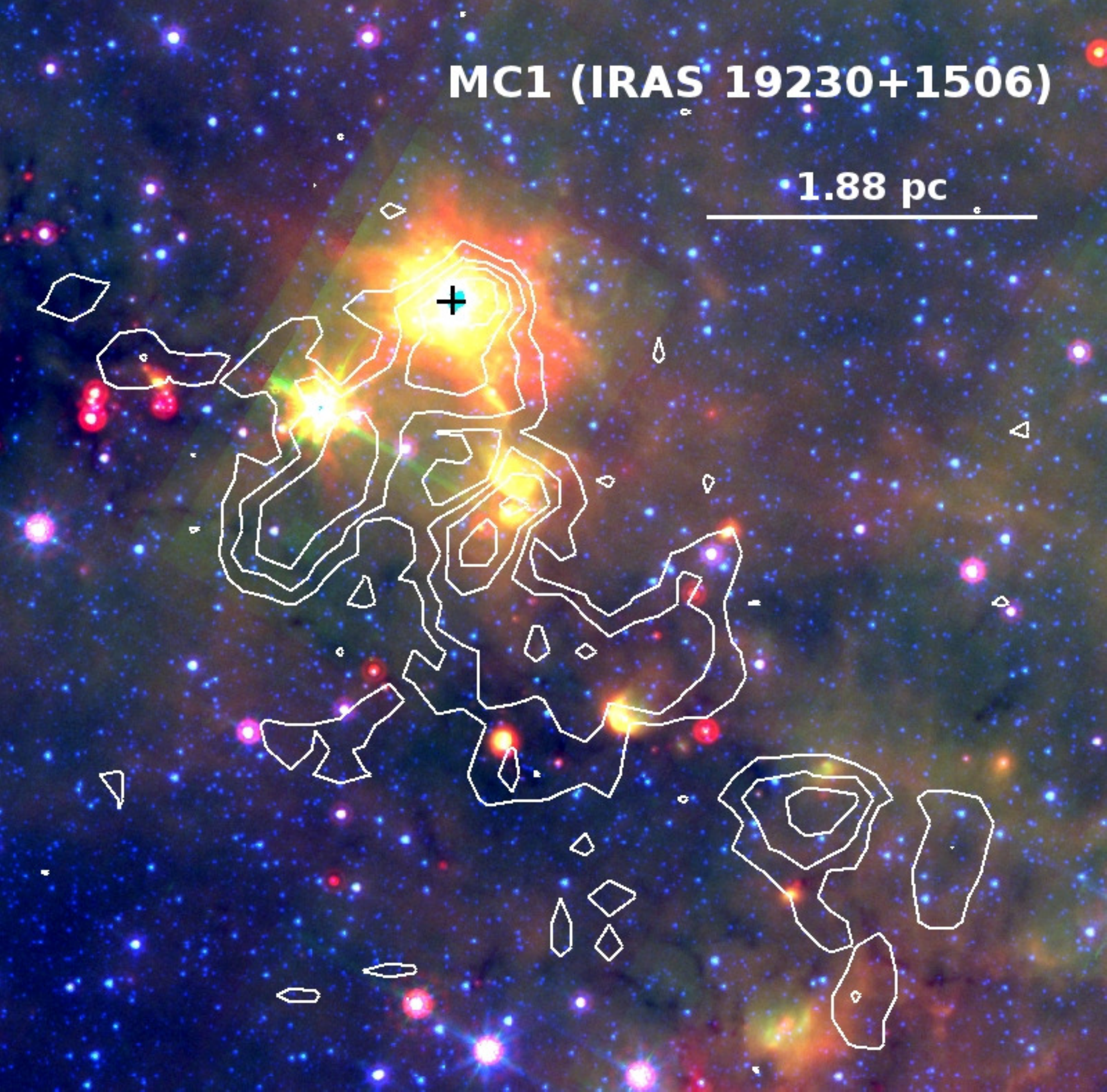}\hspace*{3mm}\includegraphics[width=8.5cm]{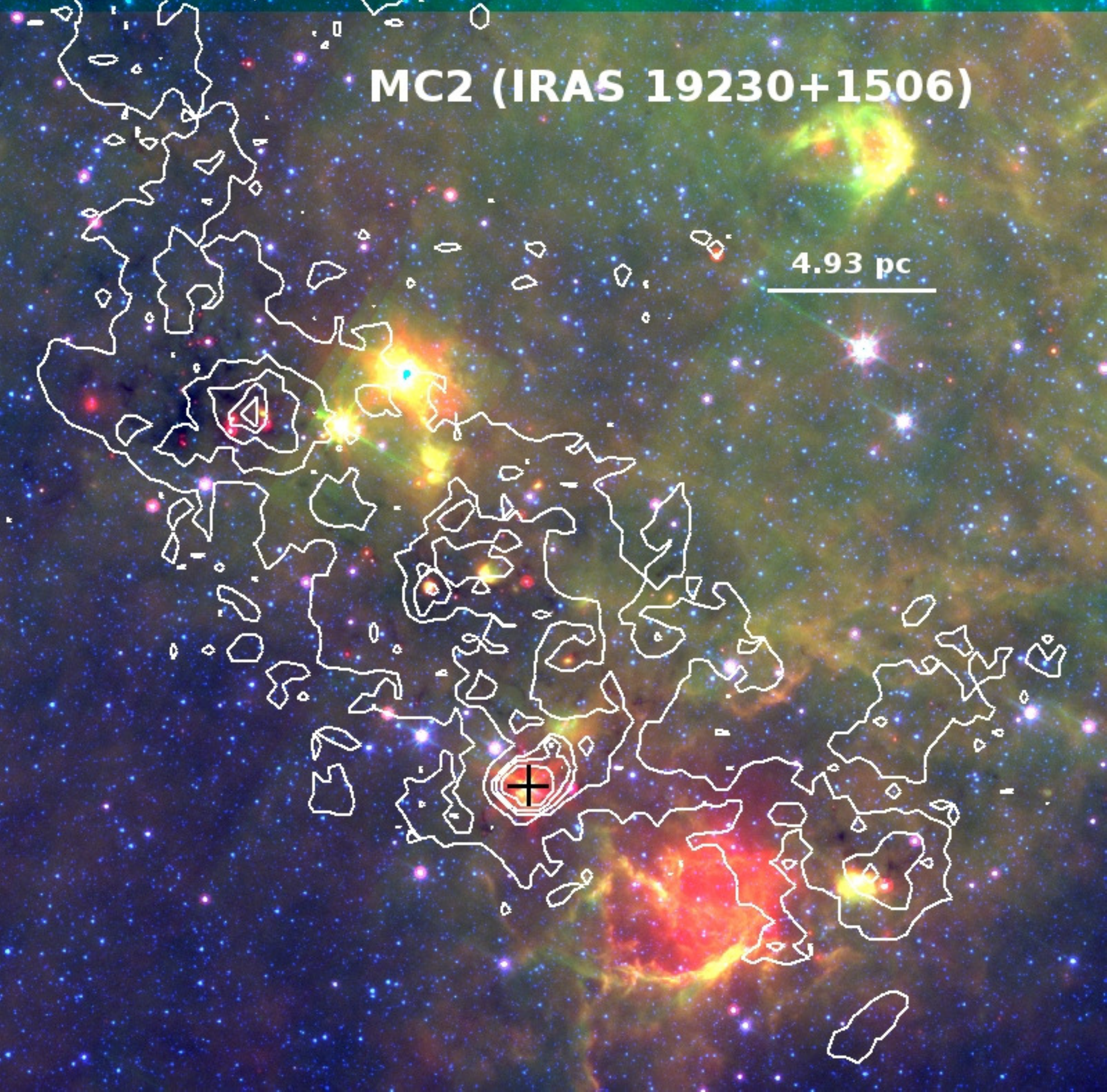}\\
\vspace*{2mm}
\includegraphics[width=8.5cm]{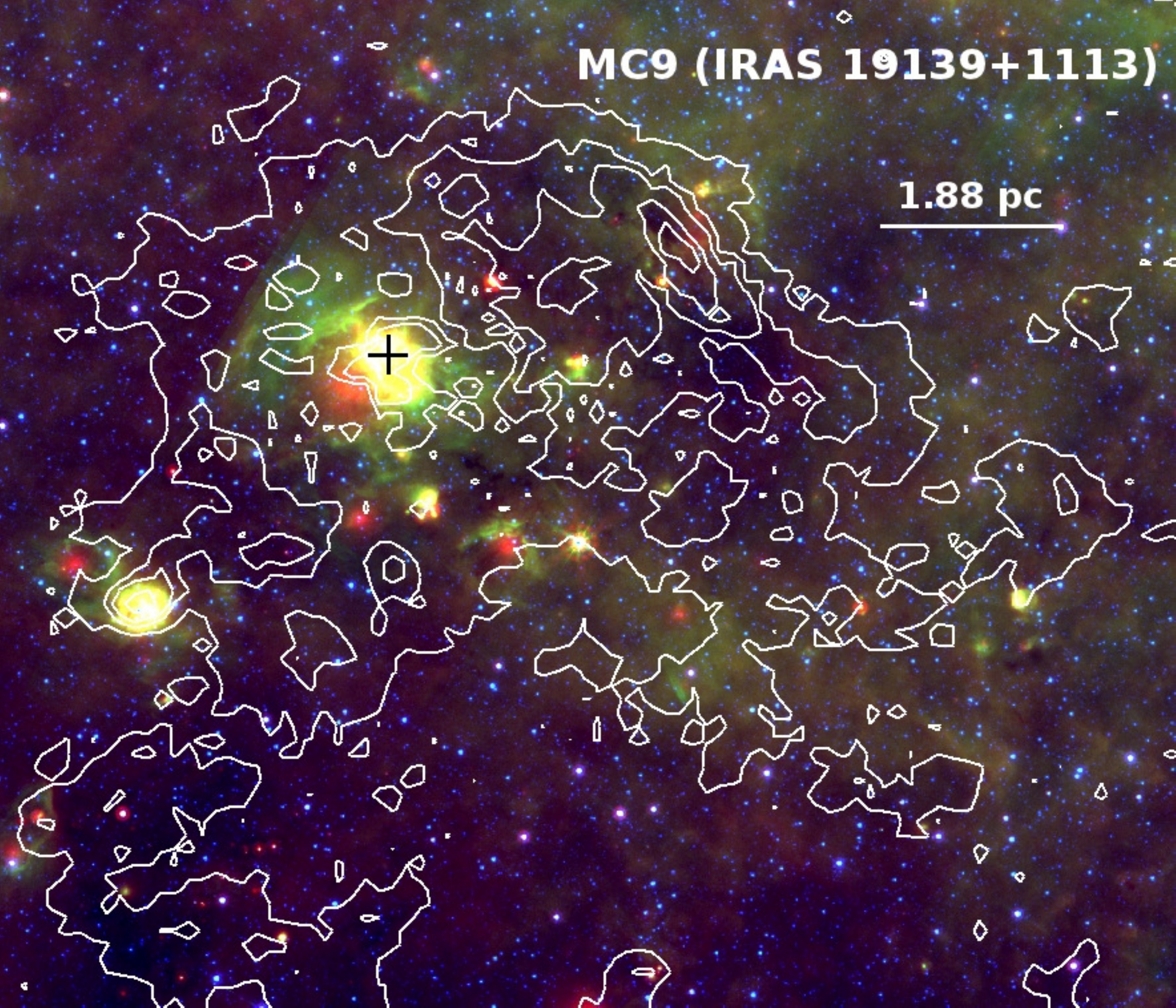}\hspace*{3mm}\includegraphics[width=8.5cm]{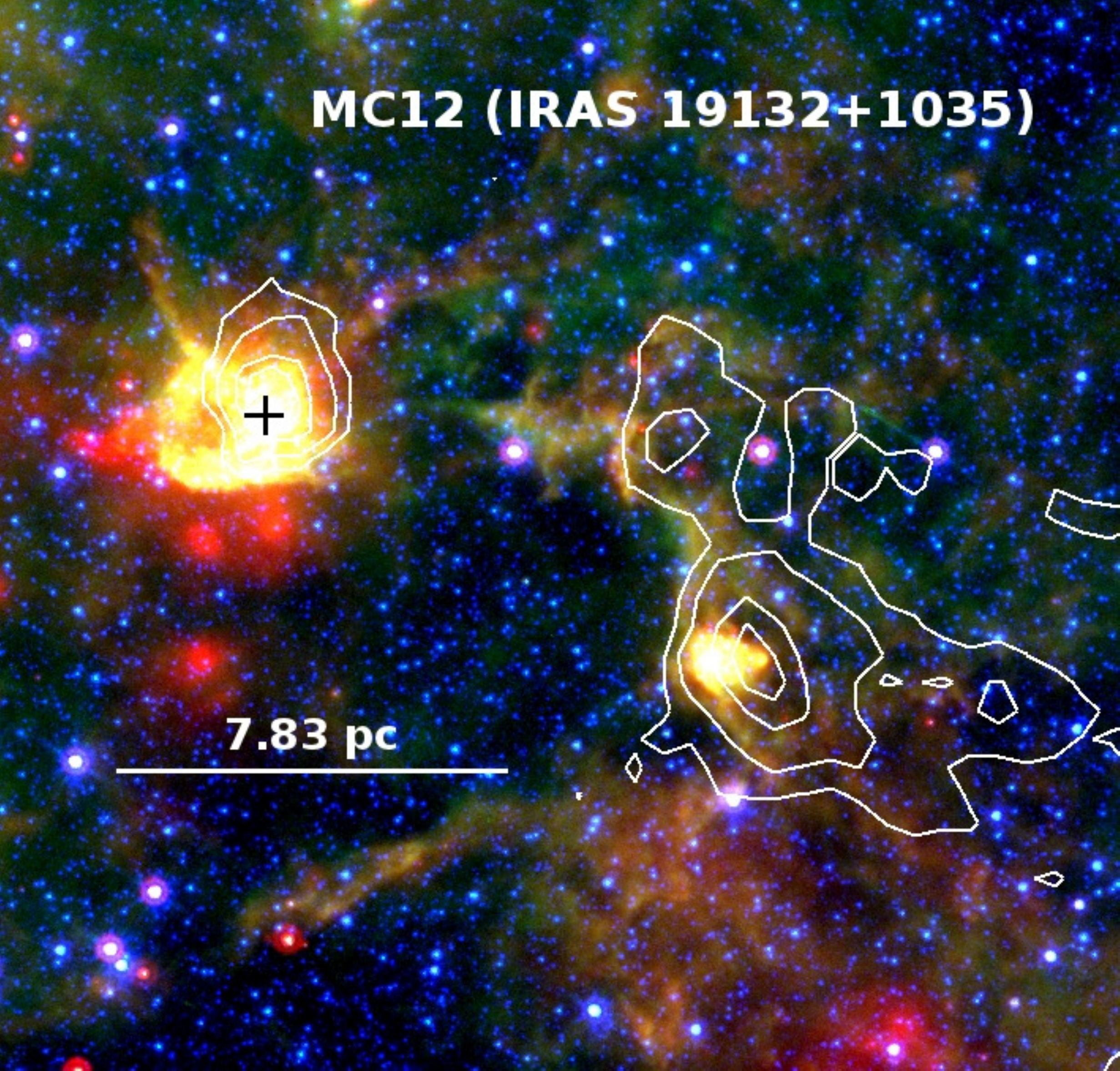}\\
\includegraphics[width=8.5cm]{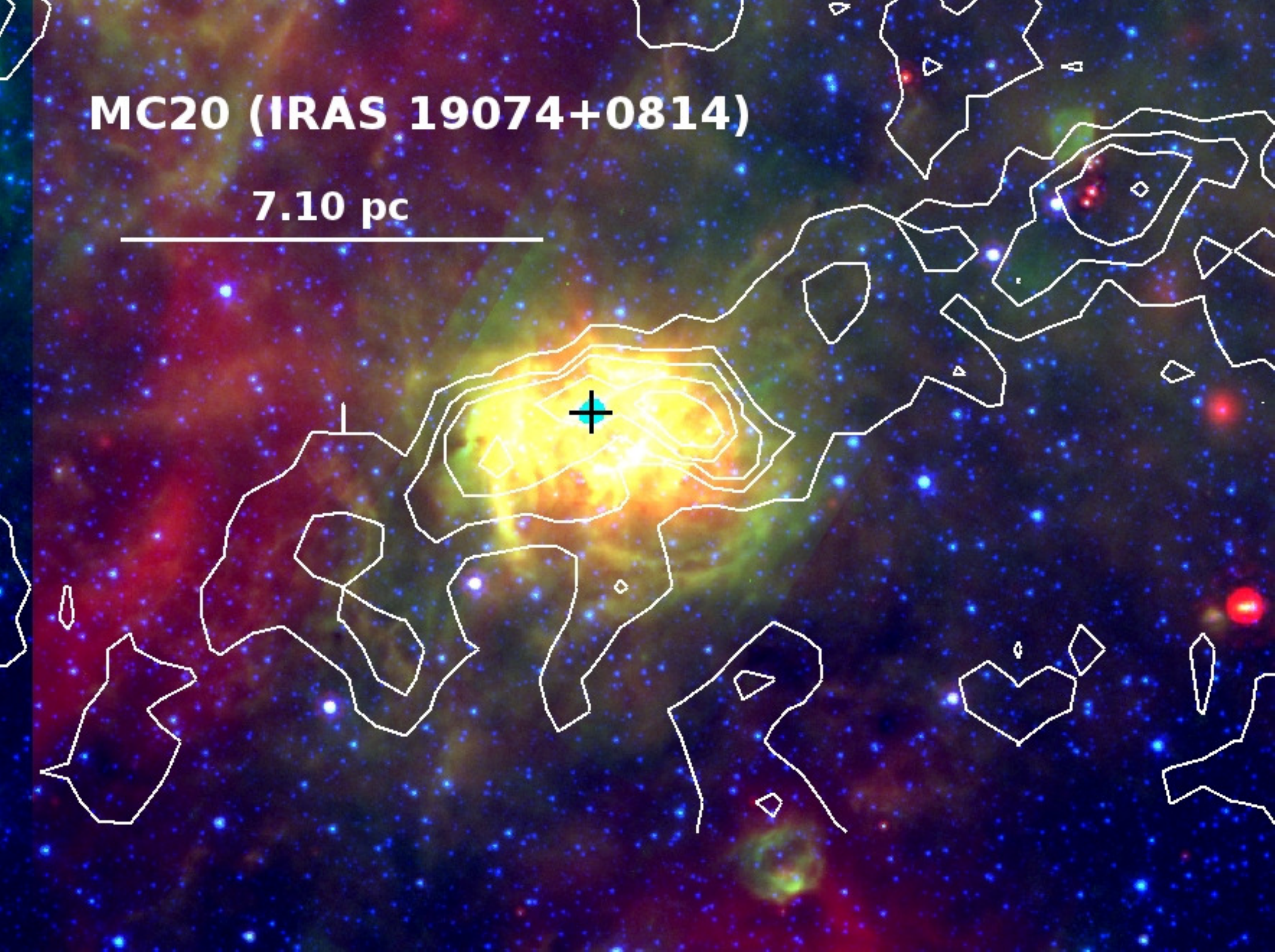}\hspace*{3mm}\includegraphics[width=8.5cm]{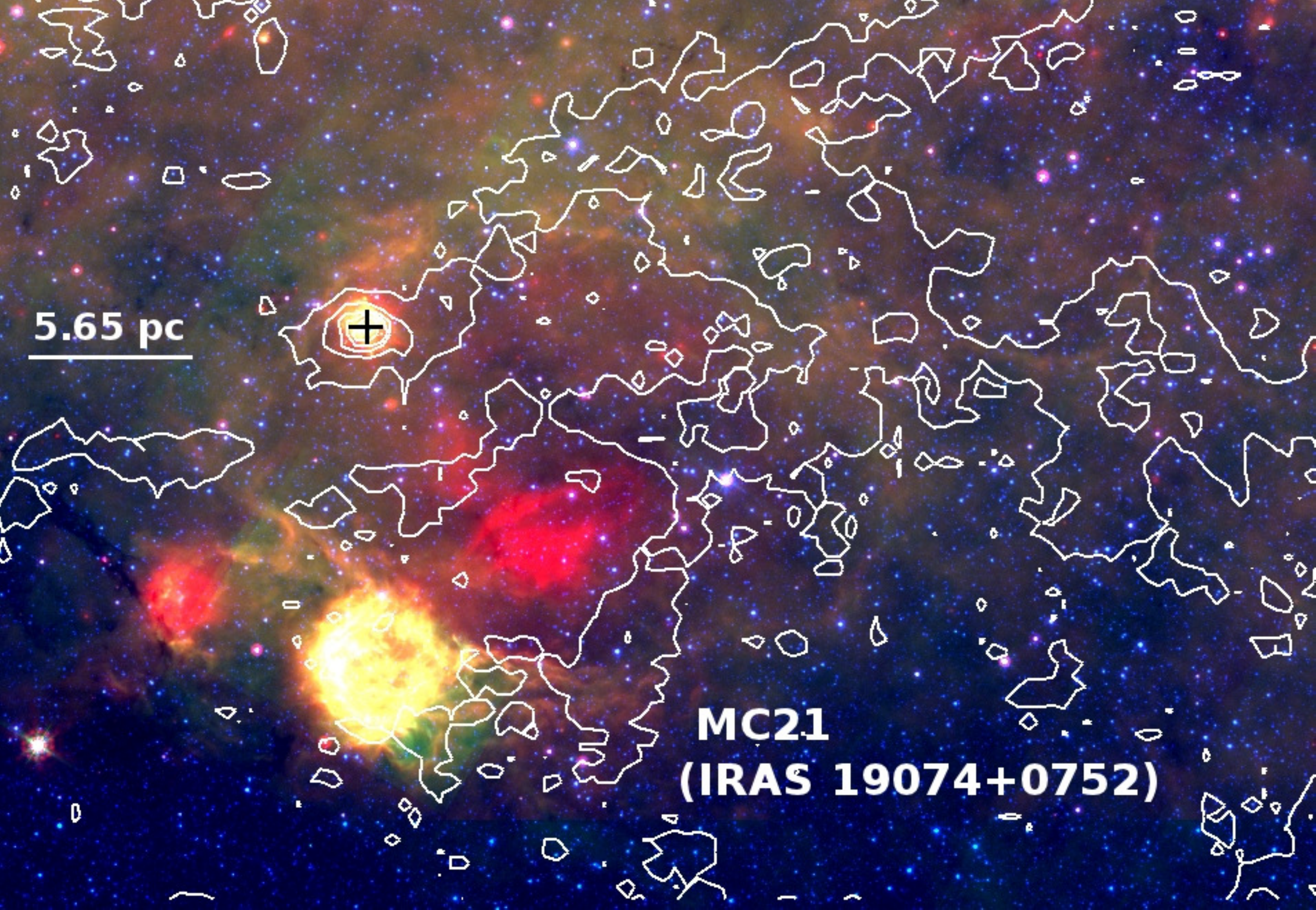}
\caption{Color composite RGB image of our sample of molecular clouds using Spitzer 
\mipslam\ (R), 8~$\mu$m (G) and 3.6~$\mu$m (B) bands. The velocity-integrated \co\ 
column densities are given in contours, with the lowest level corresponding to
N(H$_2$)=$1\times10^{21}$~cm$^{-2}$ ($A_{\rm V}\approx 1.0$~mag) and successive 
levels increasing in steps of $5\times10^{21}$~cm$^{-2}$. The IRAS source position 
is marked with the cross symbol. Scale bar corresponding to 5$^\prime$ is
shown as a horizontal bar.}
\end{figure*}


\begin{figure*}
\vspace*{2mm}
\includegraphics[width=8.5cm]{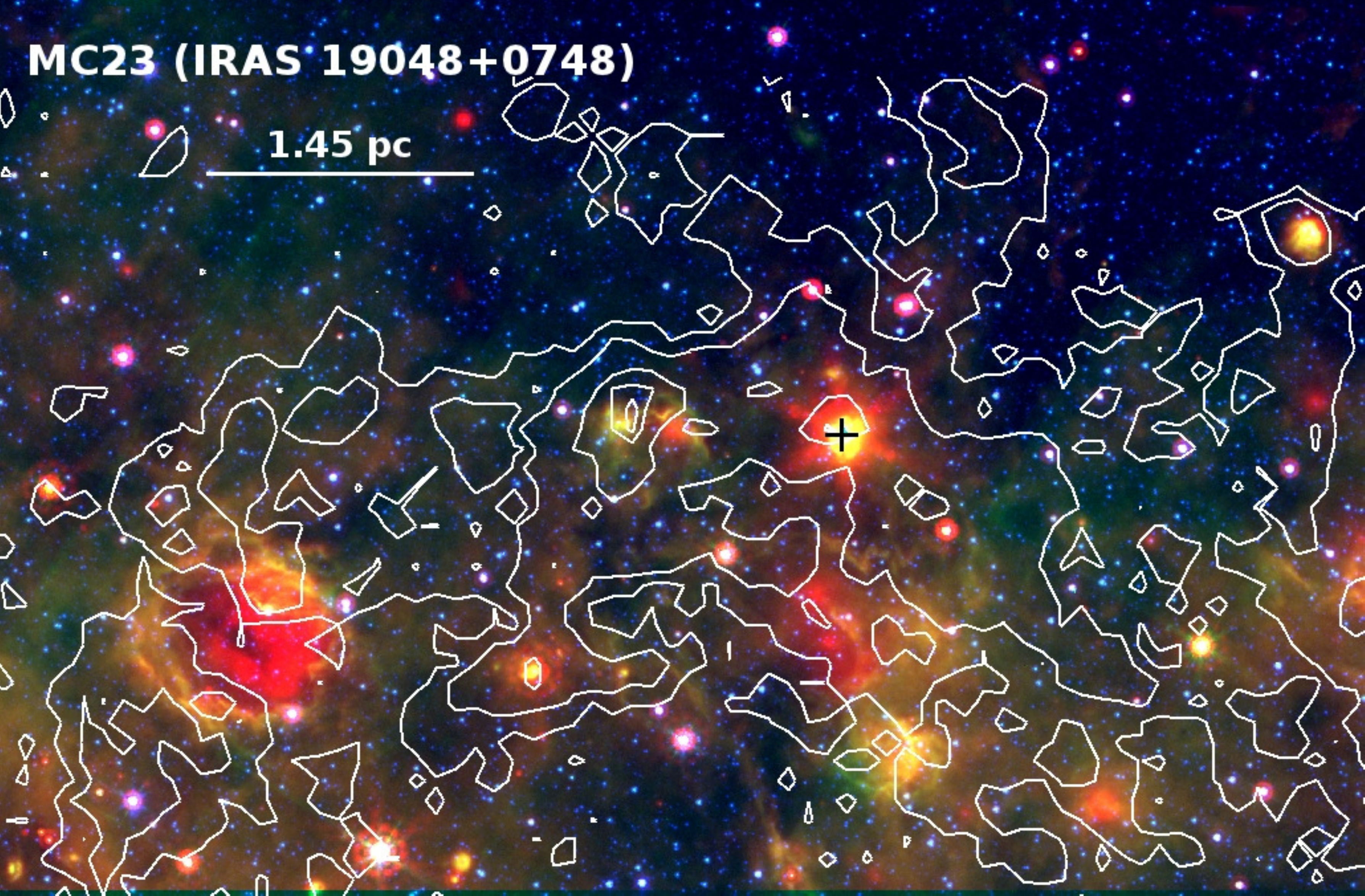}\hspace*{3mm}\includegraphics[width=8.5cm]{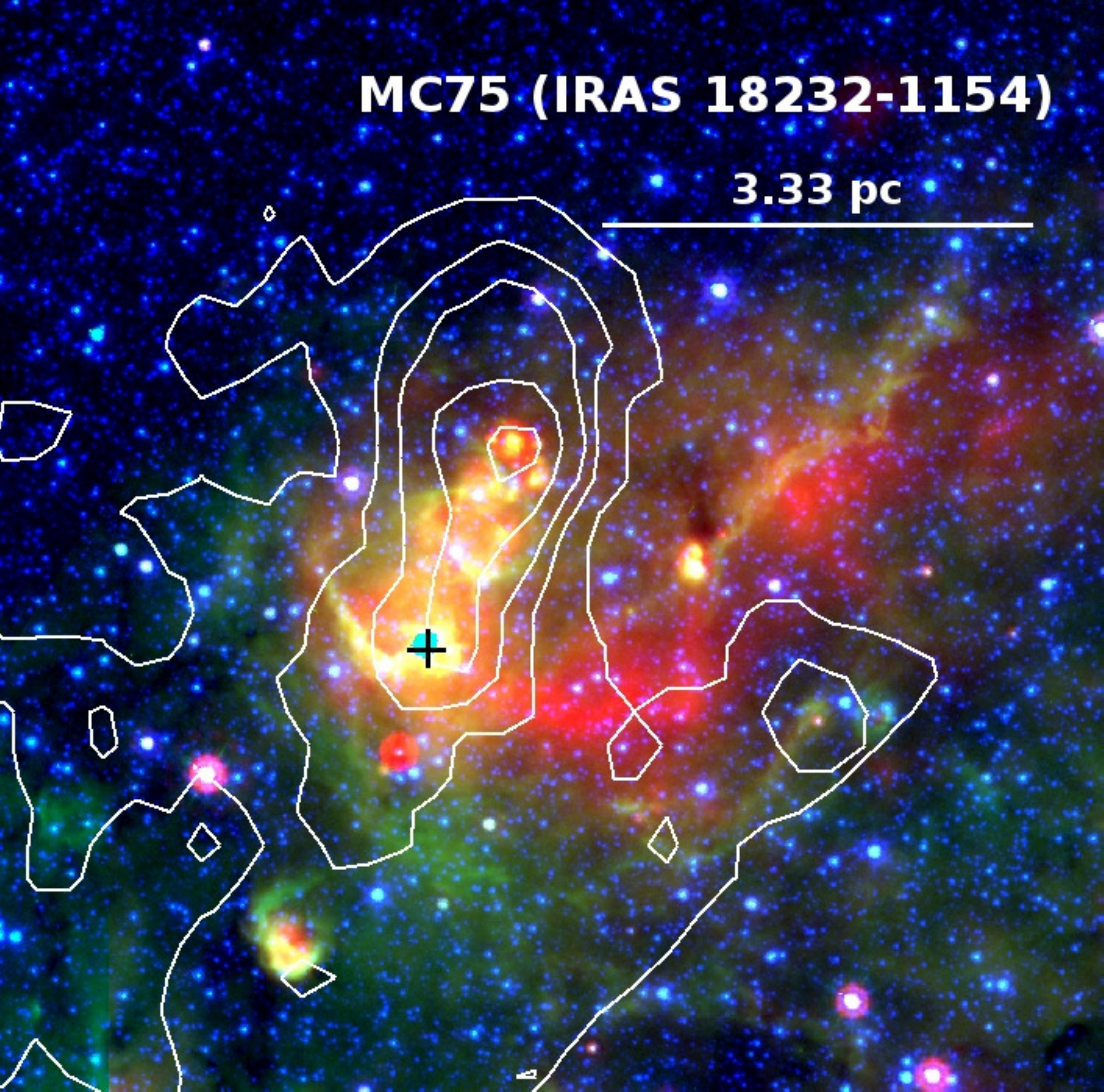}
\vspace*{2mm}
\includegraphics[width=8.5cm]{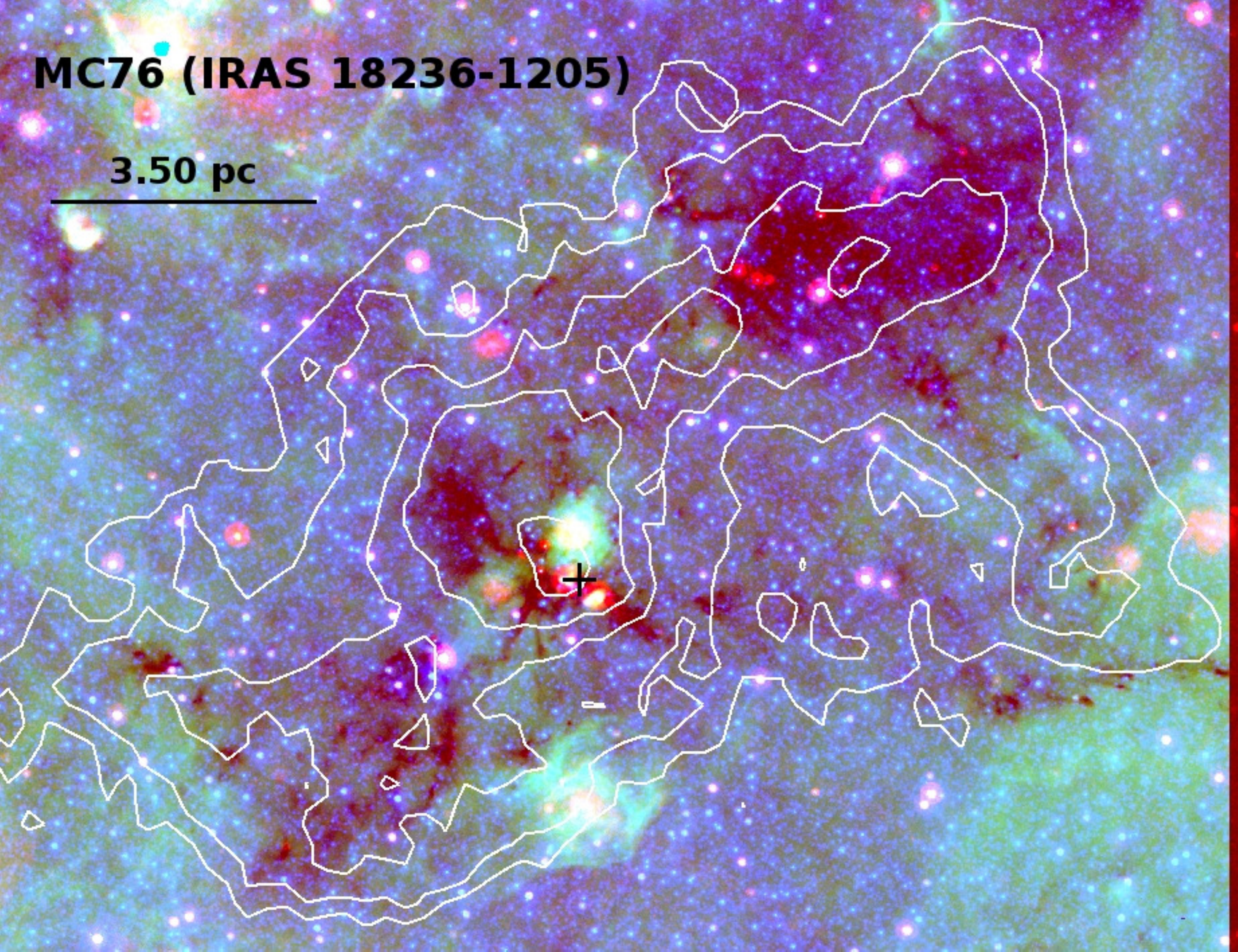}\hspace*{3mm}\includegraphics[width=8.5cm]{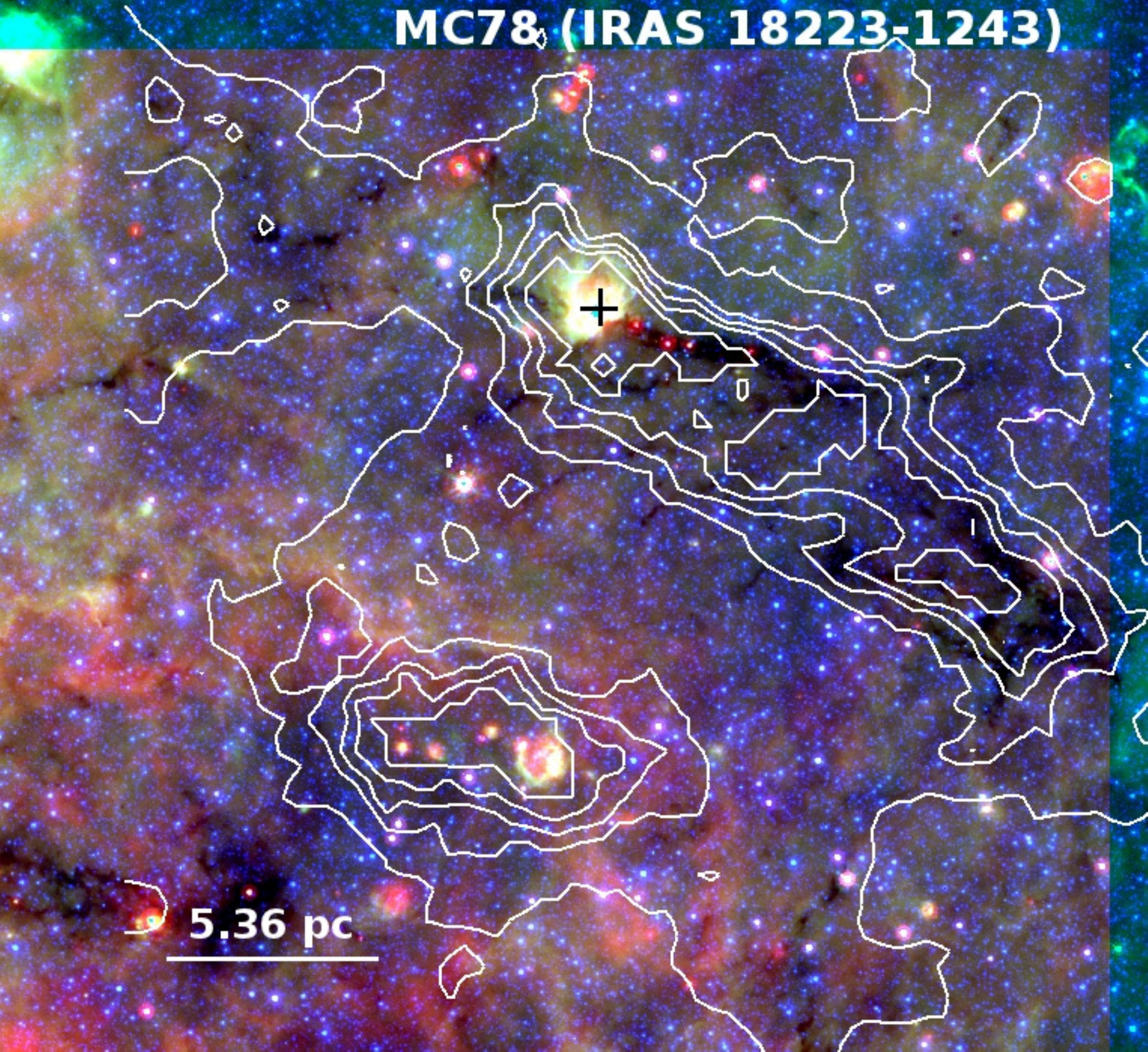}
\includegraphics[width=8.5cm]{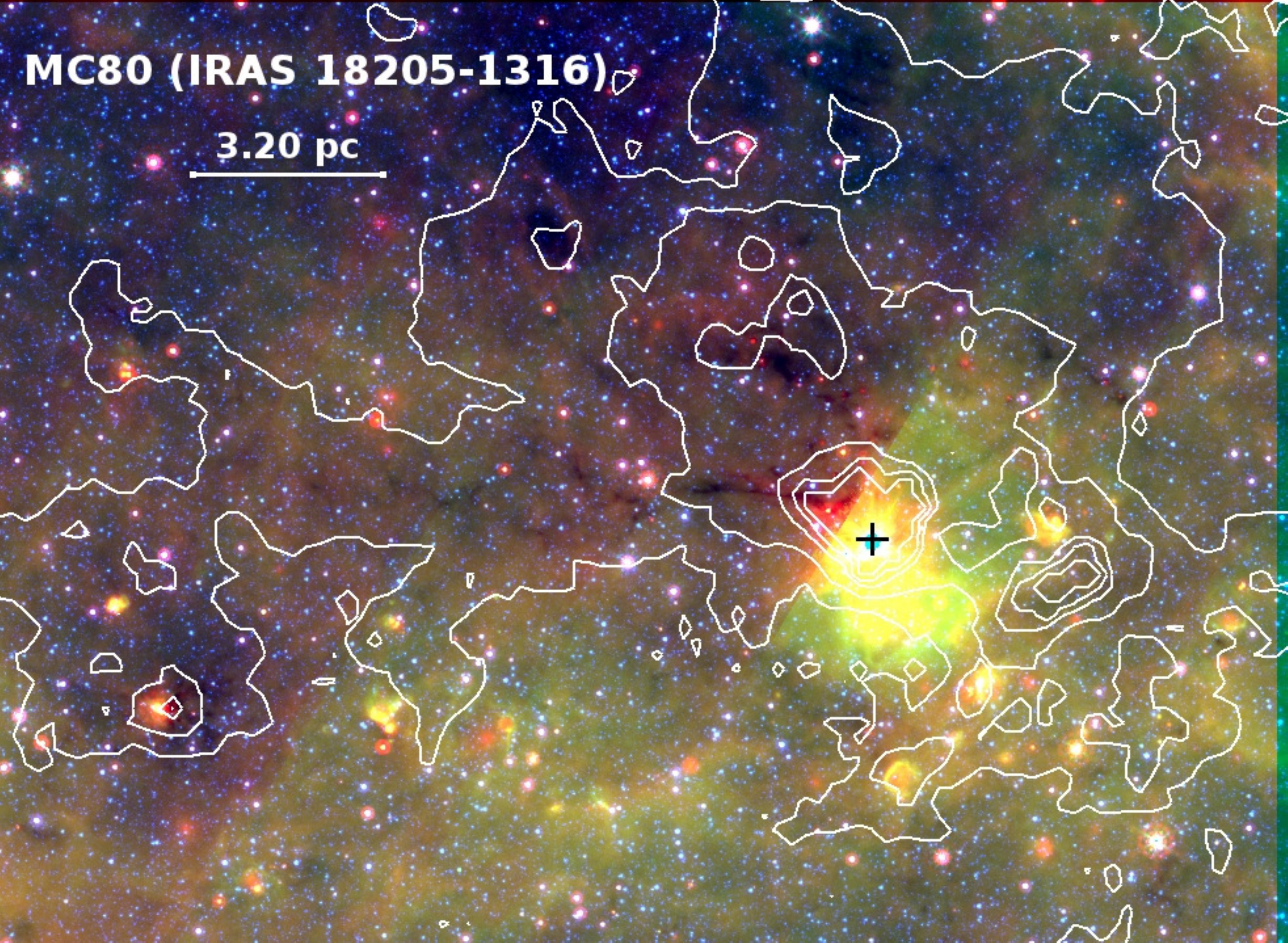}\hspace*{3mm}\includegraphics[width=8.5cm]{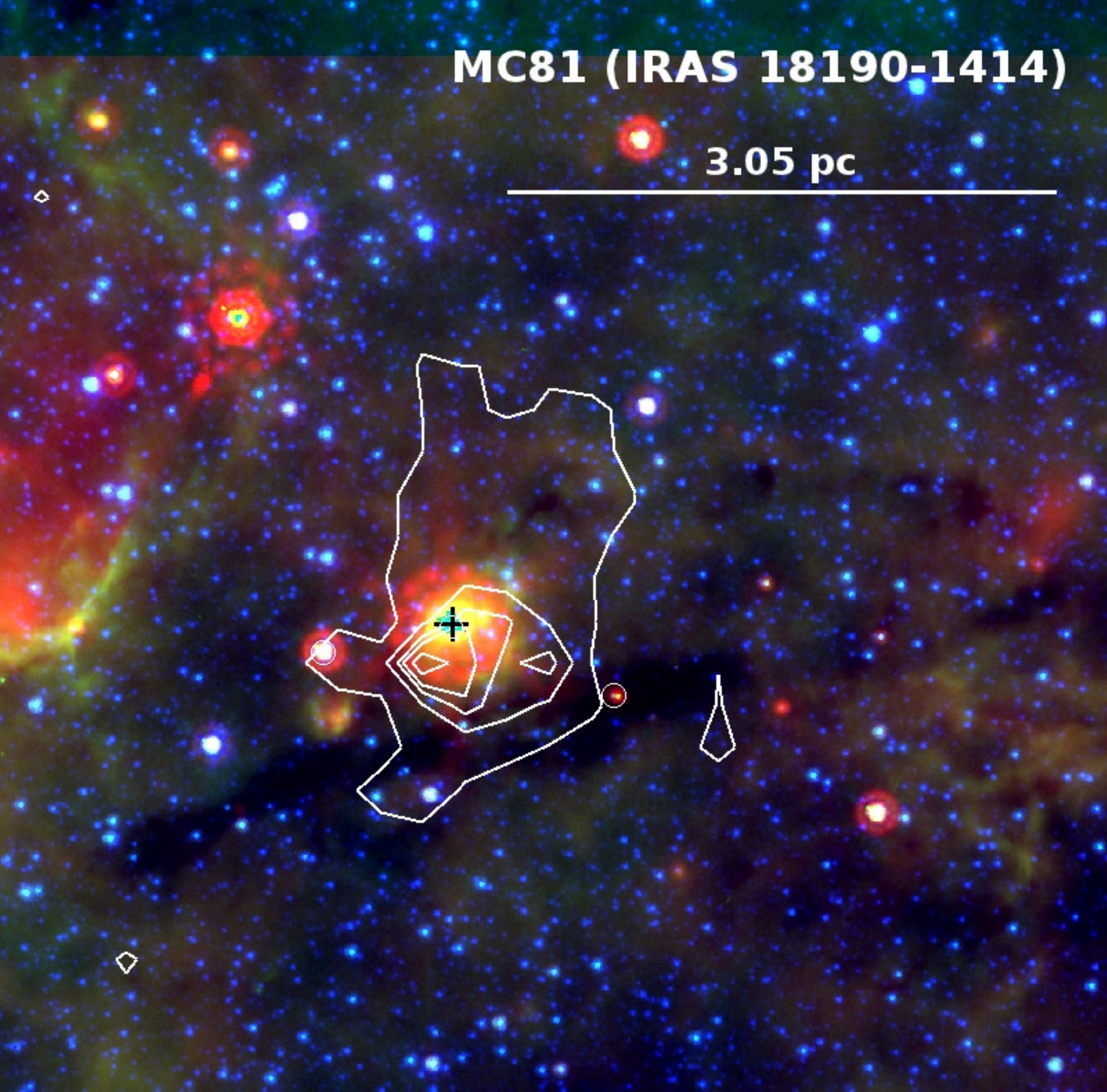}\\
Figure 1 continued. 
\label{figure1}
\end{figure*}

\begin{figure*}[!t]
\includegraphics[width=8.5cm]{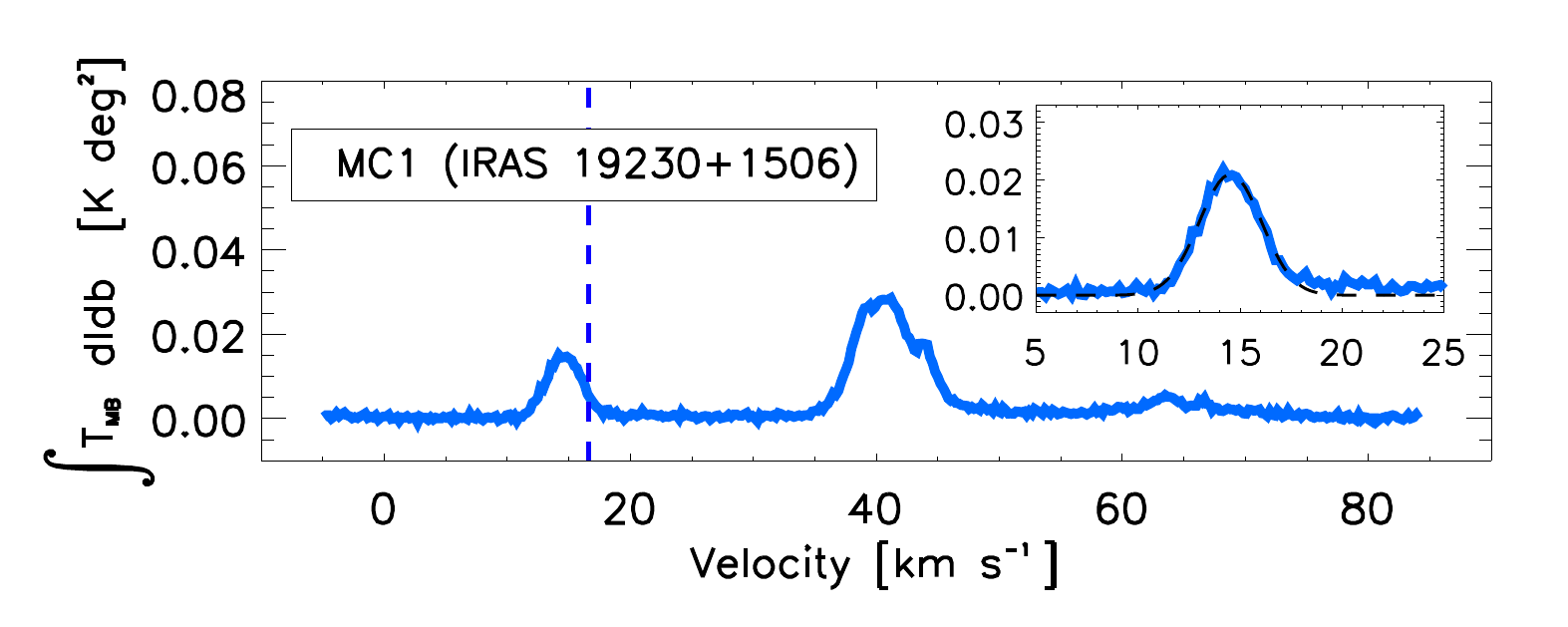}\includegraphics[width=8.5cm]{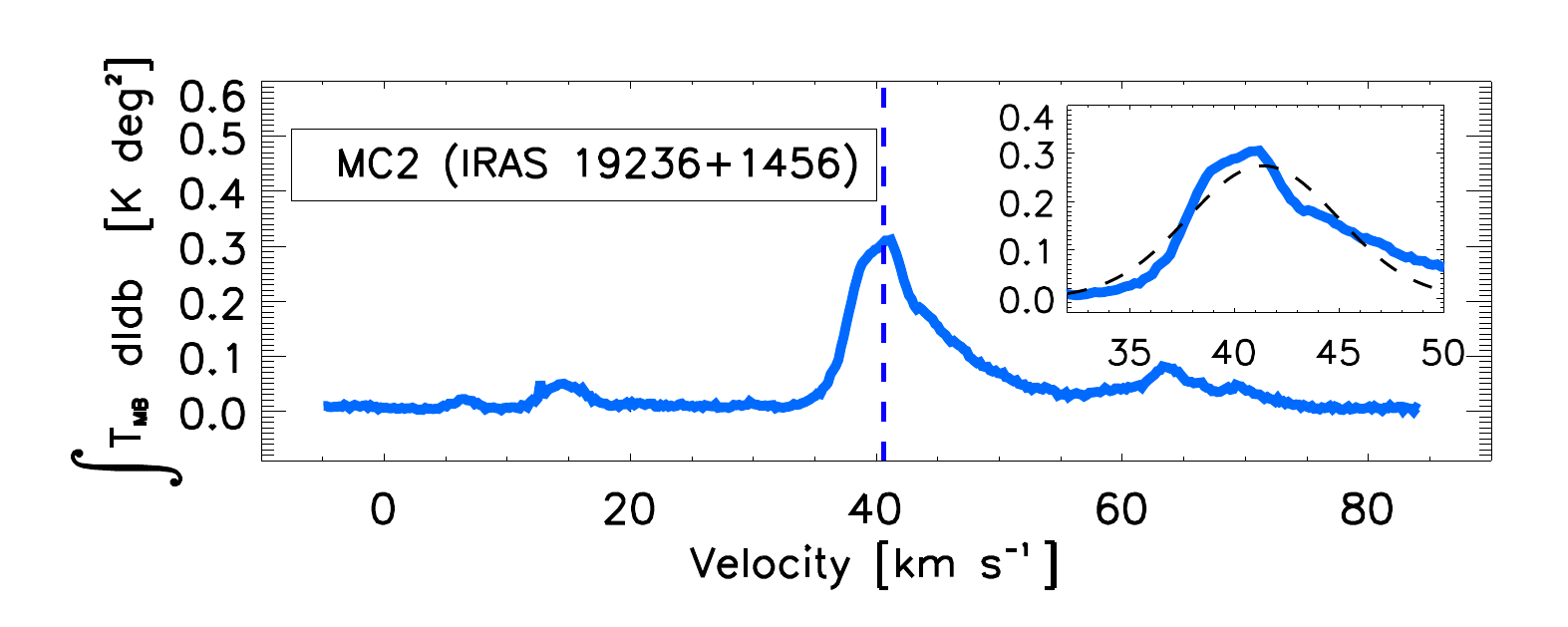}  
\includegraphics[width=8.5cm]{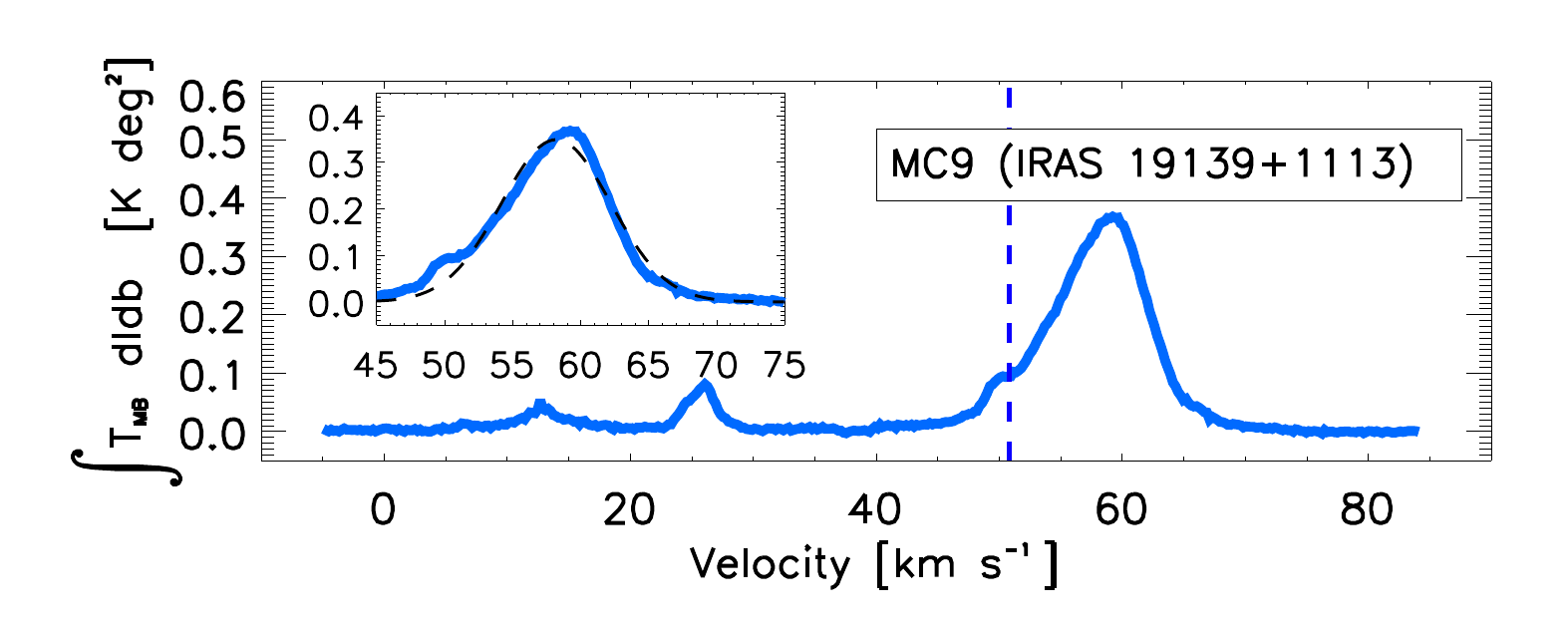}\includegraphics[width=8.5cm]{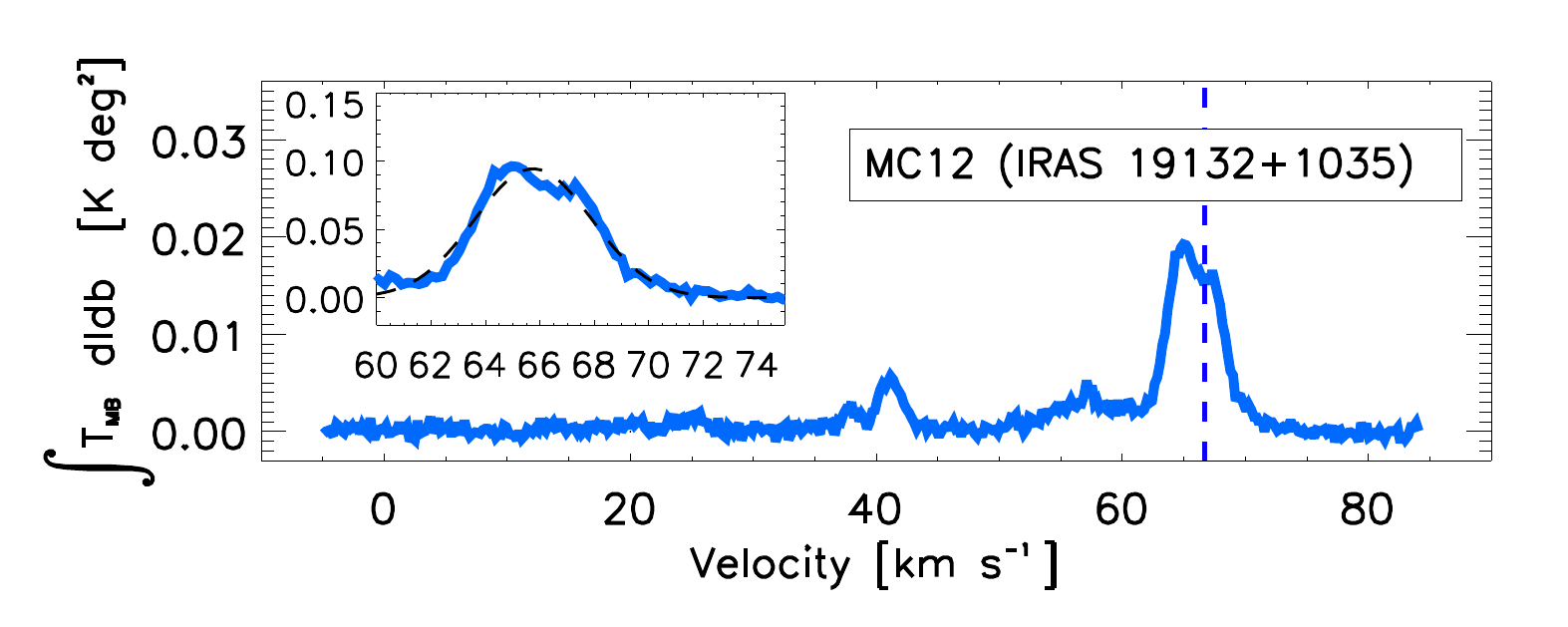}  
\includegraphics[width=8.5cm]{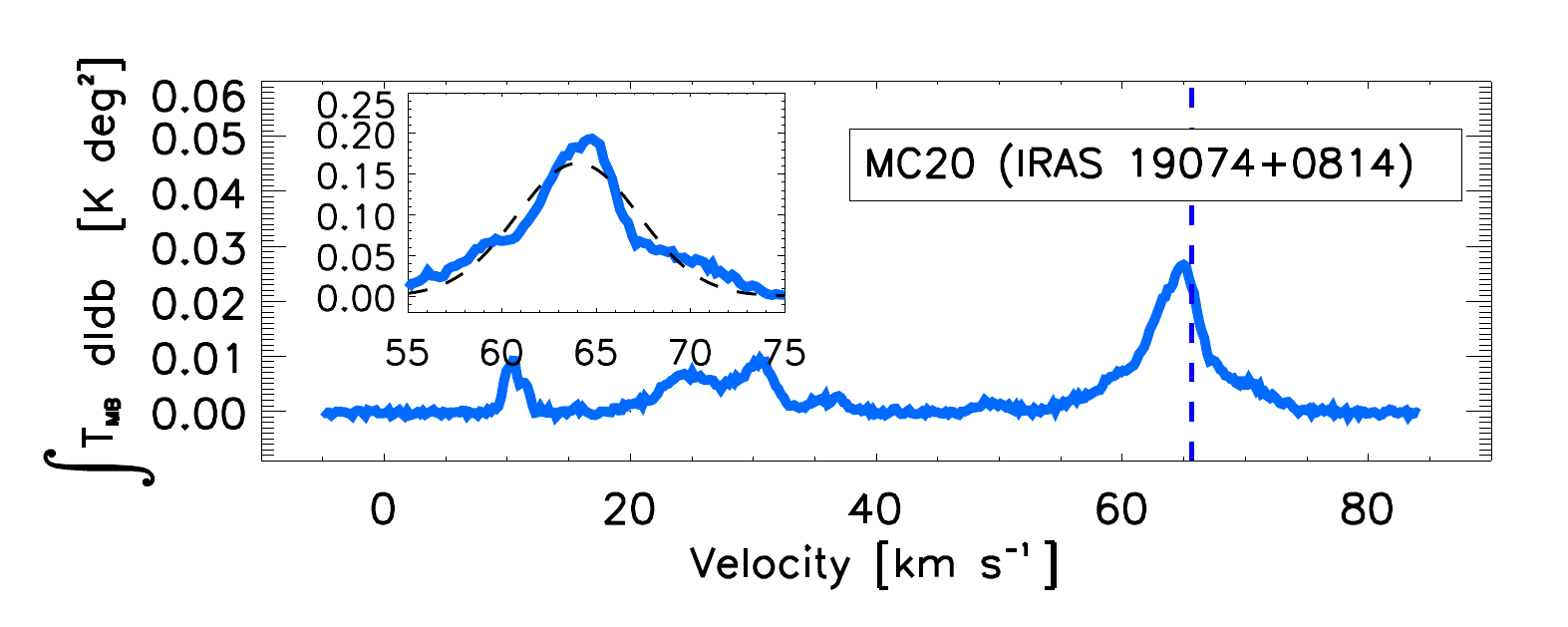}\includegraphics[width=8.5cm]{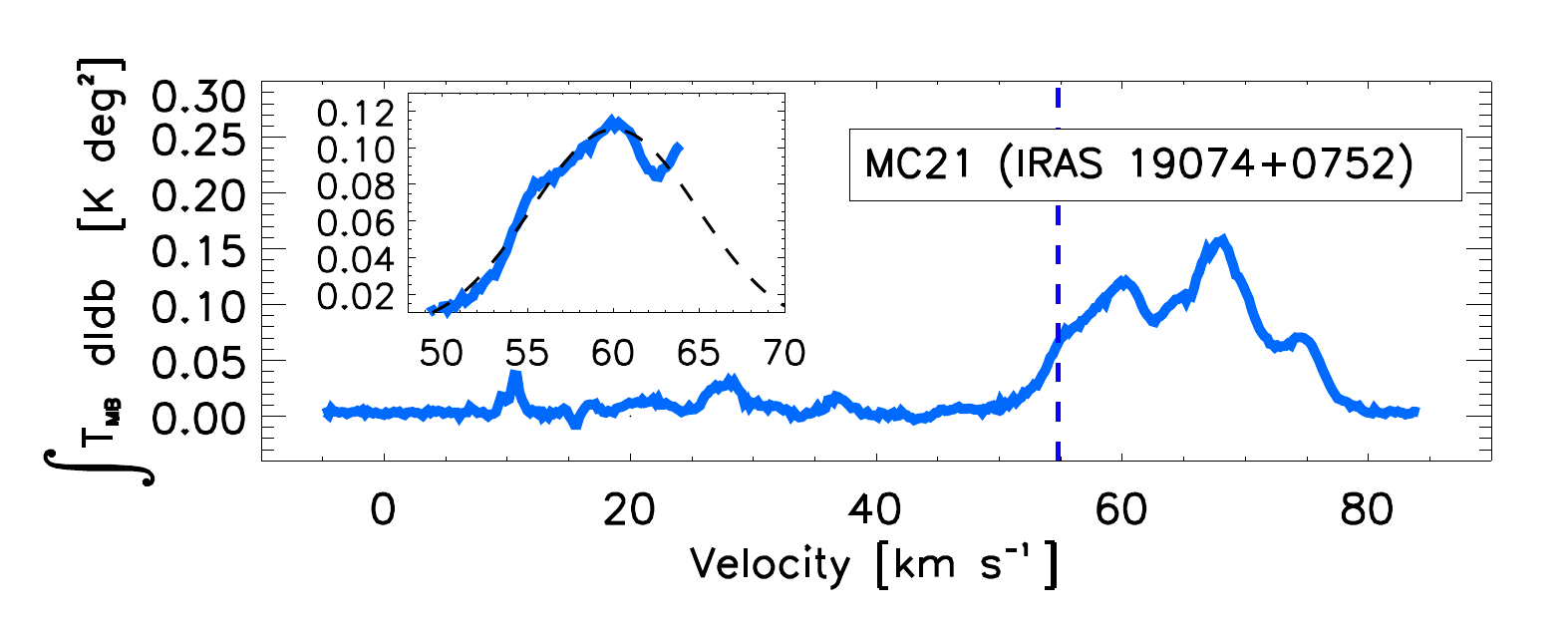} 
\includegraphics[width=8.5cm]{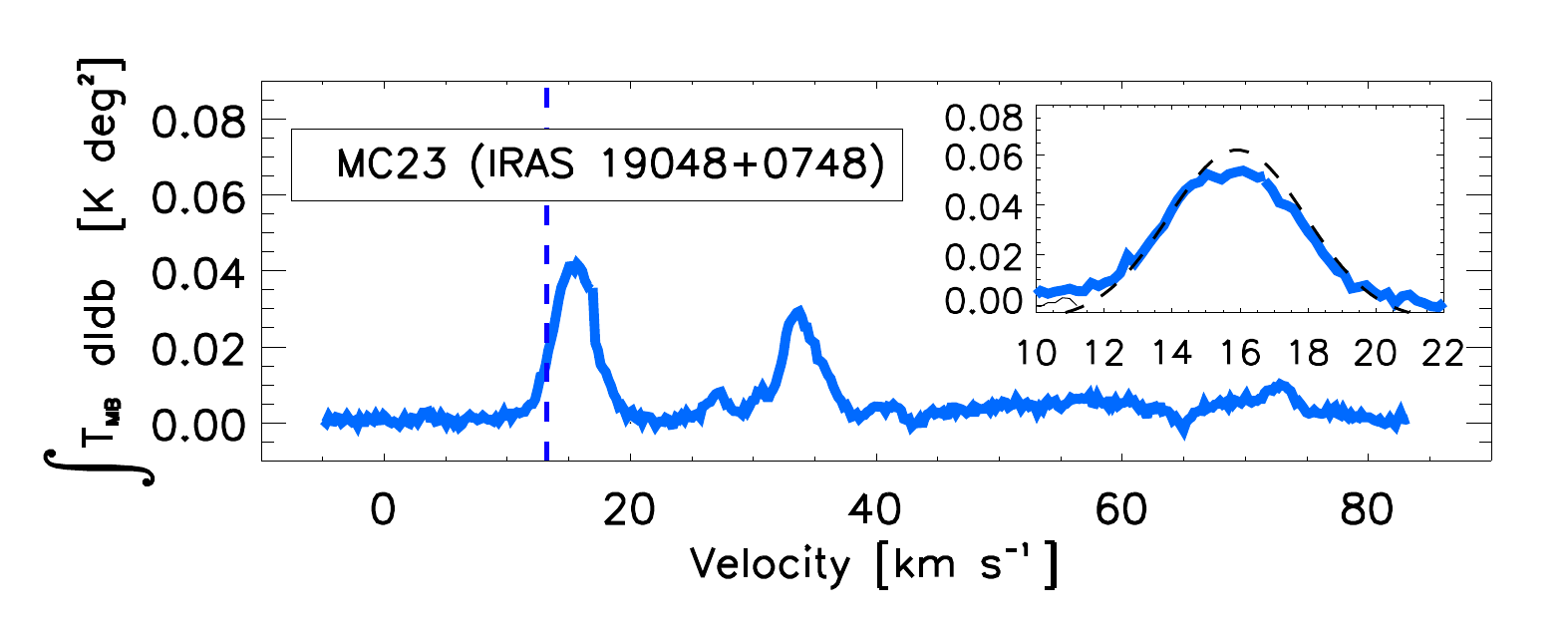}\includegraphics[width=8.5cm]{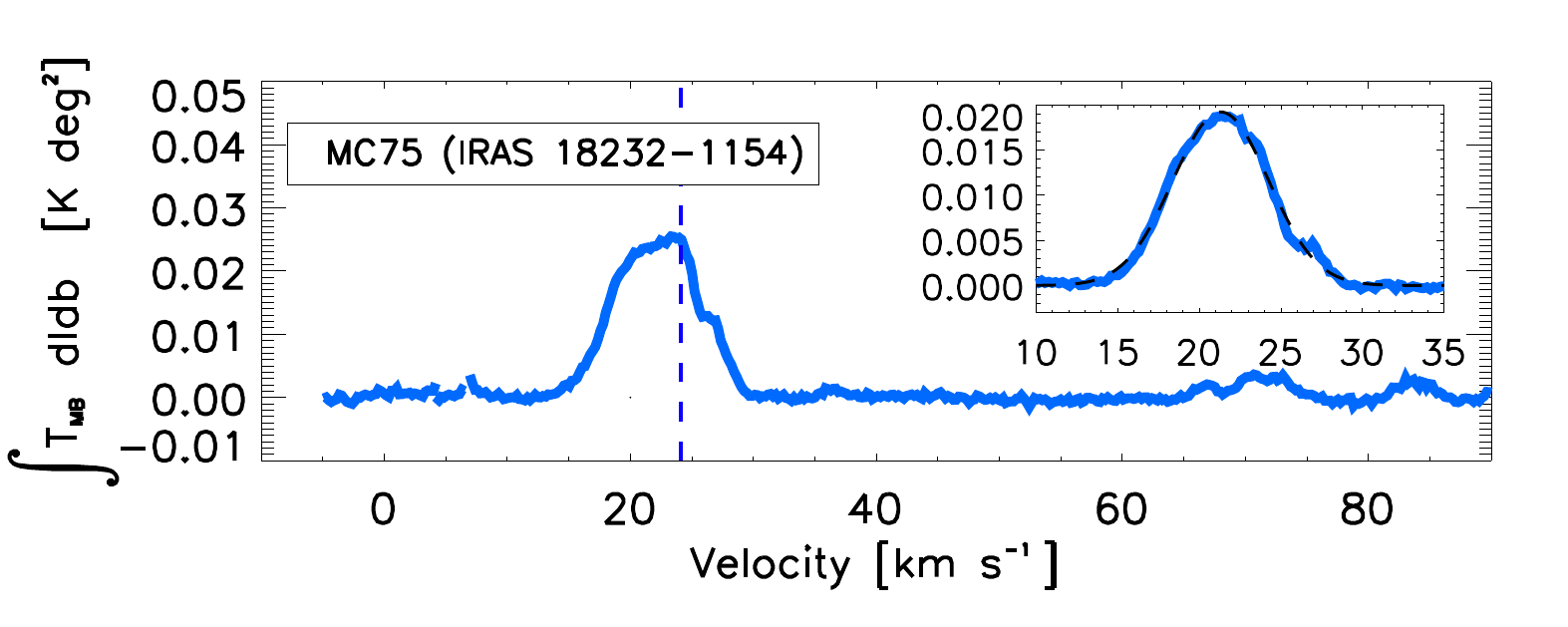} 
\includegraphics[width=8.5cm]{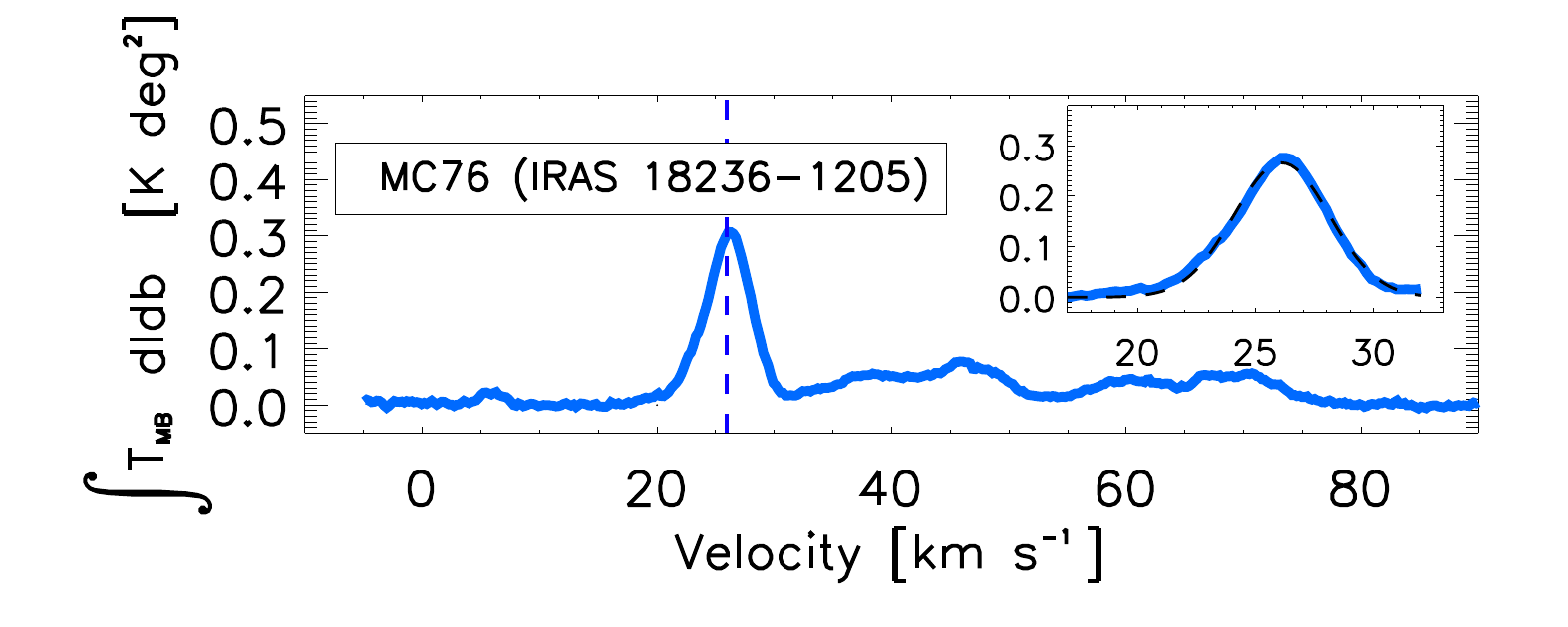}\includegraphics[width=8.5cm]{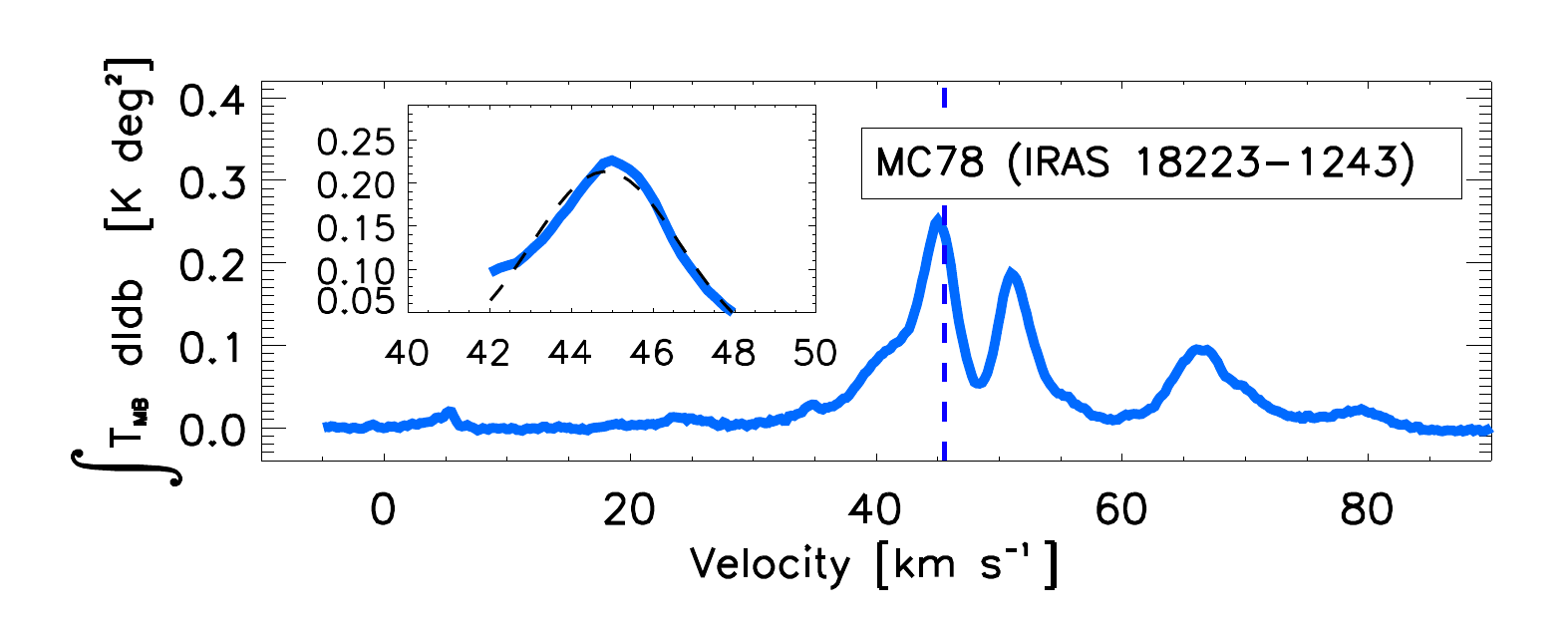} 
\includegraphics[width=8.5cm]{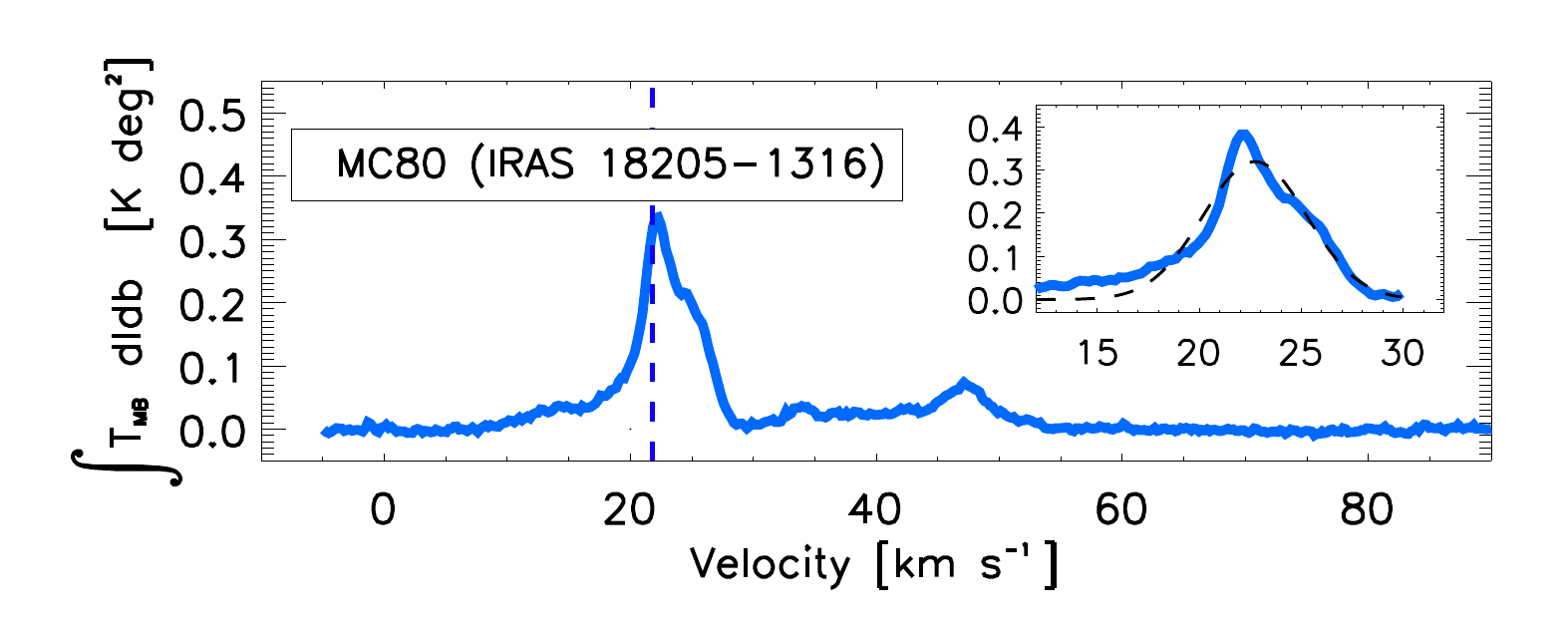}\includegraphics[width=8.5cm]{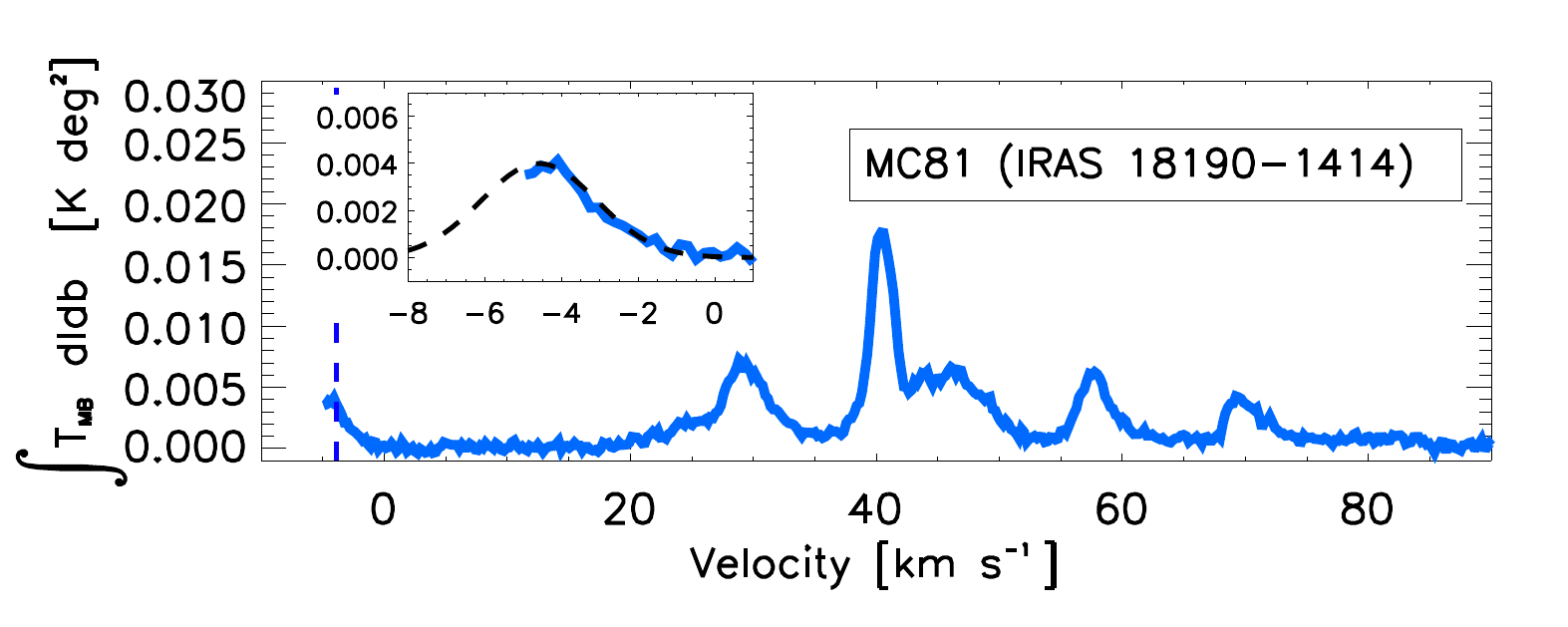}
\caption{ 
Line-Of-Sight $^{13}$CO spectra from the GRS data cube integrated over the Field of 
View (see column 5 in Table~1) of each molecular cloud. 
The vertical dashed line corresponds to the CS velocity from \citet{bronfman+96}. 
The Gaussian fit to the $^{13}$CO profile is shown in the inset 
with a dashed black line. The $^{13}$CO peak velocity and line width values obtained 
from the fit are listed in the column 12 in the Table~1.}

\label{figure2}
\end{figure*}

\begin{table*}[!h!t]
\begin{center}
\centering 
\caption{General properties of MCs associated to IRAS sources with UCHII colors. }
\begin{tabular}{lccrcrrcrrcc}
\hline
\hline
  &    &   &   &      &   &  &  &  &  &  &\\
GMC  & IRAS name & Lon & Lat & Field size & L$_{\rm IRAS}$ & $V_{\rm LSR}^{\rm CS}$ ($\rm \Delta V$) &D  &Scale  & Area & $^{13}$CO Beam      &  $V_{\rm 13CO}^{\rm peak}$ ($\rm \Delta V$)  \\
  &  & [$^\circ$]   & [$^\circ$] &[$^\prime\times^\prime$]  & [$10^3$L$_{\odot}$]   & [km\,s$^{-1}$]   & [kpc]  &  [pc/$\prime$]      &[pc$^2$]   &  [pc]   &  [km\,s$^{-1}$] \\ 
(1) & (2) & (3) & (4)  & (5) & (6)  & (7) & (8) & (9) & (10) & (11) & (12)\\
\hline
MC1  & 19230$+$1506  & 50.28 & $-$0.39  & 15$\times$15 & 4.2 (1.2)  &  16.6 (3.7)  &  1.3 (0.2)  & 0.38 & 23  & 0.29 &  14.5 (3.6)  \\
MC2  & 19236$+$1456  & 50.22 & $-$0.61  & 30$\times$36 & 6.1 (1.8)  &  40.6 (4.5)  &  3.4 (0.4)  & 0.98 & 290 & 0.75 &  42.0 (8.6)  \\
MC9  & 19139$+$1113  & 45.82 & $-$0.28  & 30$\times$24 & 40.3 (13.5) &  50.8 (4.9)  &  4.8 (0.6)  & 1.38 & 59  & 0.29 &  58.2 (9.5)  \\
MC12 & 19132$+$1035  & 45.19 & $-$0.44  & 18$\times$12 & 21.1 (6.8) &  66.7 (2.2)  &  5.4 (0.6)  & 1.57 & 92  & 1.20 &  66.0 (5.1)  \\
MC20 & 19074$+$0814  & 42.43 & $-$0.26  & 15$\times$9  & 65.1 (20.3) &  65.6 (5.0)  &  4.9 (0.6)  & 1.42 & 57  & 1.09 &  65.0 (7.8)  \\
MC21 & 19074$+$0752  & 42.11 & $-$0.44  & 36$\times$27 & 15.6 (4.6) &  54.8 (3.0)  &   3.9 (0.5) & 1.13 & 386 & 0.87 &  60.0 (10.7)  \\
MC23 & 19048$+$0748  & 41.75 &    0.09  & 27$\times$12 & 0.7 (0.2)  &  13.2 (2.4)  &   1.0 (0.1) & 0.29 & 17  & 0.22 &  16.5 (4.7)  \\
MC75 & 18232$-$1154  & 19.49 &    0.15  & 9$\times$12  & 16.8 (5.2) &  24.1 (4.0)  &  2.3 (0.3)  & 0.67 & 55  & 0.51 &  21.5 (8.0)  \\
MC76 & 18236$-$1205  & 19.36 & $-$0.02  & 25$\times$12 &  7.0 (2.1) &  25.9 (7.6)  &  2.5 (0.3)  & 0.70 &  82 & 0.54 &  26.0 (4.8)  \\
MC78 & 18223$-$1243  & 18.66 & $-$0.06  & 27$\times$24 & 22.2 (6.8) &  45.5 (2.9)  &  3.7 (0.4)  & 1.07 & 262 & 0.82 &  45.0 (3.4)  \\
MC80 & 18205$-$1316  & 17.96 & 0.08     & 36$\times$24 & 1.3 (0.4)  &  21.8 (2.8)  &   2.2 (0.3) & 0.64 & 171 & 0.49 &  23.0 (5.7)  \\
MC81 & 18190$-$1414   & 16.94 & $-$0.07 & 6$\times$9   & 5.4 (1.9)  & $-$3.9 (2.7) &  2.1 (0.3)  & 0.61 & 15  & 0.47 &  $-$4.1 (3.5)  \\
\hline
\end{tabular}
\tablecomments{ 
Brief explanation of columns: 
(1) The name of the molecular cloud containing the IRAS source;
(2) Name of the IRAS source; 
(3--4) Galactic longitude and latitude in degree; 
(5) Approximate angular size of the clouds; 
(6) Bolometric luminosity and their error based on distance of the IRAS source 
using the flux from the IRAS-PSC (http://irsa.ipac.caltech.edu/Missions/iras.html);
(7) Velocity with respect to the Local Standard of Rest of the CS source and their 
$\rm \Delta V$ from Bronfman et al. (1996); 
(8) Kinematical distance (with a percent error of 12\%) to the dense clump associate to the IRAS source from \citet{faundez+04} ;
(9) Physical scale in parsec for an angular scale of $1^\prime$;
(10) Physical area of the cloud above a column density of gas equivalent to $A_V=1$~mag;
(11) Physical size of the $46^{\prime\prime}~^{13}$CO beam at the distance of the cloud; 
(12) Peak velocity of the $^{13}$CO and their line width obtained in this work.
}
\end{center}
\end{table*}

\subsection{Definition of the MC associated to an IRAS source}

In order to define the parent molecular cloud that harbors the
high-mass star-forming regions, we used $^{13}$CO(J=1-0)
emission data from the Galactic Ring Survey (GRS) database
\citep{jackson+06}. The survey data
have velocity resolution of 0.21~km~s$^{-1}$, a typical (1$\sigma$) rms
sensitivity of $\sim$0.13~K, a main beam efficiency $\eta_{\rm mb}$=0.48, 
and a beam of 46\arcsec\ \citep{jackson+06}.
The \co\ emission spectra for the Line of Sight (LOS) to the
selected IRAS sources are shown in Figure~\ref{figure2}, where the observed velocity
of the CS(J=2-1) emission line \citep{bronfman+96} is marked with a 
dashed vertical line. 
In the inset we show the results of a Gaussian fit to the observed \co\ profile.
The best fitting value of the central velocity and the \co\ line width (Full Width at Half
Maximum - FWHM or $\rm\Delta$V) are given in the last column of Table~1.
In majority of the cases (10/12), there is a peak in $^{13}$CO spectra 
within $\lesssim$3~\kms\ with respect to the CS velocity, whereas in 
the two clouds (MC9 and MC21) the difference is $\sim$7~\kms. Nevertheless 
in all clouds the difference between the $^{13}$CO and the CS velocities 
is less than the FWHM of the fitted $^{13}$CO profile. 
This association guarantees the coexistence of dense cores 
traced by the CS line with the dense molecular structures traced by the \co\ line. 
All clouds are located in the 1$^{\rm st}$ quadrant, where the foreground 
molecular emission, if present, would have produced a prominent ($>3\sigma$) 
molecular component to the left of the CS velocities in the plotted spectra. 
We could verify the absence of such a prominent component in eleven out of our 12 clouds,
thus ensuring that the selected cloud is the nearest cloud along the LOS.
The exception is MC81, whose CS velocity lies at the lower extreme of the velocity
range covered by the GRS, which prevents us to infer the presence/absence of
foreground molecular clouds. In this particular case, we used the $^{12}$CO 
emission profile (See \citep[Figure~1 and Figure~\ref{figure2} in][ respectively]{clemens+86,dame+01}
to ensure the absence of foreground molecular clouds along the LOS.
For each MC, the integrated \co\ emission map was created by summing at each pixel
all channels that have a velocity within 15~\kms\ of the $\rm V^{^{13}{\rm CO}}_{\rm peak}$,
and intensities $>3\sigma$.

The resulting column density map is shown by contours superposed on the 
RGB image in Figure~\ref{figure1}.  
Only the portion of the map that has a molecular gas column density above
\nhtwo$\approx 1\times10^{21}$~\cmsq\ (the lowest plotted contour in Fig.~1)
is considered part of the MC associated to the \starf~region. This limiting value 
corresponds to $\rm A_{\rm V}\sim1$~mag, which is the value used in the 
literature as a physical threshold value to define MCs \citep{bolatto+13}. 
We complemented our \co\ maps with the $^{12}$CO emission integrated maps 
of \citet{sanders+86} to obtain \nhtwo, and thereby, the mass of the clouds.\\

\subsection{Molecular cloud mass estimation}

\begin{table*}[!h!t]
\begin{center}
\centering 
\caption{Physical properties of the sample molecular clouds.}
\begin{tabular}{cccccc}
\hline
\hline
GMC   & ${\rm N_{H_2}}$ & A$_{\rm V}$ &M${\rm_{vir}}$ &  M${\rm _{LTE}}$ & M${\rm_{XF}}$ \\
      & [$\rm 10^{22}cm^{-2}$]  & [mag]  &[$10^4\,M_{\odot}$]  & [$10^4M_{\odot}$] &[$10^4M_{\odot}$]  \\
 (1) & (2) & (3) & (4) & (5) & (6) \\
\hline
 MC1  & 1.73$\pm$0.77  & 18.40$\pm$8.28 & 0.44$\pm$0.38 &  0.33$\pm$0.17 & 0.48$\pm$0.26  \\
 MC2  & 1.28$\pm$0.56  & 13.62$\pm$6.13 & 8.87$\pm$7.69 &  3.07$\pm$1.55 & 4.50$\pm$2.28  \\
 MC9  & 5.32$\pm$2.39  & 56.61$\pm$25.46 & 4.89$\pm$4.21 &  2.56$\pm$1.10 & 3.75$\pm$1.89  \\
 MC12 & 2.95$\pm$1.32  & 31.38$\pm$14.12 & 1.75$\pm$1.49 &  2.24$\pm$1.12 & 3.28$\pm$1.64  \\
 MC20 & 5.33$\pm$2.39  & 56.70$\pm$25.51 & 1.24$\pm$2.82 &  2.44$\pm$1.20 & 3.58$\pm$1.78  \\
 MC21 & 1.10$\pm$0.46  & 11.70$\pm$5.26 & 16.06$\pm$13.65 & 3.52$\pm$1.59 & 5.17$\pm$2.34  \\
 MC23 & 2.64$\pm$1.18  & 28.08$\pm$12.64 & 0.60$\pm$0.53 &  0.30$\pm$0.22 & 0.44$\pm$0.32  \\
 MC75 & 5.05$\pm$4.31  & 53.72$\pm$24.16 & 3.37$\pm$2.90 &  2.32$\pm$1.20 & 3.39$\pm$1.76  \\
 MC76 & 4.65$\pm$2.09  & 49.46$\pm$22.26 & 1.48$\pm$1.30 &  3.14$\pm$1.74 & 4.61$\pm$2.54   \\
 MC78 & 6.28$\pm$2.83  & 66.80$\pm$30.06 & 1.34$\pm$1.13 & 13.52$\pm$6.46 &19.84$\pm$9.48  \\
 MC80 & 6.72$\pm$3.02  & 71.49$\pm$32.17 & 3.04$\pm$2.61 &  9.57$\pm$5.03 & 14.04$\pm$7.38  \\
 MC81 & 1.80$\pm$0.81  & 19.19$\pm$8.62 & 0.35$\pm$0.28 &  0.23$\pm$0.11 & 0.34$\pm$0.20  \\
\hline
\end{tabular}
\tablecomments{ 
Brief explanation of columns:
(1) Molecular cloud name; 
(2) Molecular hydrogen column density determined from the $^{13}$CO line;
(3) Visual extinction in mag equivalent to the derived gas column density;
(4) Cloud mass calculated using Virial equilibrium (see eq.~1);
(5) Cloud mass calculated using Local Thermodynamical Equilibrium (see eq.~2);
(6) Cloud mass calculated using the X-factor (see eq.~4);
} 
\end{center}
\end{table*}

We computed the mass of the clouds of the sample using 3 independent 
methods: (a) the Virial theorem, (b) the Local Thermodynamical Equilibrium 
(LTE) approximation and (c) a constant conversion factor between CO and 
H$\rm_{2}$ column density or an X-Factor. To calculate the Virial mass 
(M$\rm_{\rm vir}$) of the MC sample, we use the expression \citep{may+97},
\begin{equation}
\rm
\left[ \frac{M_{\rm vir}}{10^4~M_{\odot}} \right]= 1.58\times \left[ \frac{R_e}{5~pc} \right] \times
\left[ \frac{\Delta V}{5~km\,s^{-1}} \right]^{2},
\end{equation}
where $\rm\Delta$V$^{13}$ is the \co\ line width (Full Width at Half 
Maximum - FWHM) obtained from a Gaussian fit to the observed spectra, 
$\rm R_{e}$ is the effective radius obtained from 
$\rm R_{e}=\sqrt{A/\pi}$, where $\rm A$ is the area of the cloud. 
The expression assumes spherical symmetry in density ($\rho$), with 
$\rm \rho\propto~r^{-2}$ \citep{maclaren+88}.

The LTE mass (M$\rm_{LTE}$) is obtained from 
the \co\ column density using the expression \citep{simon+01}:

\begin{equation}
\rm
\left[ \frac{M_{\rm LTE}}{10^4~M_{\odot}} \right] = 3.25\times \left[ \frac{R_e}{5~pc} \right]^2 
\times  \left[ \frac{N(^{13}CO)}{10^{17}cm^{-2}}\right],
\end{equation}
where $\rm N(^{13}CO$) is obtained by integrating the \co\ column density at
each pixel. 
To obtain the \co\ column density map, we first obtained an excitation temperature 
for the cloud following the LTE approximation of \citet{dickman78} and using 
the peak $^{12}$CO radiation temperature from the published b-V maps from 
\citet{sanders+86} in the corresponding longitude range, and assumed that 
computed excitation temperature to be the same for all pixels in the \co\ map. 
The coupling between beams and antenna efficiencies of $^{12}$CO and \co\ 
were taken into account.

A third method for the determination of the molecular mass consists in 
using the X-Factor (XF). This factor is defined as the
ratio between the column density of the molecular hydrogen (\nhtwo)  
and the CO luminosity W(${\rm ^{12}CO}$) 
\citep[${\rm XF}\equiv {N(\rm H_2)}/{W({\rm ^{12}CO});}$][]{pineda+08}.
In order to compare with previous results of Galactic studies on 
the Schmidt Law \citep{heiderman+10}, we adopted the value of  
XF=${\rm 2.8\pm0.7\times10^{20}}$ [$\rm cm^{-2}~K^{-1}~km^{-1}~s$] from 
\citep{bloemen+86}. Following \citet{goto+03}, we assumed 
W($\rm^{12}$CO)/W($\rm^{13}$CO)=57, resulting in the following expression for
the column density of \nhtwo:

\begin{equation}
\rm
\left[ \frac {N(H_2)} {cm^{-2}} \right] = 57\times2.8\times10^{20}\times W(^{13}CO).
\end{equation}

The molecular cloud mass (M$_{\rm XF}$) is then calculated using the equation:

\begin{equation}
\rm
\left[\frac{M_{XF}}{10^4~M_{\odot}}\right] = 
\rm 3.26\times\left[\frac{R_e}{5~pc}\right]^2 \times \left[\frac{W(^{13}CO)}{5~K~km~s^{-1}}\right]. 
\end{equation}
All the above equations are re-written from the original references by
normalizing the quantities to their typical values in our MCs.
The computed masses and other physical quantities of the clouds are listed in Table 2. 
The errors on the tabulated physical quantities are calculated by propogating
the errors on $\rm R_e$ due to distance, $\rm ~\Delta V~and~N(^{13}CO)$.

\section{Identification of YSOs associated to the molecular clouds}

We used the sources from GLIMPSE and MIPSGAL programs of the 
Spitzer\footnote{Spitzer Space Mission: {\it https://www.spitzer.caltech.edu}}
Space mission to identify the YSOs in our sample of MCs. 
GLIMPSE sources are directly extracted from the GLIMPSE PSC catalog
\citep{benjamin+03}. We selected only those objects in the PSC that
have detectable emission in all the four GLIMPSE bands.
MIPSGAL PSC \citep{gutermuth+15} was not available at the time when we started this
work. We extracted MIPSGAL sources using the \mipslam~ images\footnote{The
BCD MIPSGAL images are provided by MIPSGAL survey online database,
{\it http://irsa.ipac.caltech.edu/data/SPITZER/MIPSGAL/}} \citep{carey+09}
following the method described in the next subsection. 
The GLIMPSE and MIPSGAL sources are merged to form a complete catalog of all YSOs.
Any GLIMPSE source that is within the beam of FWHM=6\arcsec\ of the \mipslam~image
is considered a genuine counterpart of the \mipslam ~source.
We aim to classify each detected YSO following the IR spectral index 
criterion of \citet{lada87}, see \S3.4.
The sensitivity of the MIPSGAL and GLIMPSE surveys ensures that our catalog
contains all Class I sources of mass $\rm \gtrsim0.3$~\msun, Class II sources 
more massive than $\sim$1~\msun ~and Class III  sources more massive than 
$\sim$5~\msun ~for a cloud at an average distance of 3~kpc.
These detection limits 
are obtained by converting the flux sensitivities (5$\sigma$ corresponding to a point source) of GLIMPSE~\citep{churchwell+06}  
and MIPSGAL\footnote{MIPS Instrument Handbook version 3: {\it https://irsa.ipac.caltech.edu/data/SPITZER/docs/\\
mipsinstrumenthandbook/MIPS-Instrument-Handbook.pdf}} 
to bolometric luminosities at a distance of 3~kpc, and then using the Class-dependent
luminsoity to mass relation described in \S4.2.1.
It may be noted that most of the bolometric luminosity of Class I sources 
in their very early phase is outside the GLIMPSE bands. These sources,
are expected to be the brightest ones at \mipslam. Thus, the addition of MIPSGAL
sources allows us to select the high-mass tail of the mass function, as will
be discussed in \S4.

 \begin{table*}[!h!t]
 \centering 
\caption{Selected MIPSGAL YSOs in our sample of MCs.}
 \begin{tabular}{llll|lll}
 \hline
 \hline
           &\small  MIPSGAL       &\small  Selected  &\small  Selected       & NC                    &\small MS 2MASS  &\small HLYSOs\\
GMC  &\small Sources  &\small  Sources   &\small  Sources [\%] &\small  Sources &\small Sources      &\small  Sources\\
 (1)     &  (2)     &  (3)   &  (4)         &  (5)   &  (6)  & (7) \\
\hline 
 MC1     & 16    & 16  & 100         & \ldots  & \ldots &16\\
 MC2     & 179 & 170 & 95 (5)     & 25       &9&124\\
 MC9     & 172  & 141& 82 (18)   & 73       &31&117\\
 MC12   & 23    &  7    & 30 (70)   & \ldots  &\ldots&7\\
 MC20   & 46    & 36   & 78 (22)   & 13      &3&33\\
 MC21   & 228  & 219 & 96 (4)     & 60     &11&152\\
 MC23   &128  & 109  & 85 (15)   & 24     &21&27\\
 MC75    & 40   & 15   & 38 (62)    & \ldots&\ldots&15\\
 MC76   & 149  & 103 & 69 (31)    & 54     &45&102\\
 MC78   & 142  & 125 & 88 (12)    & 29     &15&125\\
 MC80   & 299  & 279 & 93 (7)      & 38     &20&213\\
 MC81     & 13   & 4    & 31 (69)    & \ldots  &\ldots&4\\
 \hline
Avg & \ldots  & \ldots & 72 (28)      & \ldots     &\ldots&\ldots \\
Median    & \ldots  & \ldots & 82 (18)      & \ldots     &\ldots&\ldots\\
 \hline
 \end{tabular}
\tablecomments{
(1) ID cloud; (2) Selected sources from MIPSGAL \mipslam~image; 
(3) Selected sources after MIR filter. (4) Percent of selected sources after MIR filter, 
in parenthesis the percent of rejected sources. Sources without conterpart in 
GLIMPSE are shown in column 5. Main Sequence 2MASS stars are shown in column 6. 
The final list of YSOs candidates with $\rm L_{bol}>$\lsun\ are shown in the column 7.  
} 
 \end{table*}

\subsection{Identification of \mipslam ~YSO candidates}

We used {\it SExtractor} \citep{bertin+96} on the \mipslam ~images
to detect all sources
that have S/N$\ge$5/pixel within the previously-defined boundaries of the MCs.
The {\it IRAF/phot{\,\footnote{IRAF is distributed by the NAAO, which
is operated by the Association of Universities for Research in
Astronomy Inc.}}} task was used to obtain the photometric magnitudes
of the selected sources. We carried out aperture photometry of the 
{\it SExtractor}-selected point sources using an aperture radius of 5 pixels 
(1~pix=1.25\arcsec and FWHM of \mipslam~MIPSGAL is 
6\arcsec), and the sky annulus of inner radius of 18 pixels 
with a width of 5 pixels. The measured fluxes are corrected for the flux 
outside the aperture
(infinite aperture correction) using a correction factor of 8.44 (average factor
in all MIPSGAL images). Vega flux of 7.14~Jy as suggested in the MIPSGAL 
Data Handbook (Version 3.3) was used to convert the instrumental magnitudes 
to the \mipslam ~magnitudes. 
These sources are listed in the column 2 of the Table~3.

The {\it SExtractor}-selected source list is contaminated by
sources such as foreground stars (bright main sequence and AGB) and 
background objects (galaxies and highly reddened main sequence stars).
Use of bright 24~$\mu$m sources, where the infrared excess is more than
two orders of magnitude above the photospheric emission,
relatively small sizes ($\lesssim$30\arcmin) of our sample clouds, 
and the high column density towards
the molecular clouds, minimizes the fraction of contaminating
sources in our catalog. Nevertheless, we
applied a Mid-Infrared (MIR) photometric filter to reject contaminating
sources from our catalog.

\begin{table*}[!h!t]
\begin{center}
\centering 
\caption{Photometric and physical properties of the YSOs. }
\begin{tabular}{lccrlrlrlrlrlclrll}
\hline
\hline
  &    &   &   &      &   &  &  &  &  & &  &  &  &  &  & & \\
Name  & Lon & Lat & [3.6] & e3.6 & [4.5] & e4.5 & [5.8] & e5.8 & [8.0] & e8.0 &[24] & e24 & C &\scriptsize $\beta$& $\rm L_{bol}$ &  Mass & H \\
  & $^\circ$   & $^\circ$ & mag & mag  & mag   & mag   & mag  & mag & mag  & mag   & mag   & mag  &        &    & \lsun   & \msun & \\ 
(1) & (2) & (3) & (4)  & (5) & (6)  & (7) & (8) & (9) & (10) & (11) & (12) & (13) & (14) & (15) & (16) & (17) & (18)  \\
\hline
\scriptsize MC1-M1  &  50.26111 & $-$0.49540 &   8.55  &  0.04 &   8.49  &  0.05 &   7.97  &  0.04 &   7.96 &   0.03  &  6.22  &  0.04 &  2 &  $-$2.2  &   0.33  &  1.5 &    0  \\
\scriptsize MC1-M2  &  50.26936 & $-$0.47162 &  14.87  &  0.14 &  13.19  &  0.12 &  11.67  &  0.10 &  10.91 &   0.05  &  7.68  &  0.10 &  2 &  $-$2.2  &$-$0.26  &  1.0 &    0  \\
\scriptsize MC1-M3  &  50.21999 & $-$0.45542 &  12.50  &  0.18 &  12.50  &  0.18 &   5.28  &  0.03 &   5.32 &   0.02  &  5.15  &  0.03 &  3 &  $-$3.3  &   2.53  &  4.5 &    1  \\
\scriptsize MC1-M4  &  50.32986 & $-$0.44956 &   8.32  &  0.04 &   7.99  &  0.05 &   7.55  &  0.04 &   7.49 &   0.04  &  6.95  &  0.06 &  3 &  $-$3.3  &   1.81  &  3.0 &    1  \\
\scriptsize MC1-M5  &  50.23466 & $-$0.47592 &  12.45  &  0.07 &  12.23  &  0.09 &  12.05  &  0.18 &  12.50 &   0.18  &  7.22  &  0.07 &  1 &  $-$1.3  &$-$0.07  &  1.0 &    0  \\
\scriptsize MC1-M6  &  50.22746 & $-$0.50346 &  13.51  &  0.07 &  13.18  &  0.10 &  12.50  &  0.18 &  12.50 &   0.18  &  7.05  &  0.07 &  1 &  $-$1.3  &$-$0.01  &  1.0 &    0  \\
\scriptsize MC1-M7  &  50.22213 & $-$0.49915 &  12.93  &  0.07 &  12.17  &  0.07 &  11.65  &  0.10 &  10.95 &   0.10  &  4.04  &  0.02 &  1 &  $-$1.3  &   1.20  &  2.0 &    1  \\
\scriptsize MC1-M8  &  50.23388 & $-$0.47583 &  13.86  &  0.11 &  12.27  &  0.09 &  10.98  &  0.09 &  10.33 &   0.06  &  7.22  &  0.07 &  2 &  $-$2.2  &   0.71  &  1.5 &    0  \\
\scriptsize MC1-M9  &  50.23447 & $-$0.50454 &  14.39  &  0.13 &  12.69  &  0.11 &  11.88  &  0.10 &  11.17 &   0.07  &  8.07  &  0.12 &  1 &  $-$1.3  &$-$0.41  &  0.9 &    0  \\
\scriptsize MC1-M10 &  50.22111 & $-$0.49915 &  12.80  &  0.07 &  12.18  &  0.08 &  11.51  &  0.10 &  11.14 &   0.13  &  4.04  &  0.02 &  1 &  $-$1.3  &   1.20  &  2.0 &    1  \\
\scriptsize MC1-M11 &  50.22149 & $-$0.49979 &  12.69  &  0.08 &  11.71  &  0.07 &  10.42  &  0.06 &   8.81 &   0.03  &  4.04  &  0.02 &  2 &  $-$2.2  &   1.98  &  3.0 &    1  \\
\scriptsize MC1-M12 &  50.29696 & $-$0.42889 &  12.50  &  0.18 &   6.65  &  0.06 &   6.25  &  0.03 &   6.08 &   0.03  &  5.09  &  0.03 &  1 &  $-$1.3  &   0.78  &  1.5 &    0  \\
\scriptsize MC1-M13 &  50.30923 & $-$0.42042 &   7.18  &  0.04 &   6.66  &  0.04 &   5.97  &  0.03 &   5.51 &   0.03  &  3.17  &  0.01 &  2 &  $-$2.2  &   2.33  &  4.0 &    1  \\
\scriptsize MC1-M14 &  50.22479 & $-$0.46537 &  13.01  &  0.06 &  11.48  &  0.07 &  10.38  &  0.06 &   9.51 &   0.03  &  4.98  &  0.03 &  1 &  $-$1.3  &   0.82  &  1.5 &    0  \\
\scriptsize MC1-M15 &  50.25849 & $-$0.50773 &  11.51  &  0.04 &  10.85  &  0.06 &  10.37  &  0.06 &   9.52 &   0.04  &  6.61  &  0.05 &  2 &  $-$2.2  &   0.95  &  1.5 &    0  \\
\scriptsize MC1-M16 &  50.27218 & $-$0.50201 &  11.54  &  0.15 &  11.09  &  0.13 &  12.50  &  0.18 &  12.50 &   0.18  &  3.59  &  0.01 &  3 &  $-$3.3  &   1.38  &  2.0 &    1  \\
\hline
\end{tabular}
\tablecomments{ 
This table is available in its entirety in the electronic version.\\
Brief explanation of columns: 
(1) The name of the MIPSGAL/GLIMPSE YSO in the sample. The name includes two parts, the first part
is the name of the cloud, and the second part starts with a letter M or G (M indicates detected in MIPSGAL,
and G indicates detected only in GLIMPSE), followed by the source number. The GLIMPSE-only detected sources
do not have \mipslam\ related quantities, and hence a place-holder value of $-1.00$ is assigned for these sources under
columns 12--13 and 15--18;
(2--3) Galactic longitude and latitude for the YSOs;
(4--13) GLIMPSE and \mipslam-MIPSGAL photometric magnitudes and errorrs;
(14) IR spectral classification of the YSOs. Numbers 0,1,2,3 stand for Class 0, I, II and III, respectively;
(15) $\log\beta$-values of the YSOs as defined in \S3.4;
(16--17) $\rm \log(L_{bol})$ and mass of the YSOs derivated as explained in \S4.2. The errors on $\rm L_{bol}$ and mass are roughly 30\%;
(18) A flag indicating whether the YSO is of high luminosity: 1 for YSOs with $\rm L_{bol}\ge10~L_{\odot}$, 0 otherwise.
} 
\end{center}
\end{table*}

\subsection{\mipslam ~sources without GLIMPSE counterpart}

Some of the \mipslam ~sources do not have a GLIMPSE counterpart. 
Genuine Class 0 and transitional Class 0/I sources are expected to be
of this kind \citep{andre+10}. However, the absence of a GLIMPSE counterpart doesn't
necessarily mean the absence of emission in the 3--8~$\mu$m region.
Sources could be missed in the GLIMPSE data set due to the
problems with the photometry in any/some band(s) of GLIMPSE due to 
crowding or (bright) diffuse emission in the band. The brightest of 
such sources are expected to be detected in the NIR by the 
2MASS.\footnote{Two Micron All Sky Survey (2MASS):
{\it http://irsa.ipac.caltech.edu/Missions/2mass.html}}
In order to identify these bright reddened main-sequence stars,
we performed a positional cross-match of the MIPSGAL 
sources with the 2MASS point source catalog. 
All these 2MASS-detected objects that satisfy the photometric color 
criteria of reddened main-sequence stars
($(J-H)\geq1.75\times(H-K)$), with the Main-Sequence as defined by
\citet{bessell+brett}, were removed from the sample. 
In the column 6 of the Table~3 are listed the MS stars found in each cloud.  
The sources without 2MASS counterpart, or with $(J-H)\leq1.75\times(H-K)$ 
and $(H-K)$ color excess, are considered genuine Class 0/I YSOs. 
These sources are listed in the column 3 of the Table~5 as Class 0 YSOs 
and are included in the final YSO selection (\S~3.3). 
However, a few of these sources could be highly embedded Class II YSOs
in the high-density regions of the molecular clouds \citep{megeath+12}.

\subsection{Selection criteria of YSOs}

In order to identify the candidates to YSOs from the 24~$\mu$m-detected 
sources, we used filters formed from a combination of GLIMPSE and MIPSGAL 
24~$\mu$m colors as defined by \citet{gutermuth+09}.
As a primary step, all GLIMPSE sources from the PSC with photometric errors
$\sigma<0.2$~mag in all IRAC bands are selected.
Foreground main sequence objects are expected to be bluer than
$[X]-[24]=1.0$~mag \citep{gutermuth+09}, where $[X]$ is the magnitude in the 
[4.5] and [5.8] bands. Such sources are also expected to have $[3.6]-[24]<1.0$~mag,
which is the selection criterion we used to reject all foreground main sequence stars.
Given the sensitivities of 
GLIMPSE and MIPSGAL images, all 24~$\mu$m sources without
a GLIMPSE counterpart are redder than this criterion, and hence are genuine
YSOs. Foreground AGBs are relatively bluer and brighter than YSOs,| and have colors
$\rm [3.6]-[8.0]<1.5$ and $\rm [8.0]<6.0$~mag. \citep{marengo+08}. 
The candidate sources that satisfied this criterion are removed from our
sample. Given the relatively small field sizes and low Galactic latitudes of our
sample sources, our sample of YSOs is not expected to be contaminated
by background galaxies. In total, the rejected number of objects amounts
to $\sim$28\% (average; see column~4, Table~3). All objects remaining after applying these
filters are considered ``bona fide'' YSOs. 
These sources, including the Class 0 YSOs, their galactic 
coordinates (columns 2 and 3) and GLIMPSE and \mipslam~MIPSGAL 
photometric magnitudes (columns 4-8) are listed in the Table~4.

\subsection{Classification of YSOs}

In order to characterize the YSOs and describe their nature,
we have used the infrared spectral index ($\alpha$) as was
defined in early works by \citet{lada+84} and \citet{lada87}. This index has
been extensively used to study the nearby ($<$1~kpc), predominantly
low-mass star-forming regions 
\citep{hartmann+05,harvey+06,alcala+08,gutermuth+08}.
Though originally defined to describe the evolutionary phases
of low-mass stars, the $\alpha$ index is applicable for 
high-mass YSOs as well \citep{deharveng+12,ellerbroek+13,saral+15}.
We defined $\alpha$ for each 24$~\mu$m-source using the GLIMPSE 
and MIPSGAL data that cover a wavelength range from 3.6$~\mu$m to 24$~\mu$m. 
We followed the often-used definition of \citet{green+94} to carry out the
classification. According to this classification, YSOs are Class I
if $\alpha\geq-$0.3, Class II if $-$1.6$\leq\alpha<-$0.3, and Class III
if $\alpha<-$1.6. 
The sources without GLIMPSE counterpart (Class 0/I YSOs) are 
considered Class I and a spectral index of 0.35 is assigned to them. In the 
original definition of \citet{green+94} the YSOs with $-$0.3$<\alpha<$0.3 
are classified as Class FLAT. In this 
work, we included these $\alpha-$values into the Class I classification. 
The $\alpha-$values as defined by \citet{green+94} requires \mipslam\   
detection and hence we used the slightly less reliable photometric color 
criteria of \citet{gutermuth+08} to classify GLIMPSE YSOs without 
MIPSGAL counterpart.  For this sample, 
12--45\% of the MIPSGAL sources are Class I with a median value 
of 22\%, where the upper percentages are affected by the small 
statistics of YSOs in some clouds; whereas that only 9--31\% (with a 
median value of 23\%) is Class I YSOs in GLIMPSE sources. On the 
other hand,  majority of the GLIMPSE sources are Class II (20--86\%, 
with a median value of 77\%). 
The classification of the YSOs are shown in the column 9 of the Table~4.

\section{Mass of the YSOs}

Having obtained a sample of YSOs, we now use their merged
photometry in GLIMPSE and MIPSGAL data to obtain a Mass Function (MF). 
The first step for achieving this is
to obtain the bolometric luminosities (\lbol) of the sample YSOs. 
Knowledge of Class is crucial in determining the bolometric luminosity of 
embedded YSOs. Even so, detection at 24~\mum\ is key for a reliable 
determination of \lbol, as the bolometric correction factor for GLIMPSE-only sources 
is highly uncertain. 

\subsection{Mass of the GLIMPSE sources}

The c2d Survey \citep{evans+03} is one of the legacy programs of the Spitzer 
Space Mission that has provided uniform photometric data in the IRAC and MIPS 
bands of Spitzer of YSOs in well-known nearby star-forming regions. 
These regions were part of follow-up studies at far-infrared 
(60~$\mu$m to 500~$\mu$m) by Herschel and submm/mm by 
ground-based facilities such as SIMBA/SEST, SCUBA/JCMT and MAMBO/IRAM. 
YSOs in Gould's Belt (GB) also had been the target of multi-band surveys.
Detailed analysis of the resulting
Spectral Energy Distributions (SEDs) extending from the MIR to 
millimeter wavelengths has allowed the determination of stellar 
masses for individual YSOs. This in turn has allowed the 
construction of the mass function of YSOs \citep[e.g.][]{enoch+06,andre+10}.
Most of these low-mass \starf~regions are consistent with a 
log-normal Initial Mass Function (IMF) 
as defined by \citet{chabrier03} with a characteristic mass of 
0.5~\msun~\citep{kennicutt+12}. Following these 
studies, we assume a log-normal mass function for the low luminosity 
YSOs, which are mainly GLIMPSE sources, and assign a mean YSO mass of 
0.5~\msun ~for each source in this population.

\subsection{Mass of the \mipslam-detected sources}

\subsubsection{Bolometric luminosity}

Stars are still embedded in the MC during their PMS phase. The material 
in the immediate vicinity of the star absorbs almost all of the radiation 
emitted from the PMS star, re-radiating it in the infrared, including at 
the 24~\mum\ band. We define a quantity $\beta=L_{24}/L_{\rm bol}$, 
where $\beta$ is the fraction of the bolometric luminosity that is emitted in the 
24~\mum-band. The emission at 24~\mum\ arises from the warm 
circumstellar dust and depends on the dust temperature and the
mass of the heated dust envelope. During the evolution from Class 0/I
(protostars) to Class II (PMS object), the dust temperature of the
circumstellar material is expected to increase due to the increase
of the ``photospheric'' temperature,
which would lead to an increase of the fraction $\beta$.
The accretion of the envelope material to the disk also contributes
to the increase of the dust temperature \citep{stahler+05}. However,
the envelope mass decreases at a faster rate due to effects of outflows, 
radiative pressure and stellar winds, with the net result being a decrease 
of $\beta$ along the PMS phase. A detailed modeling of $\beta$ 
variation is beyond the scope of this work. Hence, we looked for an 
empirical relation between $\beta$ and the SED evolutionary phases 
of the YSOs using a recent complete IR plus submm photometric data set 
for YSOs.

\begin{figure}
\centering
\includegraphics[width=9.0cm]{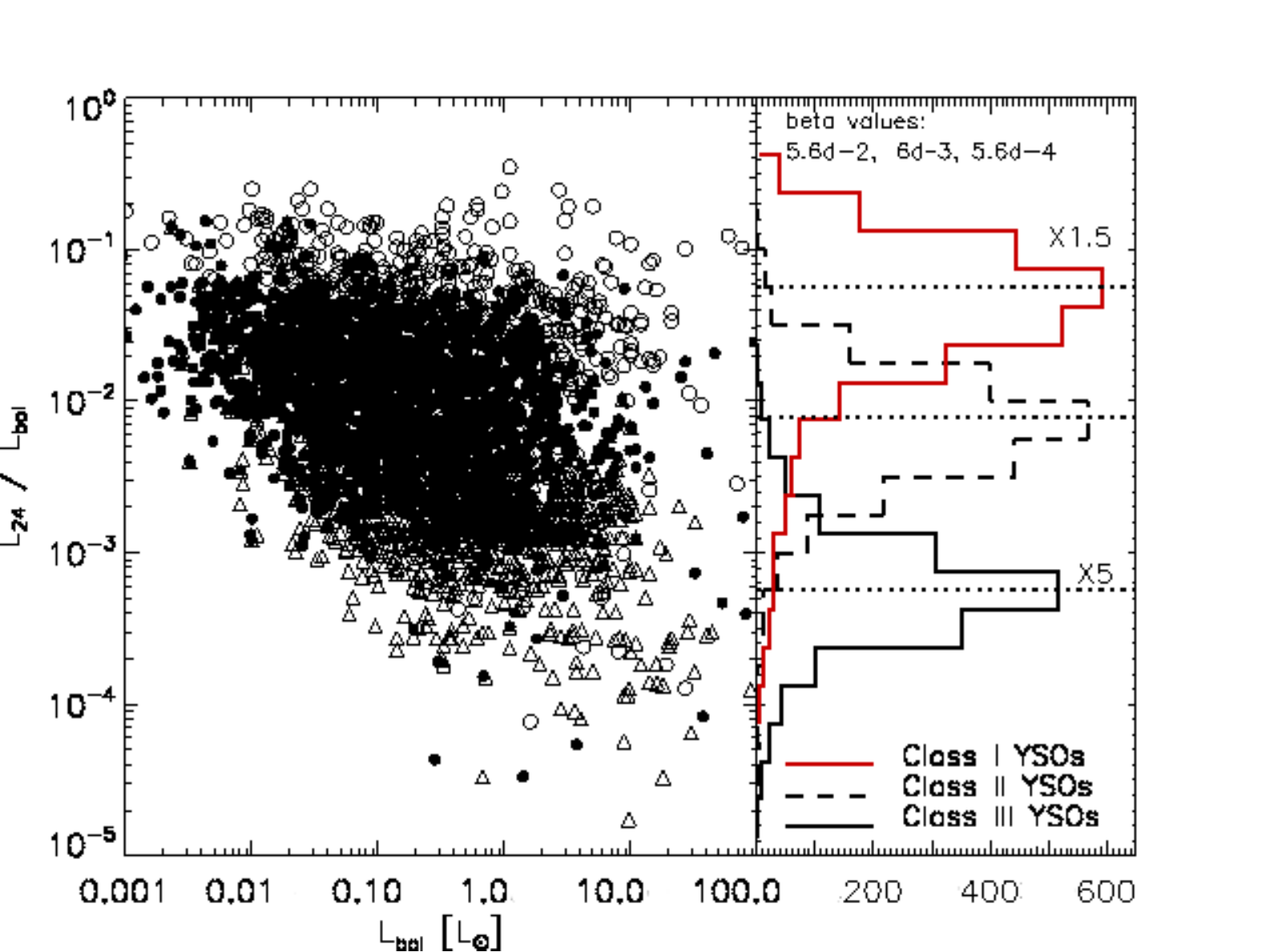}
\caption{
$\beta=L_{24}/L_{\rm bol}$ vs L$_{\rm bol}$ diagram for YSOs from c2d and GB data 
surveys (left). The Class III, Class II and Class I YSOs are shown by triangles, filled circles 
and open circles, respectively. The histograms to the right show the
distribution of ${\rm \beta}$ for each Class, where the numbers in each Class 
are normalized to roughly match the number of Class II YSOs, by multiplying
by a factor 1.5 and 5 (denoted as X1.5 and X5). The median value of each distribution is
shown by dotted line and point out in the top of pannel, which is used as a typical value 
for each Class.
}
\label{figure3}
\end{figure}

We used the YSOs from the c2d and GB recent data sets, that have both the 
\mipslam\ and bolometric luminosities catalogued, to obtain $\beta$ empirically 
using the evolutionary Classes I--III. To start with, we classified the objects
in this data sets into evolutionary SED Classes I, II and III following
the criteria of \citet{green+94}. In Figure~\ref{figure3}, we show 
the $\beta-L_{\rm bol}$ diagram for the c2d+GB YSOs from 
\citet{evans+03} and \citet{dunham+15}. This population is dominated by Class II 
YSOs (filled circles), although there is a considerable fraction of Class I
(open circles) and Class III YSOs (triangle symbols). The Class I sources 
have the highest $\beta$-values, whilst the Class II and III
span values between $10^{-1}$ and $10^{-5}$. The distribution
of $\beta$ values for each Class is shown as histogram in the
right panel, where the peak values traced by dashed lines are used as 
typical values for each Class. These $\beta$-values are $\rm 5.6\times10^{-2},
~6.0\times10^{-3}~and~5.6\times10^{-4},$ for Class I, II and III, respectively.

For our embedded sources of Class 0/I, which are detected only in \mipslam,
we used the relation $L_{\rm bol}=3.31\times L_{\rm MIR}$ of \citet{kryukova+12}
for spectral index of 0.35.
The mid infrared luminosity, $L_{\rm MIR}$, is the luminosity in the 
1--24~$\mu$m range, which is calculated from the \mipslam\ flux ($S_{24}$ in Jy)
using the relation, $L_{\rm MIR}/L_\odot=[0.05+0.81~S_{24}]\times D^2$, 
where $D$ is the distance to the cloud in kiloparsec. For deriving the
latter relation, we have substituted the flux terms in the 2MASS and GLIMPSE 
bands by the 3$\sigma$ upper limits in these bands. 

\begin{figure}[ht!]
     \begin{center}
\includegraphics[width=0.99\columnwidth]{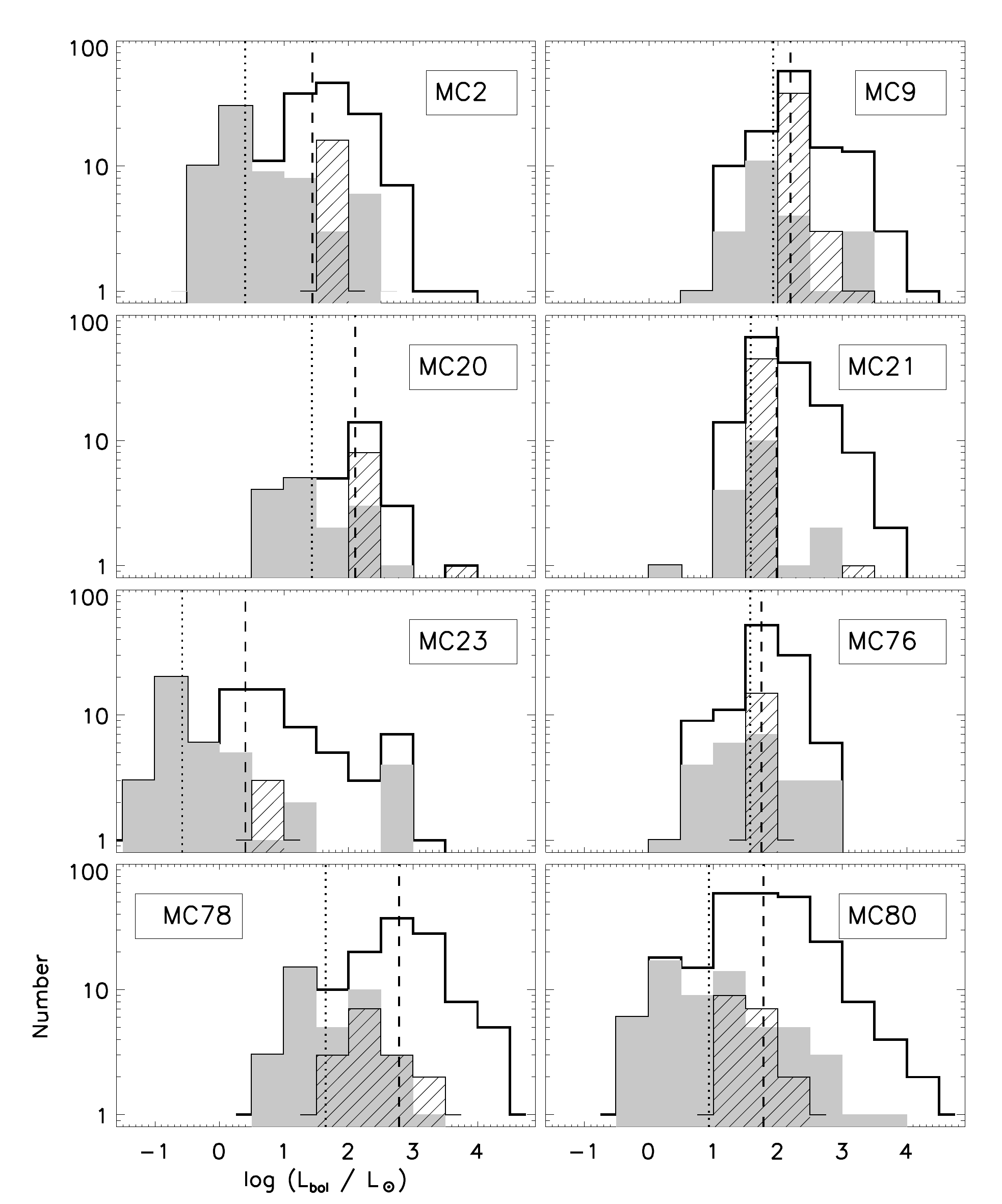}
    \end{center}
    \caption{
Bolometric Luminosity Function of MIPS-detected YSOs in 8 of our 
molecular clouds, separated by Class as indicated in the plot 
(All: solid histogram, Class I: shaded, Class 0/I: hatched).
Median values for All (dashed) and Class I (dotted) YSOs are shown 
by vertical lines. Sources with L$_{\rm bol}>10~{\rm L}_{\odot}$
are considered as high luminosity YSOs.
}
\label{figure4}
\end{figure}

We constructed the bolometric luminosity function (LF) for all the \mipslam-detected
YSOs of our sample using the $\beta$ value corresponding to its Class.
The resulting bolometric LFs for eight of our MCs that contain at least 20 
luminous YSOs are shown in Figure~\ref{figure4}.  
These are plotted 
separately for Class I (shaded area), Class 0/I (hatched area) and all YSOs.
The median luminosities of MIPSGAL-selected Class I YSOs (vertical dotted line) 
in different clouds varies between 0.3--80~\lsun, with a global median of $\sim$8~\lsun.
It is interesting to note that the median luminosity for all sources (vertical dashed line)
is higher than that for Class I YSOs in every cloud. This implies that
the most luminous sources (and the most massive sources) in the MCs are 
systematically more evolved (Class II and III).


\subsubsection{Luminosity-Mass relation}

\begin{table*}[!h!t]
\begin{center}
\centering 
\caption{Class and mass distribution of YSOs in our sample of MCs} 
 \begin{tabular}{ccccccccccccccc}
 \hline
 \hline
   & \multispan{5}{High-Luminosity}   &\multispan{4}{Low-Luminosity}   &   &\multispan{2}{ Stellar Mass} &   &  \\
\hline
GMC & CI& C0  & CII & CIII  & \footnotesize HLYSOs   & CI  & CII & CIII  &\footnotesize LLYSOs  & { Total}  & HLYSOs  & LLYSOs  & HLYSOs & LLYSOs  \\
    &         &  &    &       &         &     &     &       &        & { YSOs}  &[$M_{\odot}$] & [$M_{\odot}$] &\msun/N &\msun/N \\
(1)  & (2) & (3) & (4) & (5) & (6) & (7) & (8) & (9) & (10) & (11) & (12) & (13) & (14) & (15)\\
 \hline
 MC2   & 17 & 16 & 71  & 15   & 119  &  74   &  352   & 11     & 437    & 556   & 440$\pm$13  & 219$\pm$5  & 3.7   & 0.5   \\
 MC9   & 22 & 42 & 21  & 32   & 117  &  93   &  633   & 18     & 744    & 861   & 630$\pm$19  & 372$\pm$5  & 5.4   & 0.5  \\
 MC20 & 11 & 10 & 6    & 2     & 29    &  40   &  211   & 12      & 263    & 292  & 144$\pm$9  & 132$\pm$3  & 4.9   & 0.5 \\
 MC21 & 17 & 46 & 30  & 59   & 152 &  49   &  448   & 23      & 520    & 672   & 665$\pm$17  & 260$\pm$5  & 4.4  & 0.5  \\
 MC23 & 6 & 3     & 14  & 4     & 27    &  162 &  398  & 5         & 565    & 592  & 102$\pm$7  &  283$\pm$5  & 3.8   & 0.5  \\
 MC76 & 19 & 15 & 57  & 8     & 99 &  121  &  433  & 13      & 567    & 666   & 371$\pm$12  & 284$\pm$5  & 3.7  & 0.5  \\
 MC78 & 33 & 14 & 11  & 61   & 121 &  149  &  435  & 16      & 600    & 721   & 795$\pm$24  & 300$\pm$5  & 6.6  & 0.5  \\
 MC80 & 29 & 18 & 52  & 112 & 211 &  118  &  325  & 7        & 450    & 661   & 882$\pm$20  & 225$\pm$5 & 4.2  & 0.5  \\
\hline
 MC1   & 7 & 0     & 3     & 2     & 12    &  14    &  91     & 13     & 118    & 130  & 25$\pm$3    & 59$\pm$3   & 2.0  & 0.5  \\
 MC12 & 2 & 0     & 1     & 4     & 7      &  51    &  106   & 6       & 163    & 170  & 28$\pm$3    & 82$\pm$3   & 3.5  & 0.5  \\
 MC75 & 5 & 0     & 4     & 5     & 14    &  131  &  428   & 23     & 582    & 596  & 62$\pm$6    & 293$\pm$5 & 4.6  & 0.5  \\
 MC81 & 1 & 0     & 1      & 2    & 4       &  3      &  2       & 5       & 10      & 14   & 13$\pm$2    & 5$\pm$1      & 3.3  & 0.5  \\
\hline
 \end{tabular}
\tablecomments{
Brief explanation of columns:
(1) GMC name; 
(2-6) distribution of high luminosity YSOs into Class I (CI),  Class 0 (C0),
Class II (CII) and Class III (CIII). The column labeled  HLYSOs contains 
the total number of HLYSOs; 
(7-10) distribution of low luminosity YSOs into different Classes. The column 
labeled  LLYSOs contains the total number of LLYSOs;
{ (11) Total number of HL and LL YSOs};
(12-13) total mass of all high and low luminosity YSOs ;
(14-15) mean mass of a high and low luminosity YSO;
Data for the four MCs with less than 20 high luminosity YSOs are given in the bottom four rows.
} 
\end{center}
 \end{table*}

Observed bolometric luminosity of an accreting star comes from 3 physical
processes: (1) accretion, (2) nuclear burning, and (3) Kelvin-Helmholtz 
contraction. The contribution from the last two processes is well understood 
and is calculated theoretically in traditional PMS evolutionary tracks 
\citep{palla+stahler93,tognelli+11}. 
For stars with mass $>$1.5~\msun, the luminosity from accretion is not
dominating the observed \mipslam ~flux, and that the nuclear luminosity 
is a good approximation to the observed values \citep{hillenbrand+white04}.
For example, \citet{myers14} obtain, depending on the model 
assumptions, accretion luminosities between 250--1500~\lsun\ for a 
protostar of 5~\msun, which 
is comparable to the 5~\msun~ZAMS value of $\sim$800~\lsun~ 
from \citet{tognelli+11}.
 
However, the relative contribution of accretion to the total luminosity, 
for low-mass stars (mass$<$1~\msun) in their
early Class I stage is still under debate. In general theoretically predicted
accretion rates produce luminosities a factor of $\sim5$ higher as compared
to the median (observed) luminosity of accreting stars 
\citep{kenyon+90,enoch+09}. On the other hand, observations of 
high-luminosity variable sources, such as the prototype FU Ori, suggest 
that protostars undergo periods of high accretion, in events known as 
episodic accretion \citep[e. g.][]{hartmann+kenyon96, vorobyov+basu05}.
In the extreme case, protostars may spend most of their life in low-luminosity,
low-accretion phase and accrete most of their mass during short, intense 
accretion bursts. Even with episodic bursts of accretion, protostars need
to find a way to lose as much as 75\% of the accretion energy in non-radiative 
winds \citep{offner+11}.

The sensitivity of MIPSGAL at the distances of sample clouds corresponds 
to bolometric luminosity of $\sim$1~\lsun. The median 
bolometric luminosity of Class I YSOs of our sample is $\sim$8~\lsun, 
which is similar to the value given for GLIMPSE-selected YSOs of \citet{offner+11} 
but exceeds by a factor of 5  from \citet{kennicutt+12}. 
On the other hand, 10~\lsun~ 
corresponds to luminosity of a $\sim$2~\msun ~ZAMS star in 
models of \citet{tognelli+11}. The above discussions clearly illustrate 
that reliable masses could be obtained for luminous MIPSGAL 
Class I YSOs using PMS evolutionary tracks.
For Class II and Class III sources, the accretion contribution is expected
to be even less. Hence we determined the high-mass end of the
mass function using \citet{tognelli+11} PMS evolutionary tracks.

YSOs spend different timescales in each of the evolutionary Classes I, II, 
and III. The timescale in each of these phases depends on the stellar mass.
These timescales have been studied in nearby star-forming regions in the 
c2d and GBS \citep{evans+09,dunham+15}.
The values used in this work are 0.5~Myr and 2.0~Myr for Class I 
and Class II, reported by \citet{kenyon+95} and \citet{evans+09}, respectively. 
{ These values are around to obtained in recent works by \citet{heiderman+15} 
and \citet{dunham+15}, particularly the timescale for early Class I YSOs.}
For the Class III, a timescale of 3~Myr is used (although there are longer 
PMS timescales for low-mass stars). 

In order to assign a mass to the bolometric luminosity of the YSOs, we
have used the following procedure: (a) if the source is Class III, a mass 
from the PMS evolutionary track for 3~Myr is used, (b) if the 
source is Class II, we assign a mass from the evolutionary track for 
2~Myr, while for Class I and Class 0/I YSOs, we assign a mass from the 
0.5~Myr track. 

Hereafter, we refer all YSOs satisfying $L_{\rm bol}>10\,L_{\odot}$
as High Luminosity sources (HLYSOs), with the rest of the sources, including the 
GLIMPSE-only detected YSOs referred to as Low Luminosity sources (LLYSOs). 
In Table~5, we give the Class distribution and masses of HLYSOs and LLYSOs.
In general, HLYSOs are dominated by Class I sources, whereas the Class II
sources are the most frequent among LLYSOs. The mean mass of HLYSOs in different
MCs ranges from 2 to 6.6~\msun, with a median value of 4~\msun\ for the whole sample.

\begin{figure}[ht!]
     \begin{center}
            \includegraphics[width=8.5cm]{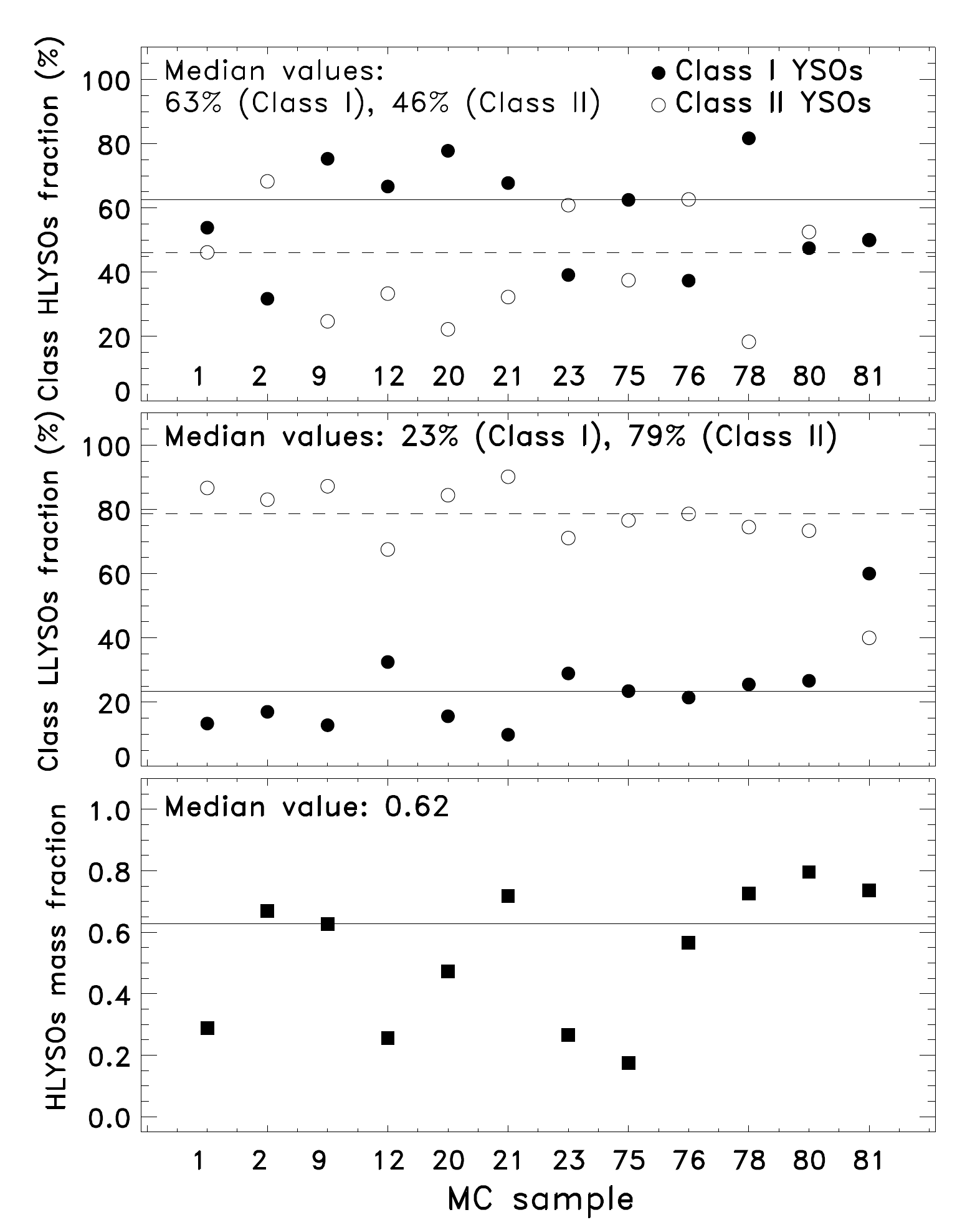}
    \end{center}
\caption{
Relative fraction of Class I (filled circles) and Class II (open circles) YSOs 
among the high luminosity (top) and low luminosity (middle) sources in
each MC identified by its number vertically below the symbol.
HLYSOs are predominantly Class I, whereas LLYSOs are dominated by Class II sources. 
The horizontal lines give the median value for each Class (solid for Class I and 
dashed for Class II). 
The relative contribution of HLYSOs to the total mass in YSOs is shown in 
the bottom panel. In our sample of MCs, HLYSOs contribute between 20--80\%
of the total mass, with the median value 63\%.
}
\label{figure5}
\end{figure}

In Figure~\ref{figure5}, we plot the fractions of Class I and Class II YSOs in
high (top panel) and low (middle panel) luminosity samples for each MC. 
HLYSOs are predominantly Class I (63\%), whereas LLYSOs are dominated by Class II 
(79\%) sources. In the bottom panel, we show the contribution of HLYSOs to the
total ``stellar'' mass for each MC. It can be noted that in 8 of our MCs,
these HLYSOs, which are systematically more massive than the LLYSOs (see the
mean mass in the last two columns of Table~5),
contribute more than 50\% of the total ``stellar'' mass. 
For the total sample, 62\% of the total mass is contributed by the HLYSOs.


\subsection{The mass function of the YSOs}

\begin{figure*}[ht!]
     \begin{center}
\includegraphics[width=0.67\columnwidth]{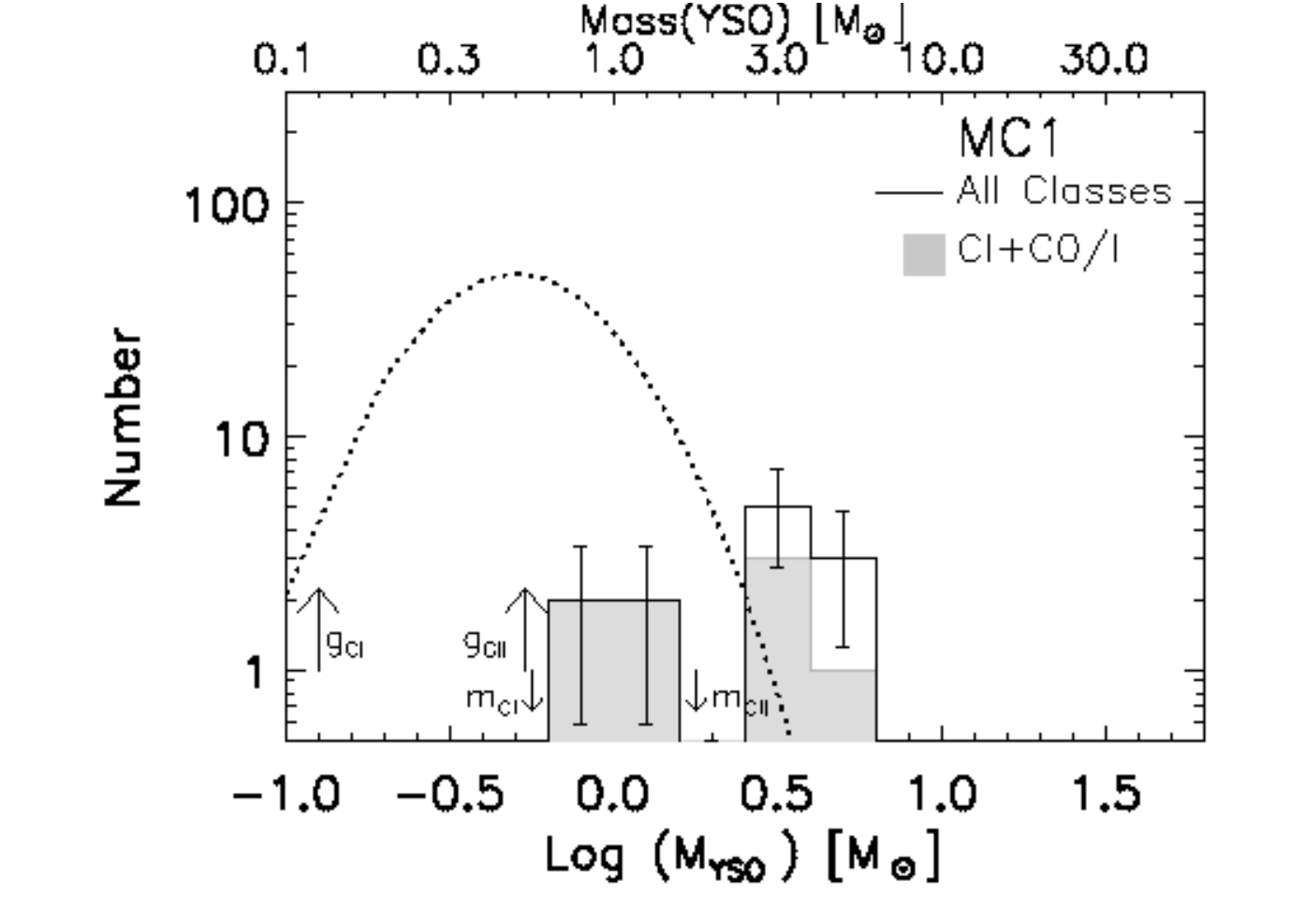} 
\includegraphics[width=0.67\columnwidth]{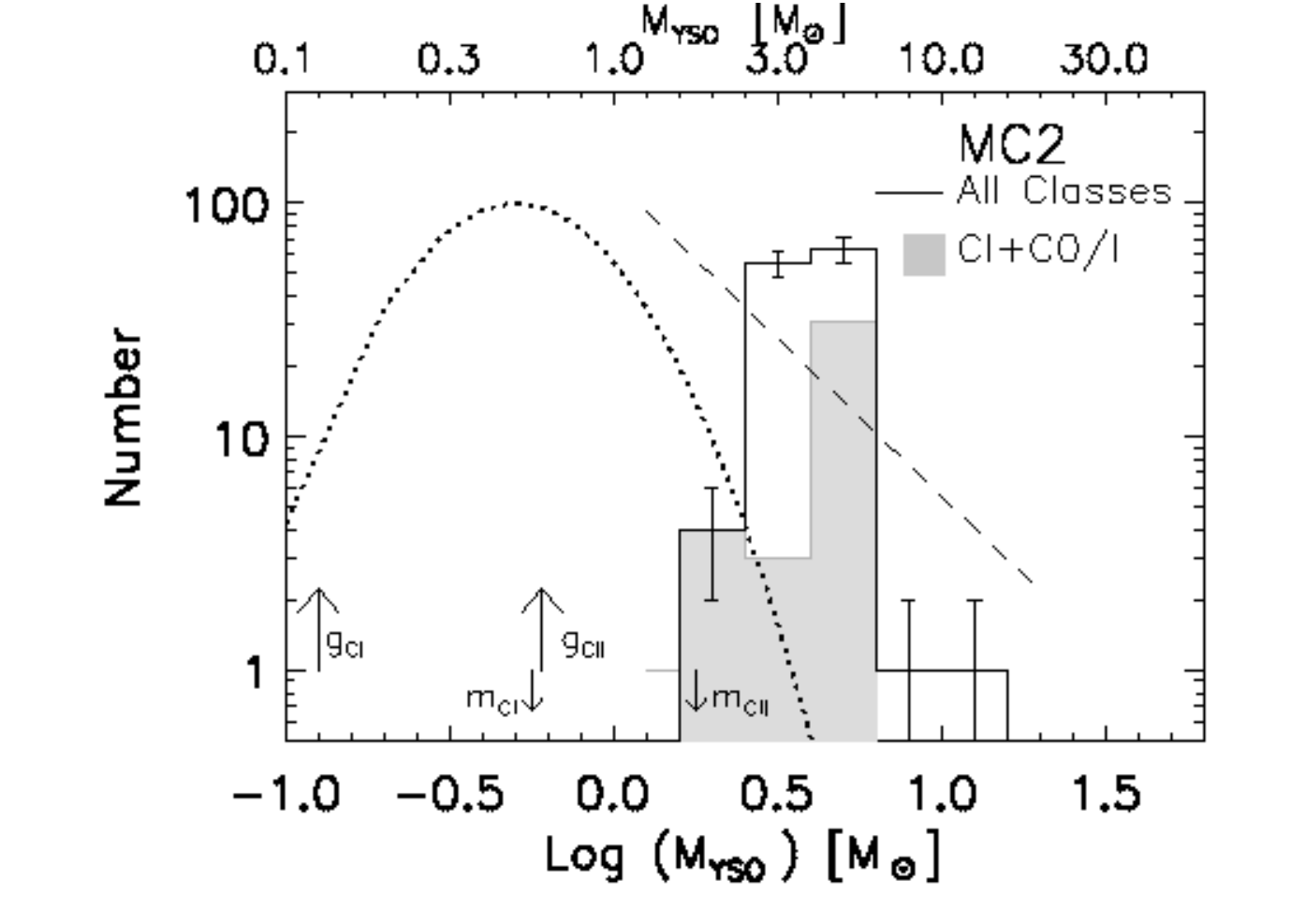} 
\includegraphics[width=0.67\columnwidth]{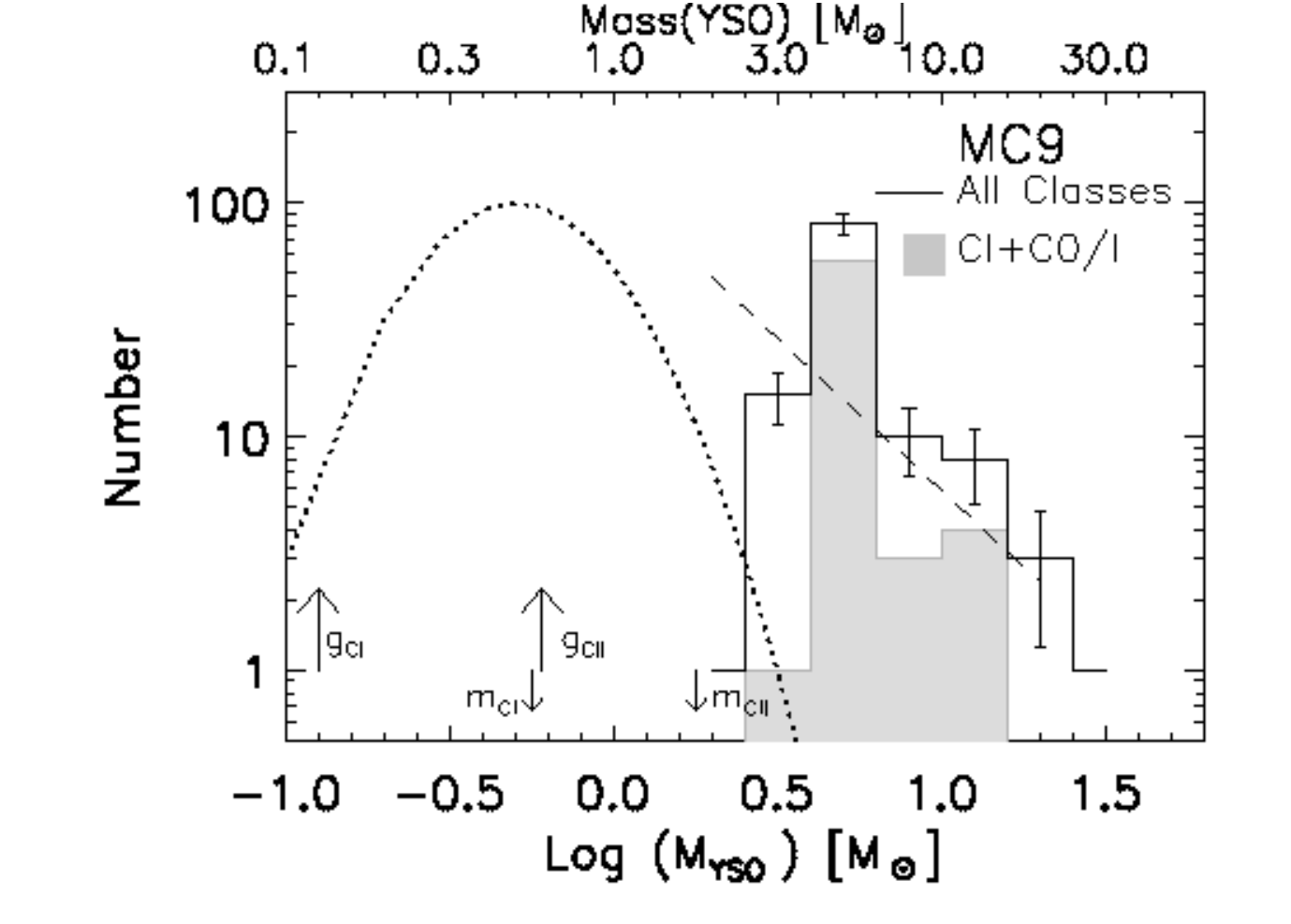} \\
\includegraphics[width=0.67\columnwidth]{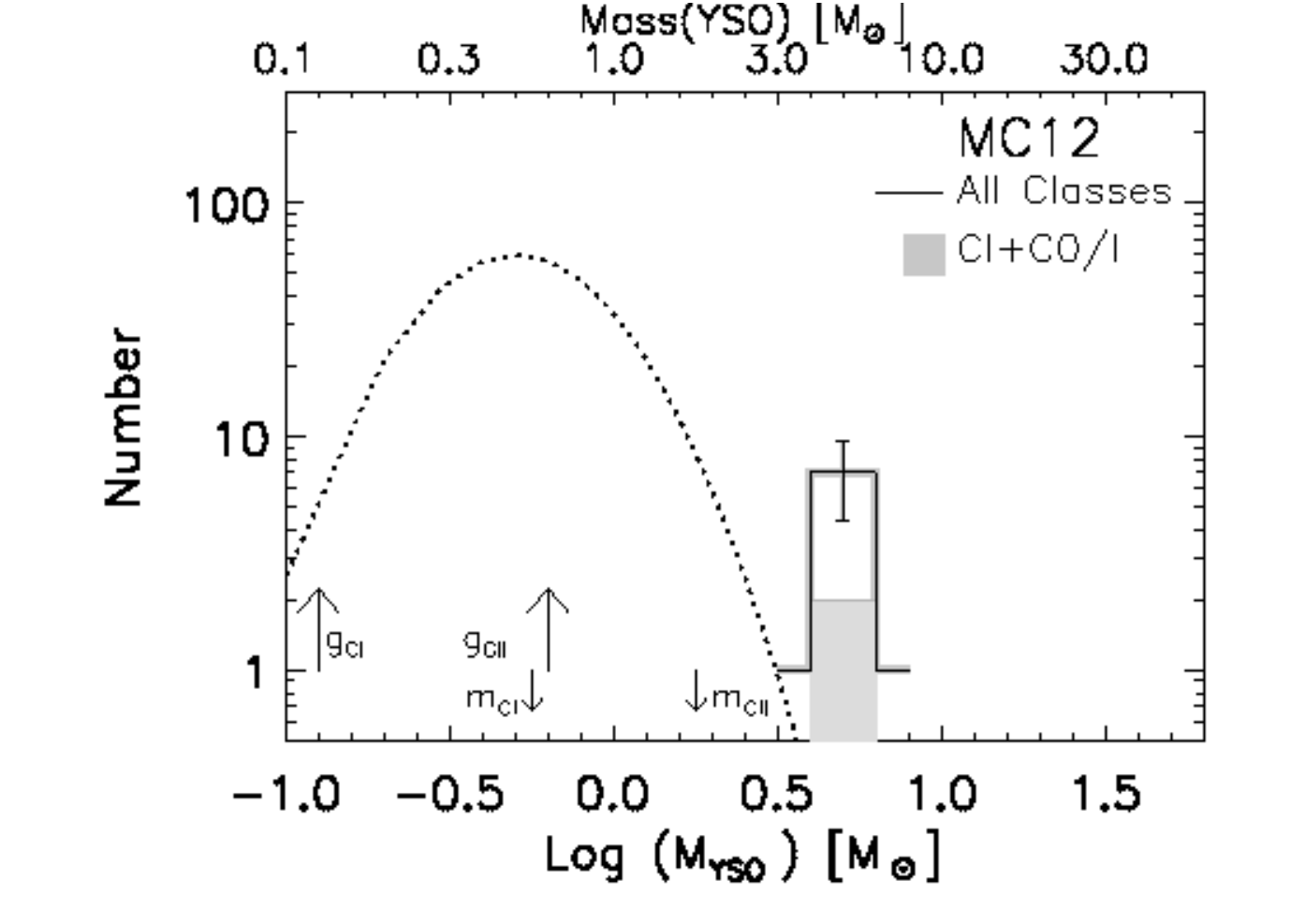} 
\includegraphics[width=0.67\columnwidth]{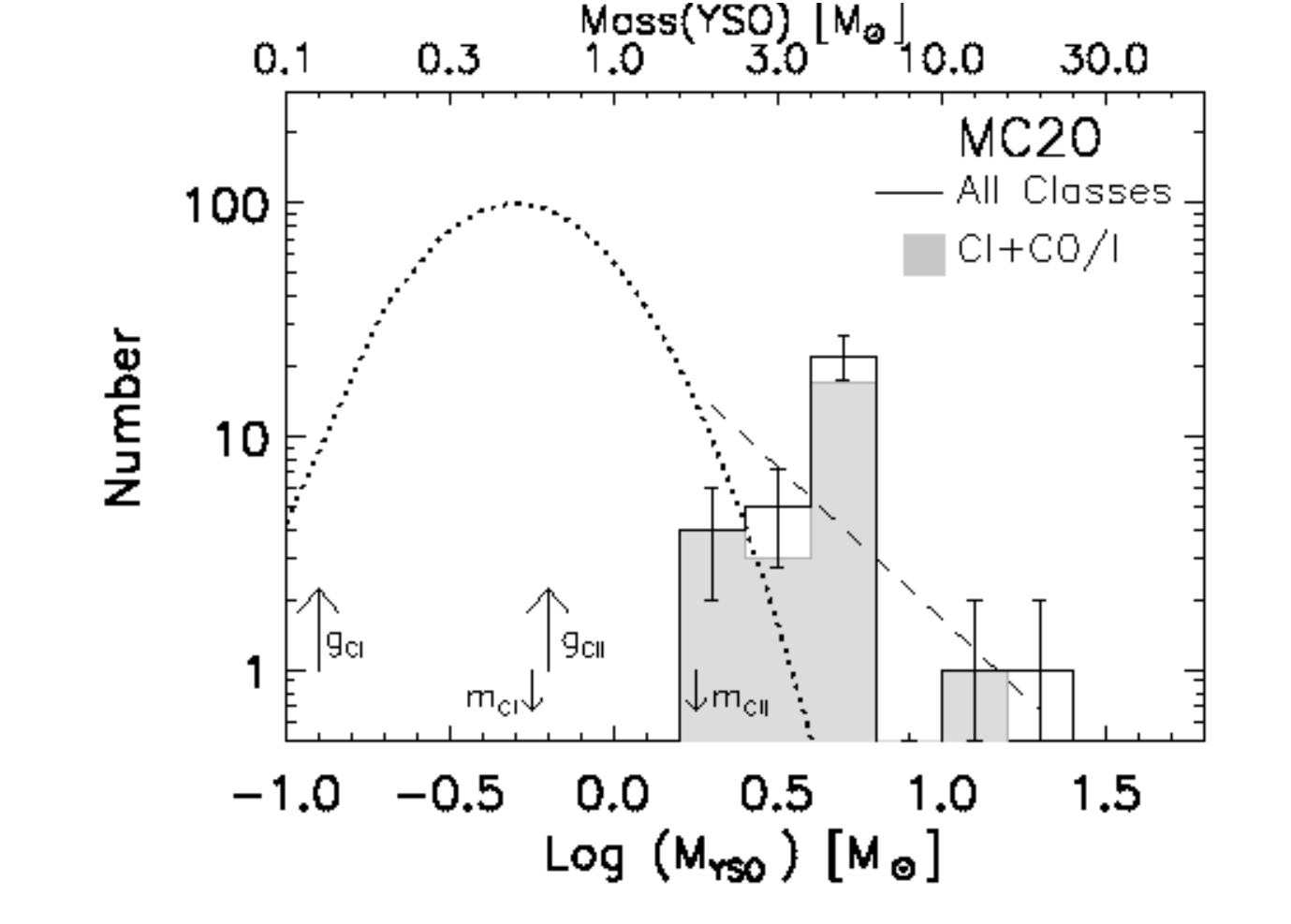} 
\includegraphics[width=0.67\columnwidth]{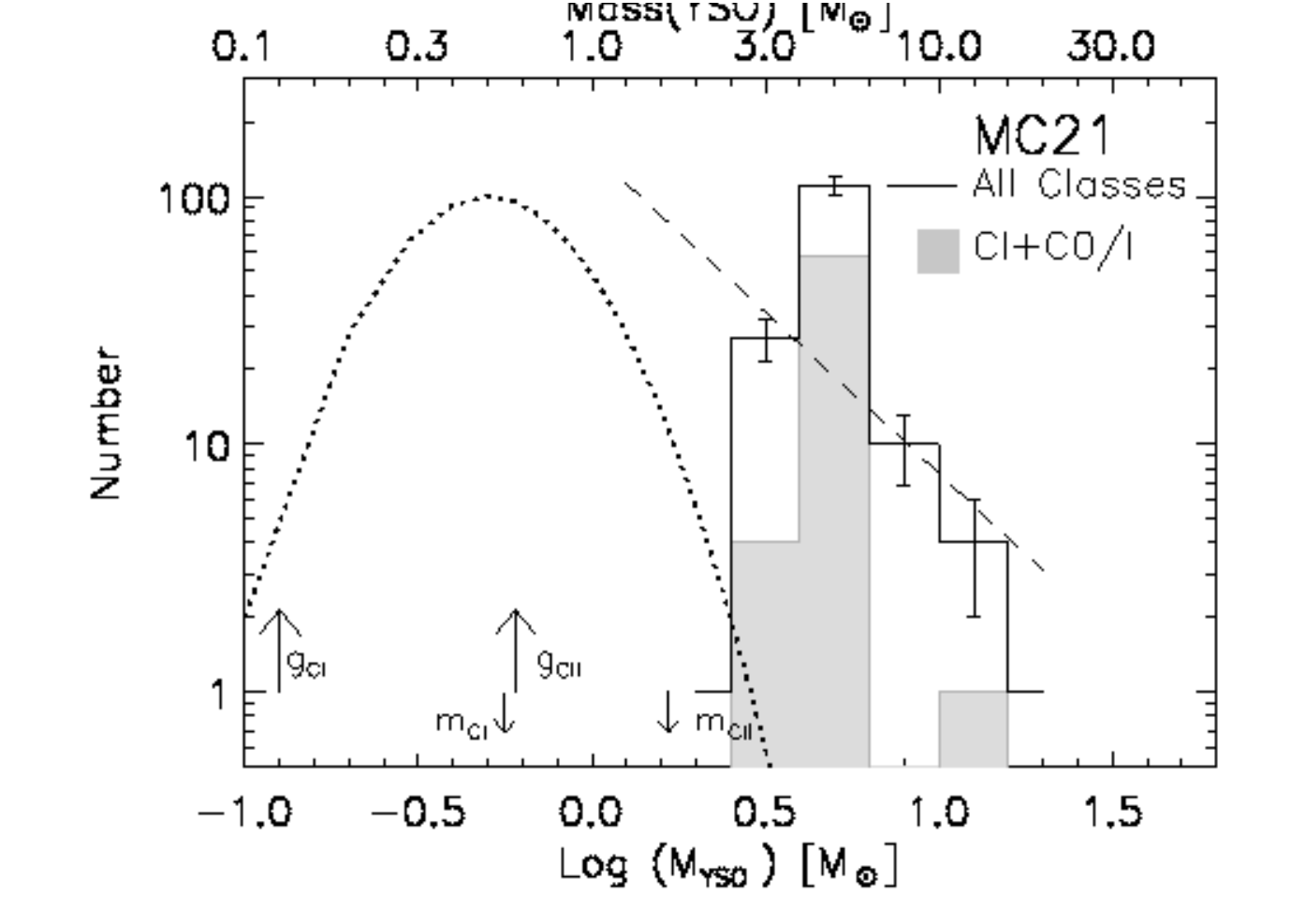} \\
\includegraphics[width=0.67\columnwidth]{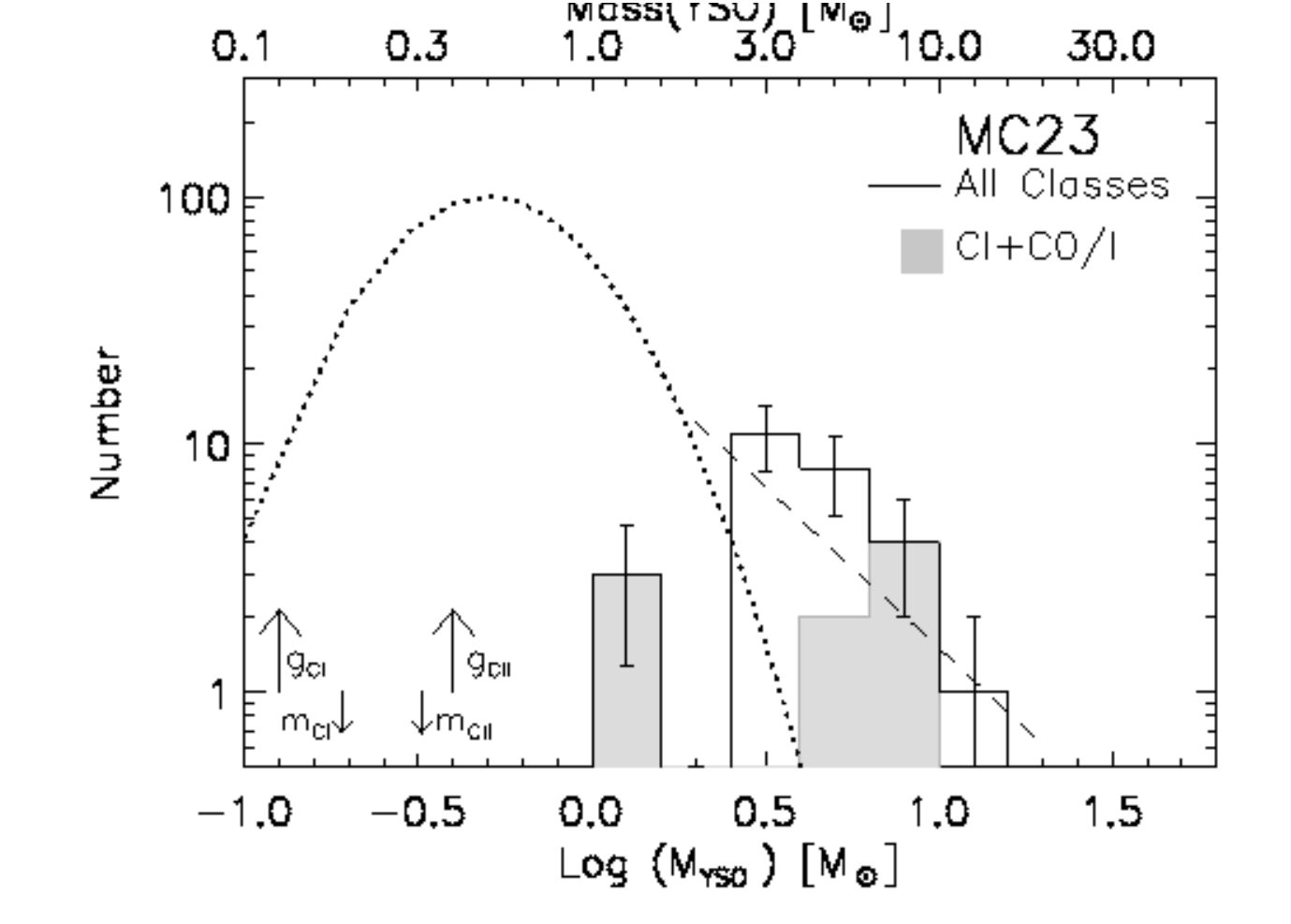} 
\includegraphics[width=0.67\columnwidth]{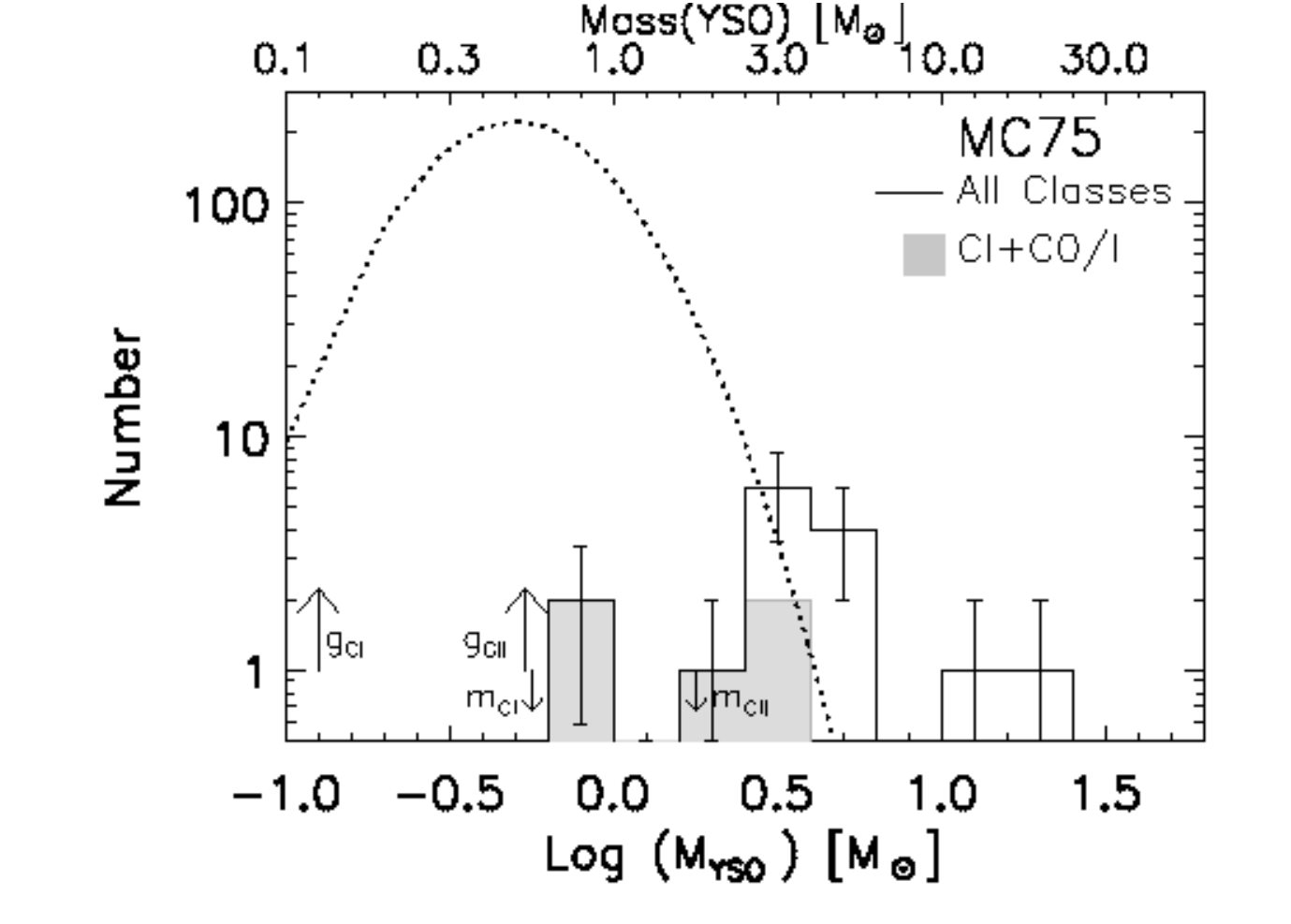} 
\includegraphics[width=0.67\columnwidth]{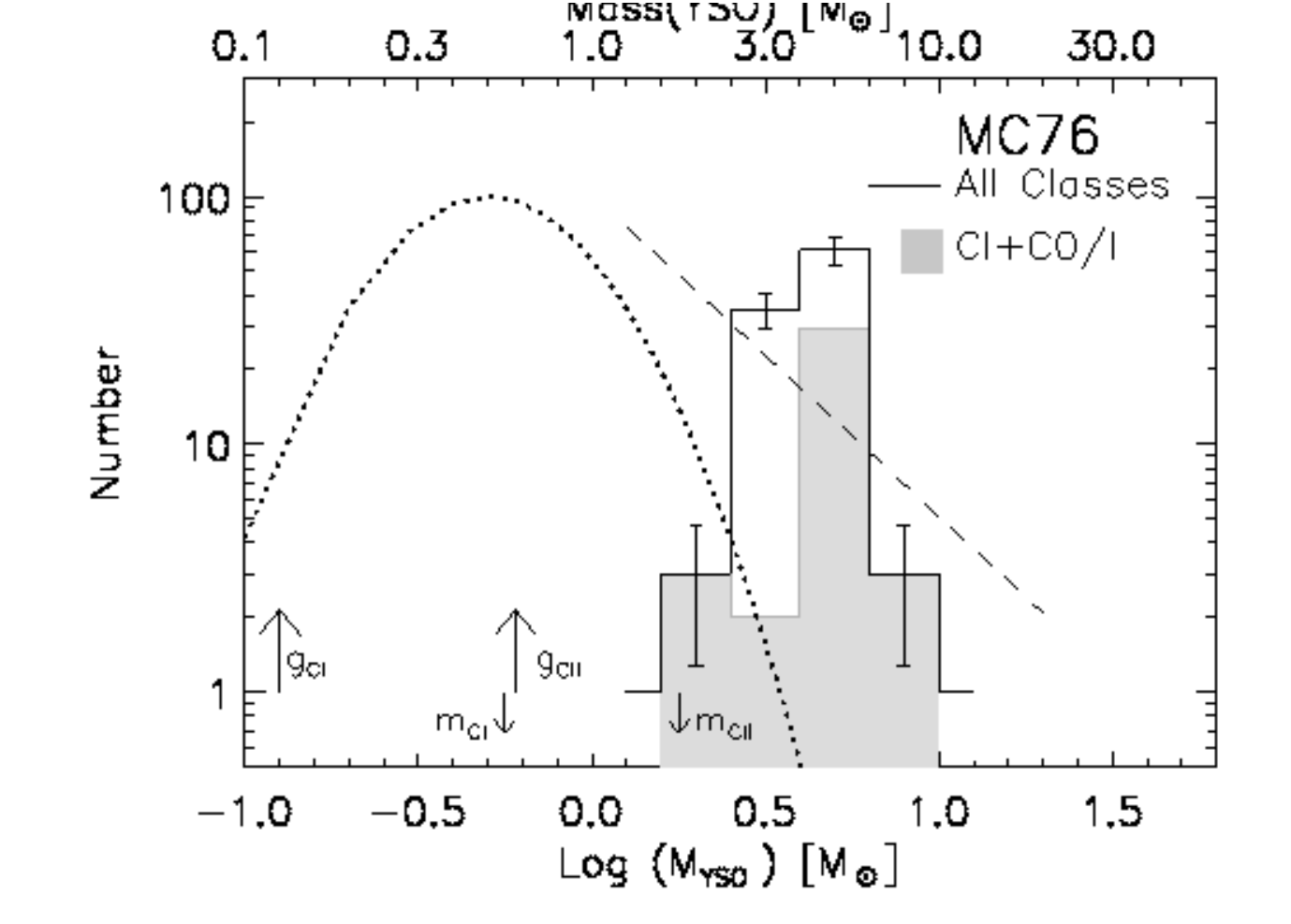} \\
\includegraphics[width=0.67\columnwidth]{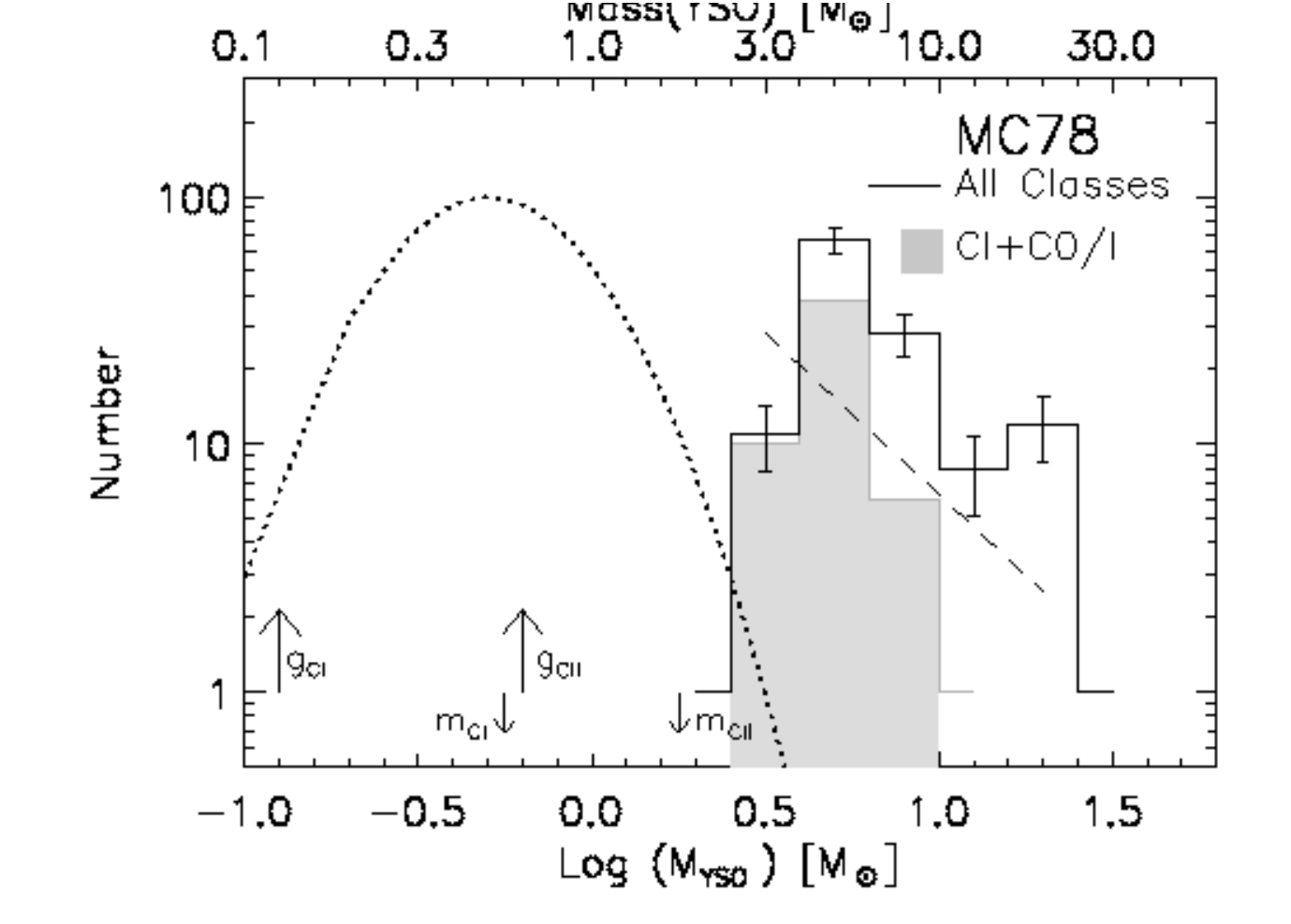} 
\includegraphics[width=0.67\columnwidth]{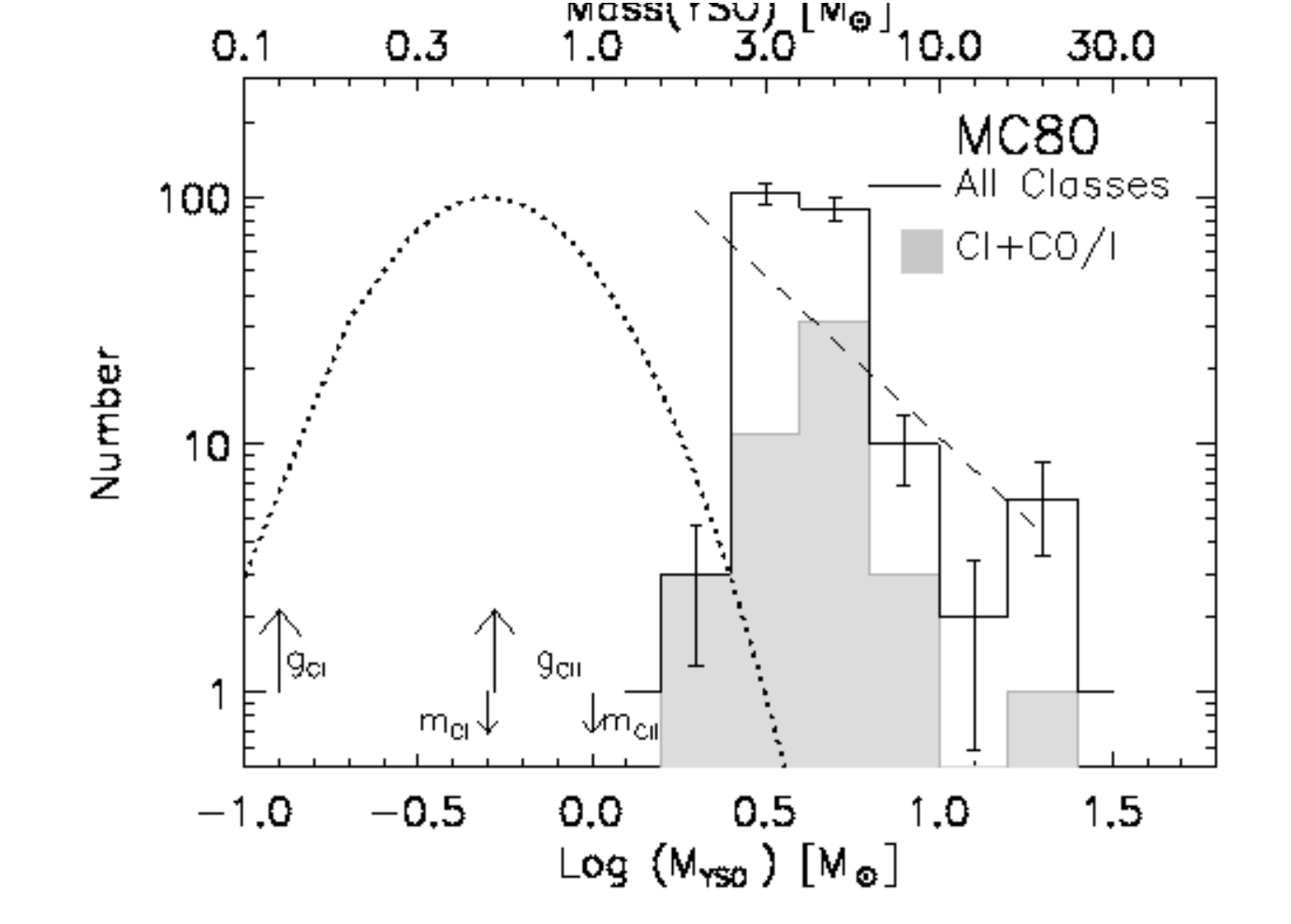} 
\includegraphics[width=0.67\columnwidth]{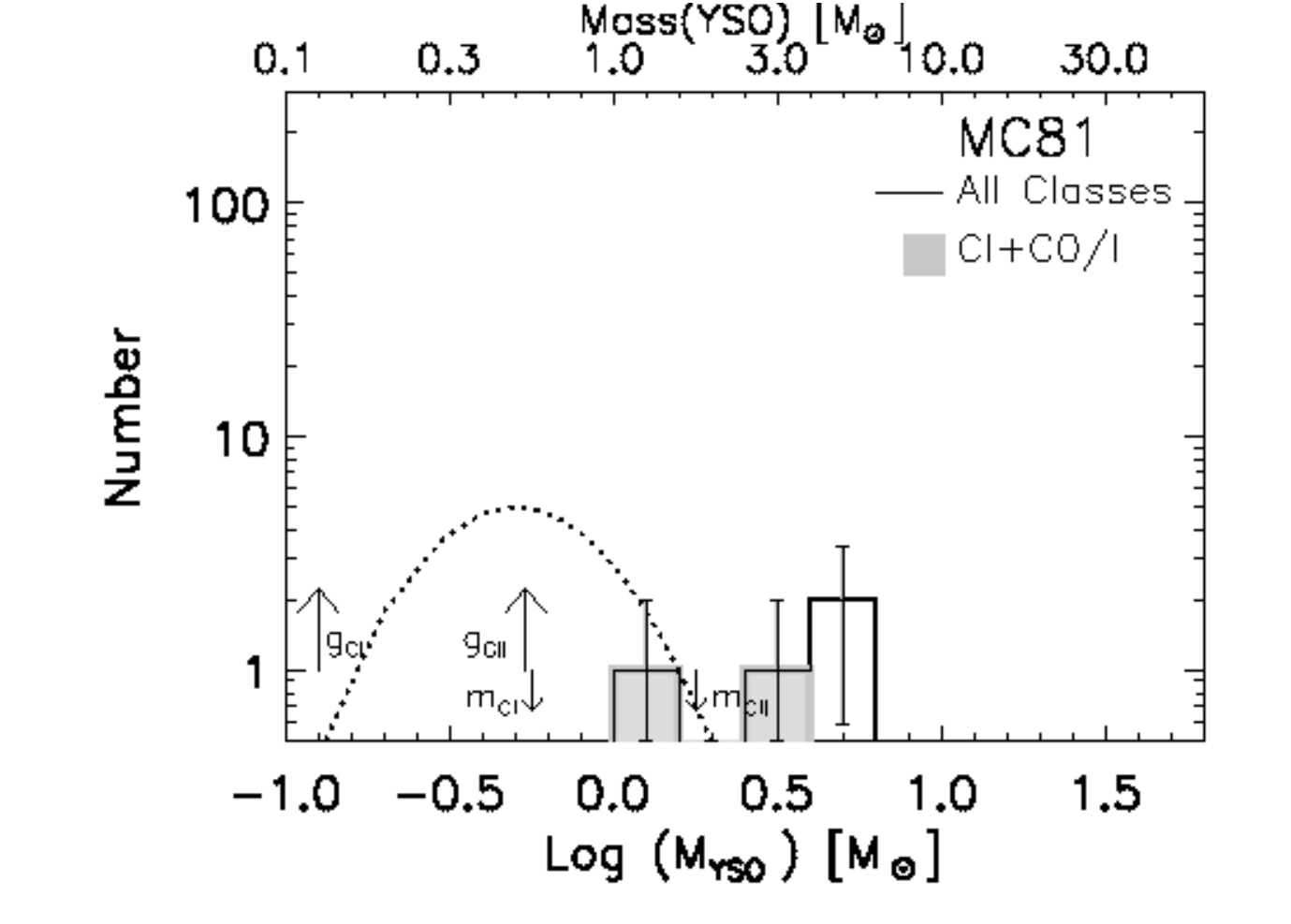} \\
    \end{center}
    \caption{
Mass Function (MF) of the HLYSOs (shaded histogram) compared to a log-normal 
\citet{chabrier03} IMF (dotted curve) that is normalized for each MC so as 
to contain the same number of sources as the observed number of LLYSOs in 
that MC. The contribution of Class I and Class 0/I sources to the histograms
is shown by the shaded part. The error bar for each bin of the histogram 
corresponds to $\sqrt{N}$ statistical uncertainties. 
The arrows marked with m$_{\rm CI}$, m$_{\rm CII}$,
g$_{\rm CI}$ and g$_{\rm CII}$ correspond to the detection limit for Class~I 
and Class~II YSOs in MIPSGAL and GLIMPSE surveys, respectively.
{ These values were obtained with the procedure described in \S3.0.}
There is clearly an excess number of HLYSOs over the \citet{chabrier03} IMF.
This excess is well represented by a Salpeter-like IMF (dashed line).
}
\label{figure6}
\end{figure*}

\begin{figure}[ht!]
     \begin{center}
            \includegraphics[angle=0,scale=0.65]{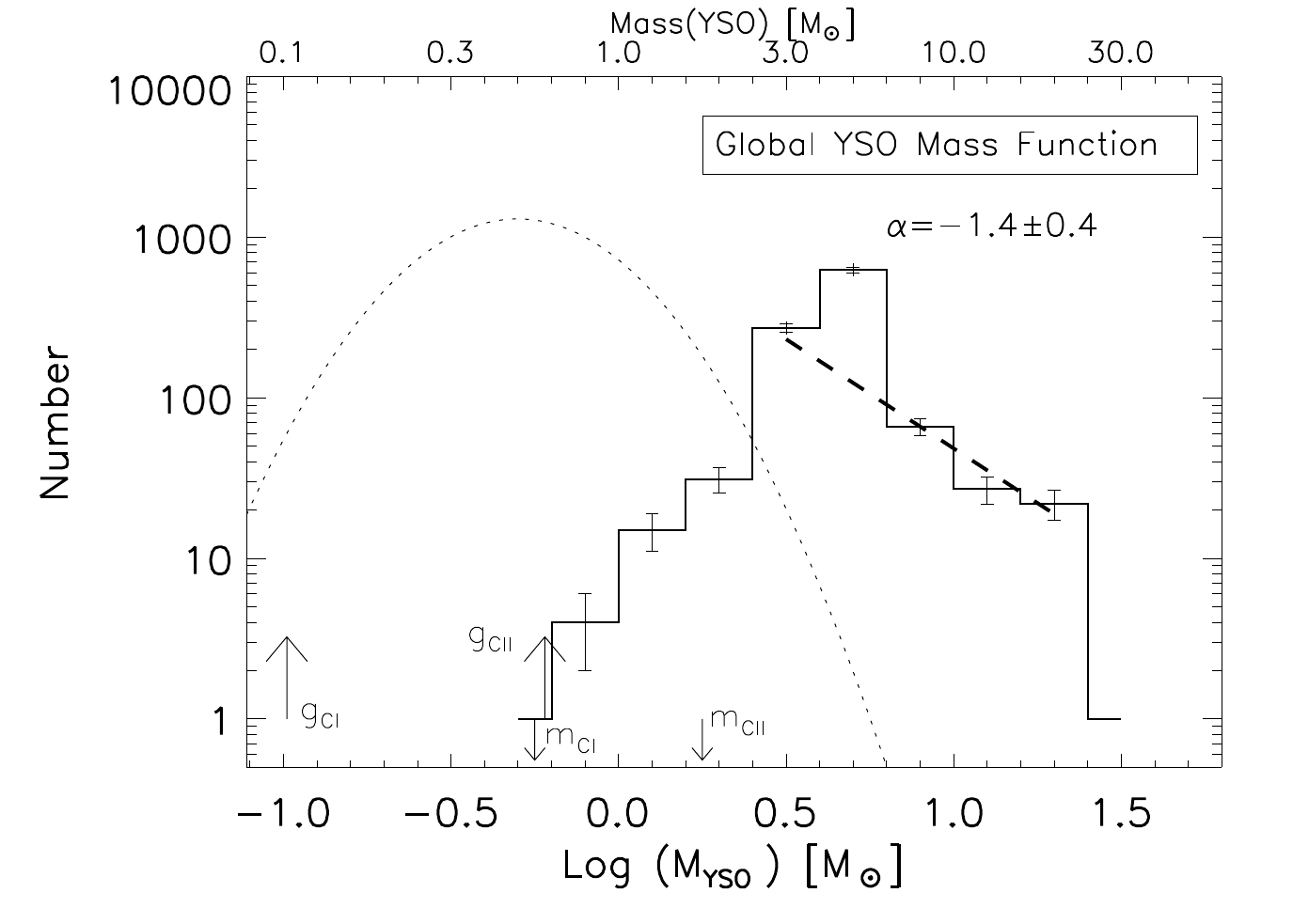}
    \end{center}
\caption{ 
Mass Function (MF) of the HLYSOs (histogram) and the log-normal IMF 
(dotted curve) of LLYSOs in the 12 clouds.  
The arrows marked with m$_{\rm CI}$, m$_{\rm CII}$,
g$_{\rm CI}$ and g$_{\rm CII}$ correspond to the detection limit for Class~I 
and Class~II YSOs in MIPSGAL and GLIMPSE surveys, respectively.
The error bar for each bin of the 
histogram corresponds to $\sqrt{N}$ statistical uncertainties.
The high-mass tail of the histogram was fitted with a single power law 
(dashed line) and resembles well to Salpeter-like IMF slope.}
   \label{figure7}
\end{figure}

Observational determination of the mass function (MF) of YSOs is 
critical for a detailed understanding of star formation. Studies of nearby 
star-forming regions have by now established the nature of the MF of YSOs 
in the $c2d$ and GB surveys \citep{enoch+06,andre+10}. Particularly, 
\citet{andre+10} found that the Protostellar Core Mass Function resembles 
very much the log normal Stellar MF \citep{chabrier03}, but with the 
characteristic mass $\sim$3 times higher.
It should be noted that the regions analyzed in these 
studies lack the high-mass star population, and hence the application of 
Chabrier function to star-forming regions containing high-mass stars, such
as the regions studied in the present work, is still questionable.
There is a growing evidence that such regions have an excess of high-mass 
stars above the Chabrier function, which can be well represented by the
Salpeter (1955) IMF \citep[see the review by][]{bastian+10}. We here 
investigate whether this is the case for the star-forming regions 
studied in the present work.

The method we have followed for the MIPSGAL sources to obtain the YSO 
masses allows us the construction of the high-mass end of the MF in all MCs.
The resultant MFs are plotted in Figure~\ref{figure6} by solid histograms.
The MFs are binned in logarithmic mass intervals of 0.3~dex. 
The error in the masses due to the unaccounted contribution of the 
accretion in deriving the YSO masses, would over-estimate
the masses at the most by one bin width.
In Figure~\ref{figure6}, we also show a log-normal Chabrier MF for the LLYSOs.
This function is centered at a characteristic mass of 0.5~\msun\ 
and normalized in such a way that the total number above 0.1~\msun\ for the
function is equivalent to the total number of observed LLYSOs in each cloud.
The plotted Chabrier MF assumes that all the YSOs more massive 
than 0.1~\msun~ are detected in the GLIMPSE survey. For each MC, we 
show the sensitivity limit for Class~I sources in the figure. It can be seen 
that almost all the Class~I sources are indeed detected by GLIMPSE in our 
sample clouds. On the other hand, we may be missing the Class II sources 
with masses lower than $\rm g_{\rm II}$, the limiting GLIMPSE mass for Class II 
sources. In all cases $\rm g_{\rm II}$ is less than the characteristic mass, and 
hence in the extreme case, the actual numbers could be a factor of two higher
than the detected YSOs.

Even after taking into account the possible error in the normalization of
the Chabrier MF, the number of expected stars more massive than 3~\msun~
is less than 1 in our sample of MCs. Thus, the observed number of high-mass 
stars is clearly above those expected from the Chabrier MF in all clouds.
On the other hand, the observed number of massive stars is consistent with
a Salpeter IMF for stars more massive than $\sim$2~\msun. 
That is shown in the Figure~\ref{figure7}, where the mass function for 
the total YSO population of the cloud sample is plotted. In this figure, the mass 
distribution of the HLYSOs is shown with the histogram while the dotted curve 
represents the mass distribution of the LLYSOs. A error weighted fit for the 
masses greater than 2~\msun\ is plotted with the dashed line, where the slope 
found ($\rm \alpha=-1.4\pm0.4$) resembles well to the Salpeter index ($\rm \alpha=-1.3$). 
This hybrid Chabrier plus Salpeter mass function that we propose for our regions, 
is in fact the favored MF in other regions with massive SF 
\citep[see the review by][]{bastian+10}.

\subsection{Age spread in the population of YSOs}

More massive a star is, lesser is the time it spends as a YSO. For example
stars more massive than 6~\msun\ reach the zero age main sequence in $<$0.2~Myr
 \citep{tognelli+11}. Thus, in a scenario where all stars formed over a period 
of time shorter than 0.2~Myr (instantaneous), more massive stars are expected to be in later 
stages of evolution. Given that our HLYSOs are systematically more massive 
than LLYSOs, we would expect a higher fraction of Class II and Class III sources
among HLYSOs as compared to the LLYSOs.

The numbers of YSOs in each Class for HLYSOs and LLYSOs are given in Table~5.
The most dominant population among both samples is the Class II, constituting
more than 50\% and 70\% for the HLYSO and LLYSO samples, respectively. 
This is expected as stars spend most of their PMS lifetime as Class II sources.
The fraction of Class III sources is marginally higher for the HLYSO sample 
as compared to that of LLYSO sample, as expected in the instantaneous SF scenario.
However, in such a scenario, we won't expect Class 0/I sources in the HLYSO sample.
Presence of massive stars in these early stages suggests that the massive star formation
process is not instantaneous in majority of our clouds.
There are stars of 3~\msun\  in Class III stage in majority of these clouds, 
implying star formation, including the formation of massive stars, has started 
at least 2~Myr ago. Thus, YSOs in our clouds have an age spread of $>$2~Myr. 

The MIPSGAL sources without GLIMPSE counterpart, namely Class I/0 YSOs are an 
interesting population, especially if these are massive sources. The nature 
of these massive (M$\rm_{YSO}>$3~M$\rm_{\odot}$) and bright sources could be 
linked to the earliest (and brightest) objects in the embedded population. 
These kind of sources are
associated to (massive) Class 0 YSOs \citep{andre+10} or transitional 
Class 0/I YSOs.  Due to their early evolutionary phase these objects
don't have (detectable) emission at NIR/MIR \citep{andre+10}
and must be (only) detected/observed at submm/mm spectral range with a 
weak contribution at \mipslam. The Class I/0 sources also could be 
confused easily as Class I/II YSOs deeply embedded in the high column density 
regions of the molecular cloud. Due to the high fraction of the Class I/0 
sources, these could be a sum of true Class I/0, and a deeply embedded 
Class I and Class II YSOs. 
Indeed, edge-on Class II YSOs with high extinction may be mis-classified 
as Class I objects \citep{robitaille+06,offner+11}.  
Nonetheless, the Class I/0 objects without GLIMPSE 
counterpart are excellent candidates to embedded prestellar and 
protostellar core population and constitute the youngest population 
(from 0.1~Myr to 0.5~Myr age; \citet{enoch+06,andre+10}) of our
sample of high-mass star formation regions. Their MF in a complete 
sample must give us a clue to the connection between the prestellar and 
protostellar MF.

\subsection{Location of High Mass YSOs in the clouds.}

\begin{figure}[ht!]
     \begin{center}
     \includegraphics[width=8.6cm]{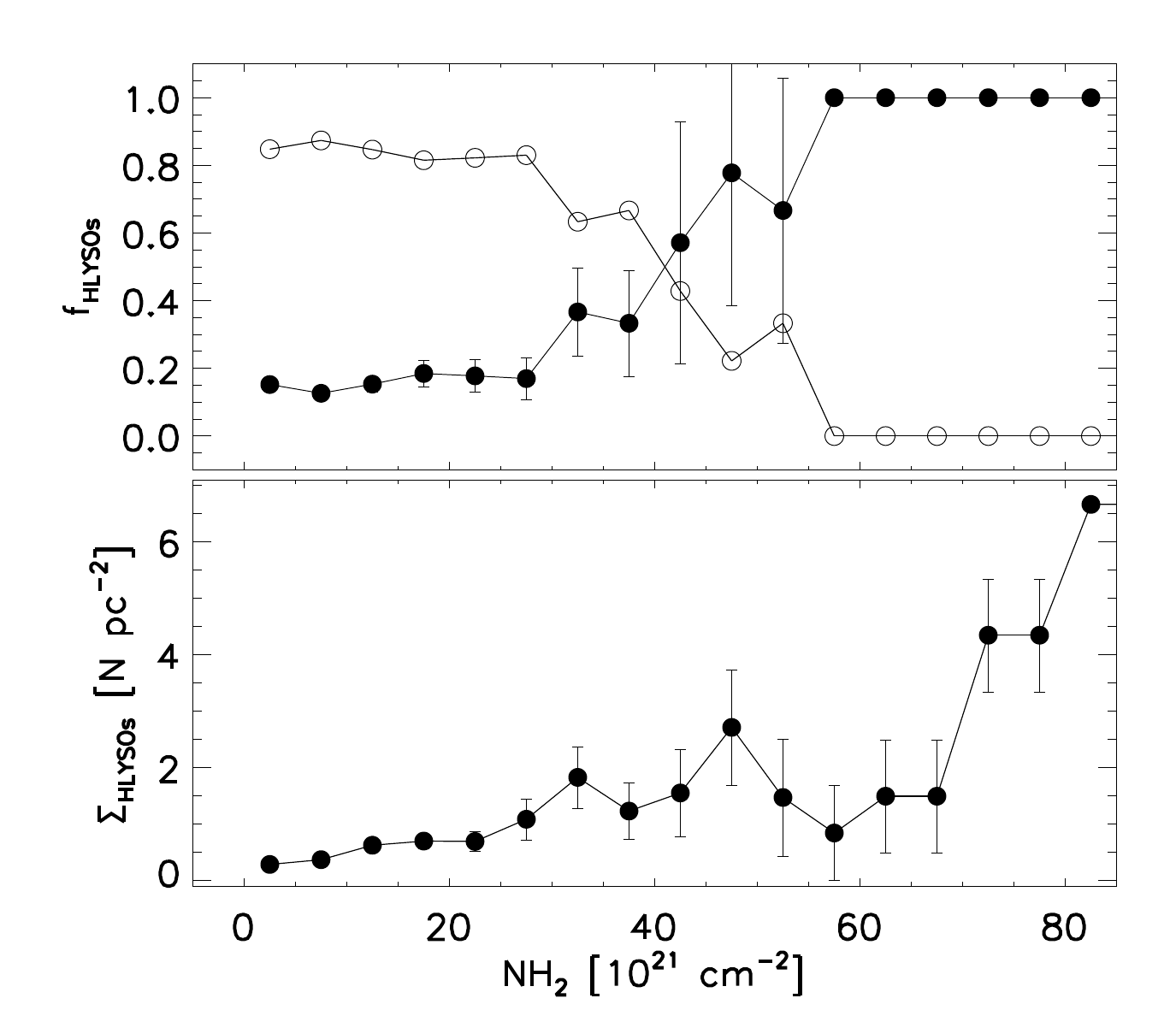}
    \end{center}
\caption{
The fraction (top pannel; filled circles) and surface density (bottom pannel; filled circles) 
of the High-Mass YSOs as a function of H$_2$~column density for our entire MC sample.
The fraction of Low-Mass YSOs is shown with the open circles in the top pannel. 
The bin width of the plot is $5.0\times10^{21}$~\cmsq.
}
\label{figure8}
\end{figure}

The HLYSOs, for definition, are brighter than 10~\lsun~(\S~4.2). 
This cut-off luminosity corresponds to a mass of 2~\msun\ for a Class I YSO,
whereas it is $\sim$3~\msun\ for the Classes II and III YSOs. This implies
that all the HLYSOs are of intermediate to high mass. For the purpose of 
the discussion in this section, we will refer these YSOs as {\it high-mass} YSOs.
Recent works \citep{krumholz+08,lopezsepulcre+10} suggest that a high density 
environment is required to form high-mass stars.  
Thus, the relative fraction of HLYSOs is expected to increase 
with the surface density of the gas. In order to explore this idea, we
plot the ratio $\rm f_{HLYSO}\equiv N(HLYSO)/(N({HLYSO})+N({LLYSO}))$
against \nhtwo\ in Figure~\ref{figure8}. This ratio contains all HMYSO population 
of the MC sample.
The high-mass YSO fraction almost remains constant at $\sim$0.15 up to 
$\sim3.0\times10^{22}$~\cmsq, above which the $f_{\rm HLYSO}$ starts to rise 
reaching unity value at $\sim5.5\times10^{22}$~\cmsq.
In this gas density range, the surface density of HLYSOs increases from
$\sim$1~pc$^{-2}$ to more than 6~pc$^{-2}$. In this plot, the fraction of LLYSOs 
($\rm 1-f_{HLYSO}$) are shown with open circles.
The error bars are obtained assuming a poisson error for each bin in the 
distribution for both  the HLYSOs and total YSOs sample. These errors in the 
counts are propagated in the ratio defined previously ($\rm f_{HLYSO}$). 
The observed tendency agrees very well with the notion that high-mass stars
require higher column densities for their formation.  

\citet{krumholz+08} addressed theoretically the issue of initial physical conditions 
for the formation of massive stars. They found a 
minimum density of 1.0~${\rm g}~{\rm cm}^{-2}$ as necessary for the formation 
of the massive stars. However, Galactic star-forming regions are found to
harbor high-mass stars at gas densities from 0.1 to 1.0 ${\rm g}~{\rm cm}^{-2}$
\citep{rathborne+09,ragan+13}. In our MCs, the high-mass fraction is 1 above 
$\sim~5.5\times10^{22}$~\cmsq, which corresponds to $\sim0.25~{\rm g}~{\rm cm}^{-2}$. 
Thus our observed values are consistent with the values observed for other high-mass Galactic clouds.

\subsection{Star formation rate and star formation efficiency}

\begin{figure}[h!]
     \begin{center}
            \includegraphics[width=9cm]{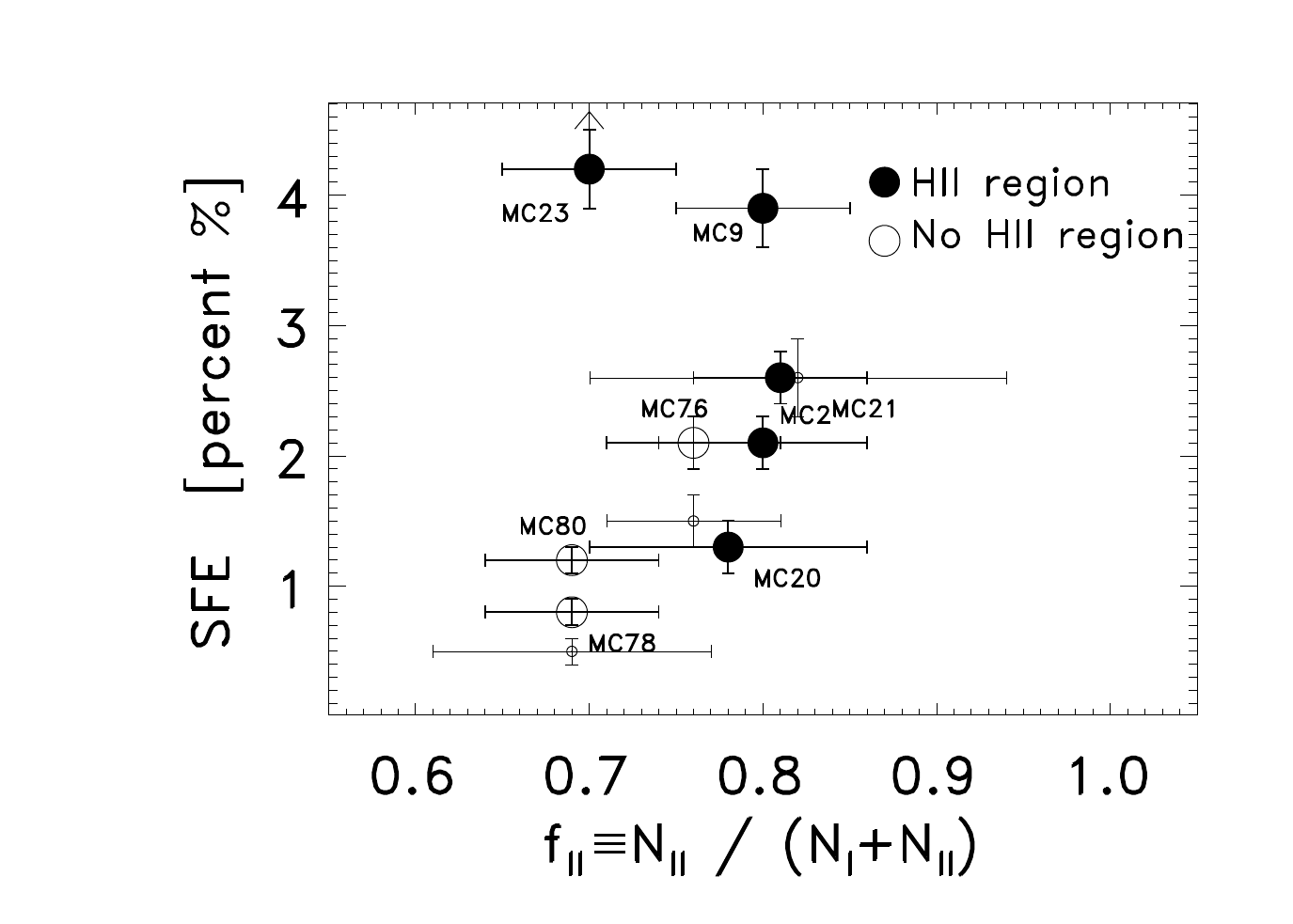} \\
    \end{center}
    \caption{
Star formation efficiency as a function of the Class II fraction in 
each of our MCs. Clouds containing IRAS sources with UCHII region detected 
are shown by filled symbols.
     }
   \label{figure9}
\end{figure}

\begin{table*}
\begin{center}
\centering 
\caption{Star formation related parameters in our MC sample. }
 \begin{tabular}{cccccccccc}
 \hline
 \hline
     &       &       &              &        &  &       &             &    &          \\
 GMC & $\rm f_{II}$ & $\rm t_{II}$ &f$_{\rm HL}$ & Radio  & Maser  &  SFR  &  $\overline\Sigma_{\rm SFR}$  &{SFE} &  $\overline\Sigma{\rm _{gas}}$\\
(1) & (2) & (3) & (4) & (5) & (6) &  (7) & (8) & (9) & (10) \\
 \hline
MC2  & 0.80$\pm$0.06  & 1.98$\pm$0.13  & 0.21$\pm$0.03  & Yes (1)  & CH$_3$O (2) & 585$\pm$294 & 2.1$\pm$1.1  &  2.2$\pm$0.1  & 155$\pm$79  \\             
MC9  & 0.80$\pm$0.05  & 1.89$\pm$0.13  & 0.14$\pm$0.01  & Yes (3)  & H$_2$O  (4)   & 1105$\pm$556 &19.1$\pm$9.6  &  3.9$\pm$0.2  & 646$\pm$332 \\
MC20 & 0.78$\pm$0.07  & 2.02$\pm$0.14 & 0.10$\pm$0.02  & Yes (5)  & CH$_3$O (6) & 357$\pm$182 & 6.4$\pm$3.3  &  1.2$\pm$0.1  & 647$\pm$320 \\
MC21 & 0.81$\pm$0.05  & 1.99$\pm$0.14  & 0.22$\pm$0.02  & Yes (7)  & \ldots              & 916$\pm$461 & 2.3$\pm$1.2  &  2.6$\pm$0.2  & 133$\pm$60  \\
MC23 & 0.70$\pm$0.05  & 1.93$\pm$0.14  & 0.05$\pm$0.01  & Yes (8)  & H$_2$O  (4)   & 370$\pm$186 & 26.8$\pm$13.5 & 11.9$\pm$0.7 & 320$\pm$240  \\
MC76 & 0.76$\pm$0.05  & 1.99$\pm$0.14  & 0.15$\pm$0.02  & No  (8)  & CH$_3$O (9) & 608$\pm$306 &  7.5$\pm$3.7 &  2.1$\pm$0.1  & 565$\pm$310 \\
MC78 & 0.69$\pm$0.05  & 1.97$\pm$0.14  & 0.17$\pm$0.02  & No  (7)  & \ldots              & 1210$\pm$610 & 4.6$\pm$2.3  &  0.8$\pm$0.1  & 763$\pm$365 \\
MC80 & 0.69$\pm$0.05  & 1.83$\pm$0.13  & 0.32$\pm$0.03  & No (11)  & H$_2$O  (4)  & 955$\pm$480 & 5.6$\pm$2.8  &  1.2$\pm$0.1  & 816$\pm$410 \\
  \hline
MC1  & 0.82$\pm$0.12  & 1.97$\pm$0.14   & 0.08$\pm$0.02  & Yes (1)    & No      (4)       & 77$\pm$39      & 3.3$\pm$1.7  &  2.7$\pm$0.3  & 210$\pm$114  \\
MC12 & 0.69$\pm$0.08 &  2.21$\pm$0.16  & 0.04$\pm$0.02  & Yes (12) & No      (4)       & 101$\pm$52   & 1.1$\pm$0.6  &  0.5$\pm$0.1  & 358$\pm$179 \\
MC75 & 0.76$\pm$0.05  & 1.99$\pm$0.14  & 0.02$\pm$0.01  & Yes (13) & No      (4)       & 312$\pm$157  & 5.6$\pm$2.8  &  1.5$\pm$0.1  & 613$\pm$320 \\
MC81 & 0.43$\pm$0.30  & 2.02$\pm$0.14  & 0.28$\pm$0.16  & Yes (14) & No      (4)        & 16$\pm$10     & 1.0$\pm$0.7  &  0.8$\pm$0.3  & 218$\pm$64  \\
\hline
 \end{tabular}

\tablecomments{ 
Brief explanation of columns:
(1) GMC name;
(2) fraction of Class II sources defined as $\rm f_{II} = N_{\rm II} / ( N_{\rm I} + N_{\rm II} )$;
{(3) timescale of Class II sources defined as $\rm t_{II} = 2.0\times ( N_{\rm I} / N_{\rm II} )\times [ f_{\rm II} / ( 1 - f_{\rm II} )]$, in $\rm [Myr]$ units;}
(4) fraction of high luminosity YSOs defined as $\rm f_{HL}=N(HLYSOs) / N(LLYSOs)$;
(5) Whether or not radio continuum at cm wavelengths (3.6~cm, 4.0~cm, 6~cm and 21~cm) detected 
in the cloud. The number in the parenthesis gives the reference of the radio observations,
as identified below under References;
(6) Name of the maser emission if detected. The last 4 MCs have been targets for the H${\rm_2}$O maser, but not detected;
The number in the parenthesis gives the reference of the maser observations, as identified below under references;
(7) SFR=M$_{\rm YSOs}/t_{\rm YSOs}$ defined in \S4, in ${\rm [M_{\odot}\,Myr^{-1}]}$ units;
(8) Surface density of star formation defined as ${\rm \overline\Sigma_{SFR}=SFR/Area}$, 
in $[{\rm M_{\odot}\,Myr^{-1}~pc^{-2}}]$ units. Area is given in column 10 of Table~1;
(9) Star formation efficiency (in \% units) defined as 
SFE$\equiv100\times$M$_{\rm YSOs}$/(M$_{\rm YSOs}$+M$_{\rm cloud}$), where M$_{\rm cloud}$ 
is the $M_{\rm XF}$ given in column 6 of Table~2;
(10) Average gas density of the MC defined as ${\rm \overline\Sigma_{gas}=M_{\rm XF}/Area}$
} 
\tablerefs{
Meaning of numbers under columns 5 and 6:
(1) \citet{urquhart+09}; (2) \citet{lim+12}; (3) \citet{rivera+10}; (4) \citet{codella+95}; 
(5) \citet{wood+89}; (6) \citet{walsh+97}; (7) \citet{sridharan+02}; (8) \citet{walsh+98}; 
(9) \citet{walsh+97}; (10) \citet{fontani+10}; (11) \citet{wang+09}; (12) \citet{condon+98}; 
(13) \citet{kurtz+94};  (14) \citet{becker+94}. }
\end{center}
\end{table*}

The values for the SFR in our sample of MCs lie between 16 and
1220~\msunmyr, with a median value of 585~\msunmyr. These values are 
comparable to the values of 715 and 159~\msunmyr\ for 
Orion A and B, respectively \citep{lada+10}. 
SFR varies over a wide range of values in the Galactic star-forming regions.
\citet{vuti+evans13} obtained SFR values from $\sim$1 to 2530~\msunmyr\ 
for Galactic high-mass star-forming regions at distances D$>$700~pc. 
The values of SFE are in the interval from 0.5\% to 
3.9\%, with MC23 being an outlier with a value of 11.9\%. These values are in 
agreement with values of Orion B (0.6\%) and Orion A (2\%)  
star-forming regions \citep{megeath+12}. However, these values
are systematically smaller than the values of $\sim$3\% to 6\% 
obtained for nearby low-mass star-forming regions using c2d and GB data 
\citep{evans+09}. 
Values of SFE $\lesssim$~1\% are commonly found 
for distributed \starf\ regions \citep{bonell+11}. Half of our 
clouds have SFE-values $\lesssim$1\%, suggesting a distributed 
SF scenario. Recent studies of young stellar clusters have revealed 
that there is a time spread in the age of the young stars in 
stellar clusters \citep{myers12,foster+14}. 
Instantaneous mode of cluster formation may apply to compact massive 
clusters such as 
Westerlund 1 with high~SFE values \citep{kudryavtseva+12}, but it 
is possible that low-mass stellar clusters and associations are built up 
over a long period of time. Thus, SFE is expected to increase with time 
as more and more stars form. We test this hypothesis below.

In the Figure~\ref{figure9}, the radio continuum emission detection  
is denoted by the filled symbol, while no detection is denoted by 
the open symbol. 
The $\rm f_{II}$ and the timescale ($\rm t_{II}$) obtained based on 
$\rm 2.0\pm1.0$ Myr \citep{evans+09}, are shown in the columns 2 and 3 of the 
Table~6. The detection or not of radio continuum emission is also included 
in the same Table (column 5).

From the statistics of the YSO population and the molecular cloud mass
for the sample of present study,  we can determine some parameters of
the star formation activity such as Star Formation Rate (SFR) and Star
Formation Efficiency (SFE). The SFR is obtained from the simple
expression \citep{evans+09}, $SFR={\rm M}_{\rm YSO} / t_{\rm YSO}$, 
where the $M_{\rm YSO}$ and $t_{\rm YSO}$ are the ``stellar'' (YSOs) total 
mass and the timescale of the YSOs, respectively. The ``stellar'' 
masses are obtained by adding the mass of the HLYSO and LLYSO populations. 
The masses of the HLYSOs are individually obtained following the method
already described.
The total mass of LLYSOs is derived assuming a mean mass of 0.5~\msun\ for every
LLYSO. This assumption is equivalent to assuming that the sources follow 
a log normal mass distribution with a characteristic mass of 0.5~\msun. 
We then used the expression $\rm M_{LLYSO}/{M_{\odot}}=0.5\times N_{LLYSO}$ 
\citep{evans+09}, where $\rm N_{LLYSO}$ is the number of LLYSOs. 
A typical timescales of 2~Myr and 0.54~Myr for the Class II and Class I, respectively, 
are used for $\rm t_{\rm YSO}$ \citep{evans+09}. 
The Star Formation Efficiency (SFE) is obtained from the expression 
${\rm SFE}\equiv{\rm M}_{\rm YSO}/({\rm M}_{\rm YSO}+{\rm M}_{\rm cloud})$,
where M$_{\rm cloud}$ is the X-Factor mass of the molecular cloud (column 6 of Table~2). 
The resultant SFR and SFE parameters are divided by the cloud area (column 10 of 
Table~1) to obtain average values of SFR (\sigsfrav) and 
gas density (\siggasav) 
for our clouds. All these quantities are tabulated in Table~6. 
\\
\\

\begin{table*}[!h!tb]
\begin{center}
\centering 
\caption{Parameters of the broken power-law model fit}
\begin{tabular}{ccrccccccc}
 \hline
\hline
GMC  &$\alpha$&A       & $\sigma_{\alpha}$&$\sigma_A$&$\chi^2$& R   &$\Sigma_{th}$ & $y(\Sigma_{th})$   & $x_i$   \\
           &              &          &                      &                     &              &          &\tiny     &\tiny    &    \\
(1) & (2) & (3) & (4) & (5) & (6) & (7) & (8) & (9) & (10) \\
\hline
MC2  &  0.49   &  $-$1.01   &  0.10        &        0.82      &  0.69      &  0.71    &   226.0      &  1.6             & (1-4)   \\
          &  2.34   &  $-$5.33  &   0.14         &       2.32      &  0.19       &  0.90     &   \ldots      & \ldots          &  (3-6)   \\
MC9  &  0.80   &  $-$1.65  &    0.10       &        0.50       &  0.30      &  0.59     &   211.0     &  1.0             &  (1-3)  \\
          &  3.05   &  $-$6.76  &    0.05       &        4.06      &  0.10        &  0.92     &  \ldots      & \ldots          &  (3-5)  \\
MC12&  0.54   &  $-$1.18  &     0.07      &       0.44       &  0.15        &  0.92     &  285.0     &  1.6             &  (1-4)  \\
          &  2.80   &  $-$6.68  &     0.35      &        3.50      &  0.90        &   0.64   &   \ldots      & \ldots         &  (3-6)   \\
MC20&  0.25   & $-$0.56   &    0.13       &        0.31      &   0.91      &    0.82   &   255.0     & 1.0             &  (1-5) \\
          &  1.31   &  $-$3.14  &    0.10       &        1.05      &   0.27       &   0.98     &   \ldots     &  \ldots        &  (3-9)   \\
MC21&   0.73  &  $-$1.19  &    0.12     &         0.59     &  0.20         &    0.66   &   180.0    & 3.2             &  (1-3)   \\
          &   1.54  &  $-$2.96  &    0.04      &        1.71      &  0.10         &    0.94    &  \ldots     & \ldots         &  (2-5)  \\
MC23&  0.26   &    0.63     &    0.03      &        0.43     &  0.10         &    0.83    &  150.0    &15.9            &  (1-3)  \\
          &  1.94   &  $-$3.02  &    0.17     &        2.17      &  0.26         &    0.61    &  \ldots     & \ldots         &   (2-4)  \\
MC75&  0.28   &   0.29       &   0.06       &       0.33       &  0.14         &   0.93     &  237.0     &7.1             &  (1-4)   \\
          &  2.68   &  $-$5.39  &   0.14       &       6.29       &  0.18         &   0.91     &  \ldots     & \ldots         &  (3-6)   \\
MC76&  0.23   &   0.01      &   0.06       &       0.63        &  0.22         &    0.97    & 360.0     &3.1             &  (1-5)  \\
          &  3.01    &  $-$7.04 &   0.19       &       7.75         &  0.34        &    0.95    &  \ldots     & \ldots        &  (4-8)   \\
MC78&  1.01   &   $-$2.01 &   0.47       &       0.27       &   2.71        &     0.98   & \ldots      &\ldots         &  (1-18)  \\
\hline
\small Total YSOs  &  0.61    &  $-$1.38      &   0.15       &      0.07        &  5.11         &    0.26    & 445.0     &1.8             &  (1-30)  \\
                                &  2.69    &  $-$6.09     &   0.33       &        0.44        &  3.65        &    0.35    &  \ldots     & \ldots        &  (20-48)   \\
\small Class I  YSOs &  0.55    &  $-$1.70   &   0.36       &       0.47        &  6.68         &   0.96    & 380.0     &0.5             &  (1-20)  \\
                                   &  3.35    &  $-$8.96   &   1.18       &       3.85        &  8.45        &    0.98    &  \ldots     & \ldots        &  (16-37)   \\
\hline
\hline
\end{tabular}
\tablecomments{
Brief explanation of columns: the first row lists the fit parameters for the low density 
range, whilst in the next row are shown the fit parameters for the high density range;
(1) GMC name;
(2) Slope for the linear fit;
(3) Y-intercept value of the linear fit;
(4) Slope error;
(5) Y-intercept error;
(6) Chi-square value for the fit;
(7) Regression coefficient;
(8) Gas surface density intersection value (\sigth) in \msunpc~ units;
(9) \sigsfr-value corresponding to the \sigth~ in \msunmyrpc~ units;
(10) Range of data points used for each linear fit.
} 
\end{center}
\end{table*}

\section{Analysis of the Star formation law}

\subsection{Star formation law within the clouds}

\begin{figure*}[ht!]
     \begin{center}
            \includegraphics[angle=0,scale=1.0]{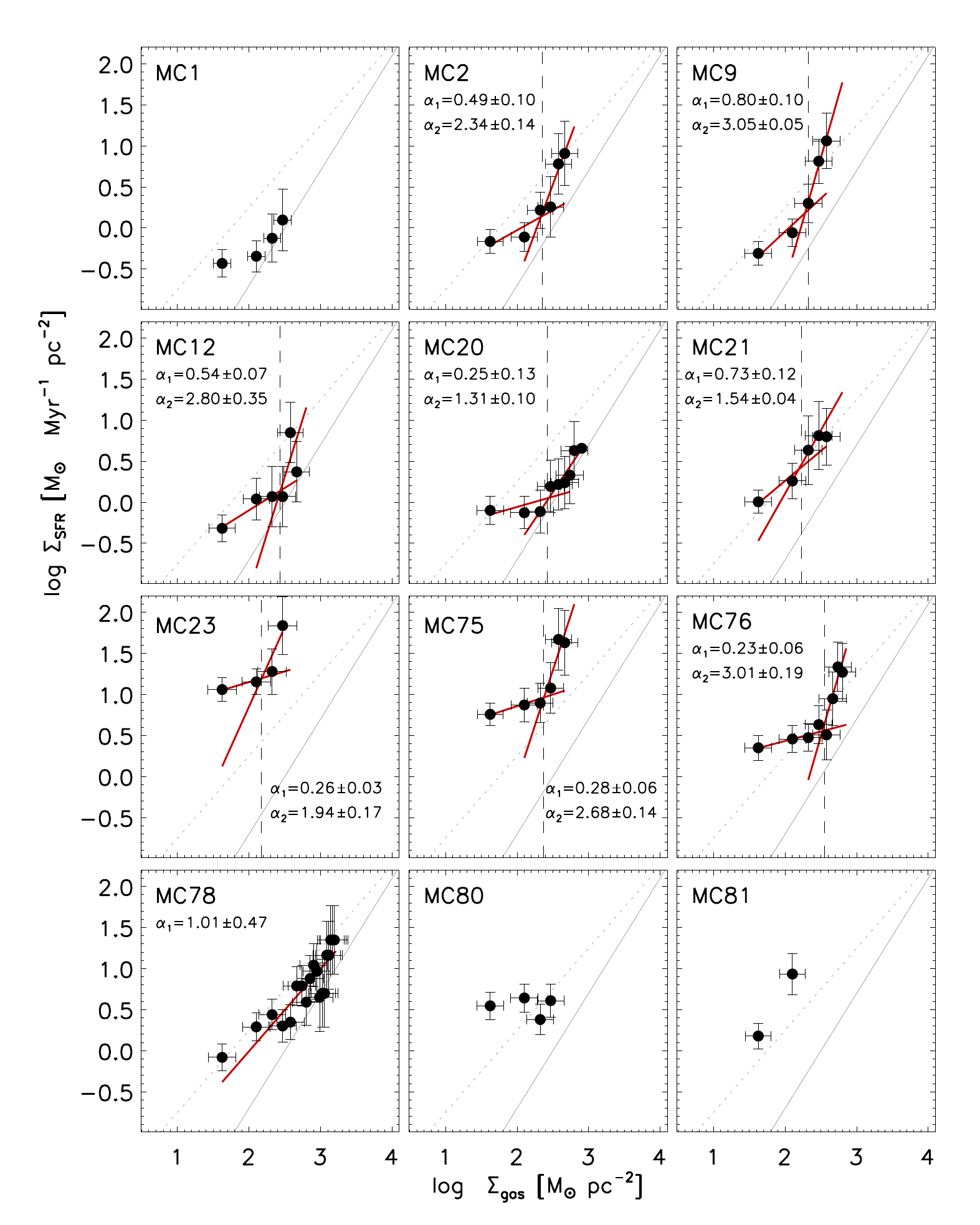}
    \end{center}
\caption{
Broken power law fits in the \sigsfr-\siggas~plane for each sample MC.
The power-law indices for the fit in the low ($\alpha_{1}$) and high gas 
density ranges ($\alpha_{2}$) are given in the top-left of the plot. The 
dashed vertical line indicates the density at the break
in \sigsfr. The Kennicutt-Schmidt law with $N=1.4$ (thin solid line) and 
the \citet{wu+05} relation (dashed line) are shown for reference.
}
   \label{figure10}
\end{figure*}

In order to study the star formation law within the molecular clouds, we 
need to obtain the surface density of star formation rate, \sigsfr, in successive
bins of gas surface density, \siggas. We used the \nhtwo\ column density maps 
derived from the \co\ data in \S2.1 in order to obtain \siggas.
The lowest \siggas\ value corresponds to the MC boundaries that are defined at column 
densities of \nhtwo=$1.0\times10^{21}$~\cmsq\ ($A_v=1$~mag; \siggas=20~\msunpc). 
A linear bin in column density is used to define
successive values of \siggas, starting at $1.0\times10^{21}$, and then taking the values 
$5.0\times10^{21}$, $1.0\times10^{22}$, $1.5\times10^{22}$, etc. 
The number of pixels in the \co\ map between 
successive column density bins are counted to determine the area occupied by the
gas at that surface density. The numbers of low and high luminosity YSOs between 
the column density contours are counted, which are then used to calculate the SFR 
following the methodology described in section \S~4.6.
The SFRs are divided by the area occupied by the contours to obtain \sigsfr\
corresponding to each \siggas.

The Schmidt law within each cloud in the range of column densities from 
\nhtwo=$1.0\times10^{21}$~\cmsq\ to $1.0\times10^{23}$~\cmsq, 
equivalent to \siggas~range from 20~\msunpc~ to 2000~\msunpc,
is shown in Figure~\ref{figure10}. For each cloud, data for each gas density bin are 
plotted with solid circles. The Kennicutt-Schmidt law with $N=1.4$ is shown by the
solid line, whereas the dotted line shows the linear relation of \citet{wu+05}
for massive clumps. In six of the twelve MCs, our points lie between the two
lines, whereas in the remaining six, most of our data points lie even above
the linear relation of \citet{wu+05}.
From a close inspection of the plot for 8 clouds (MC2, MC9, MC12, MC20, MC21, 
MC23, MC75 and MC76), it can be seen that
our data points are not consistent with a power-law fit with a single index ($\alpha$)
(linear in the log-log plane) over the entire range of gas densities. Instead, 
a broken power-law (BPL) relation seems a better fit to the 
data. Such relation is found previously by \citet{heiderman+10} in their data, 
who used separate power-laws 
of the form $y=\alpha x+A$ ($y=\log_{10}$(\sigsfr); $x=\log_{10}$(\siggas)),
one in the low density regime (index $\alpha_1$ and coefficient $A_1$) 
and the other one in the high density regime (index $\alpha_2$ and coefficient $A_2$).
We followed a similar procedure to fit our data points. We fitted two separate
lines for the low density and high density regimes and obtained the slopes 
($\alpha_1$ and $\alpha_2$) by following a weighted least-square fitting method,
with the weights defined as the inverse square of the errors. 
The best-fitted lines are shown in the figure, where we also give
the slopes $\alpha_1$ and $\alpha_2$ for each line. The regression coefficient
for each line ($R_1$ and $R_2$), as well the range of data points used for
fitting the low and high density regimes ($x_1$ and $x_2$) are given in Table~7.

A minimum of 3 points is used for the fit in both ranges of
densities with the bin(s) at the transition used in either fits.
Four MCs (MC1, MC23, M80 \& MC81) have too few points and in addition do not
offer the dynamical range in \siggas\ necessary to carry out a reliable fit
(all points lie at \siggas\ below $\sim$200~\msunpc). We hence did not
carry out power-law fitting to these clouds.
For MC78, only one point lies in the low-density regime, and hence we show
only the fit in the high-density regime. The relation is systematically flatter 
at the low density regime as compared
to that at the high density regime, with values for the former in the range
0.2 to 1.0, and the latter 1.3 to 3.0. Note that the SFR is weakly dependent 
on gas density at low densities with a power-law index $<1$ in almost cases 
(with exception MC21).
On the other hand, the power-law index at the high density regime varies 
from a similar value (1.3) to the canonical KS index of 1.4 to values as high as 3.6.

The intersection of the two lines in Figure~\ref{figure10} corresponds to 
the gas surface density value where the break occurs in the star formation 
relation. This value is marked by the dashed line in the figure for each cloud 
and their values are given in Table~7 under the column header \sigth.
The break happens at \siggas=150--360~\msunpc\ in our sample of clouds. 
The density at the break \siggas\ value is often referred to as threshold gas 
density, \sigth, in the literature. 
\citet{heiderman+10}, doing a BPL fit similar to that described here, found
a threshold value of 129~\msunpc\ in nearby low-mass \starf~regions, while 
\citet{lada+10} proposed a value of 116~\msunpc~(equivalent to 
$A_V\sim$8~mag). \citet{willis+15} used a \sigth=200~\msunpc~ to obtain 
the Schmidt law in galactic massive \starf~regions. 
Our lower threshold values are similar to those reported for other 
galactic \starf~clouds, whilst the higher values can be explained if is 
considered that the longer distances increase the value of the threshold 
density as was reported by \citet{heiderman+10}.

\begin{figure}[!ht]
     \begin{center}
            \includegraphics[angle=0,scale=0.7]{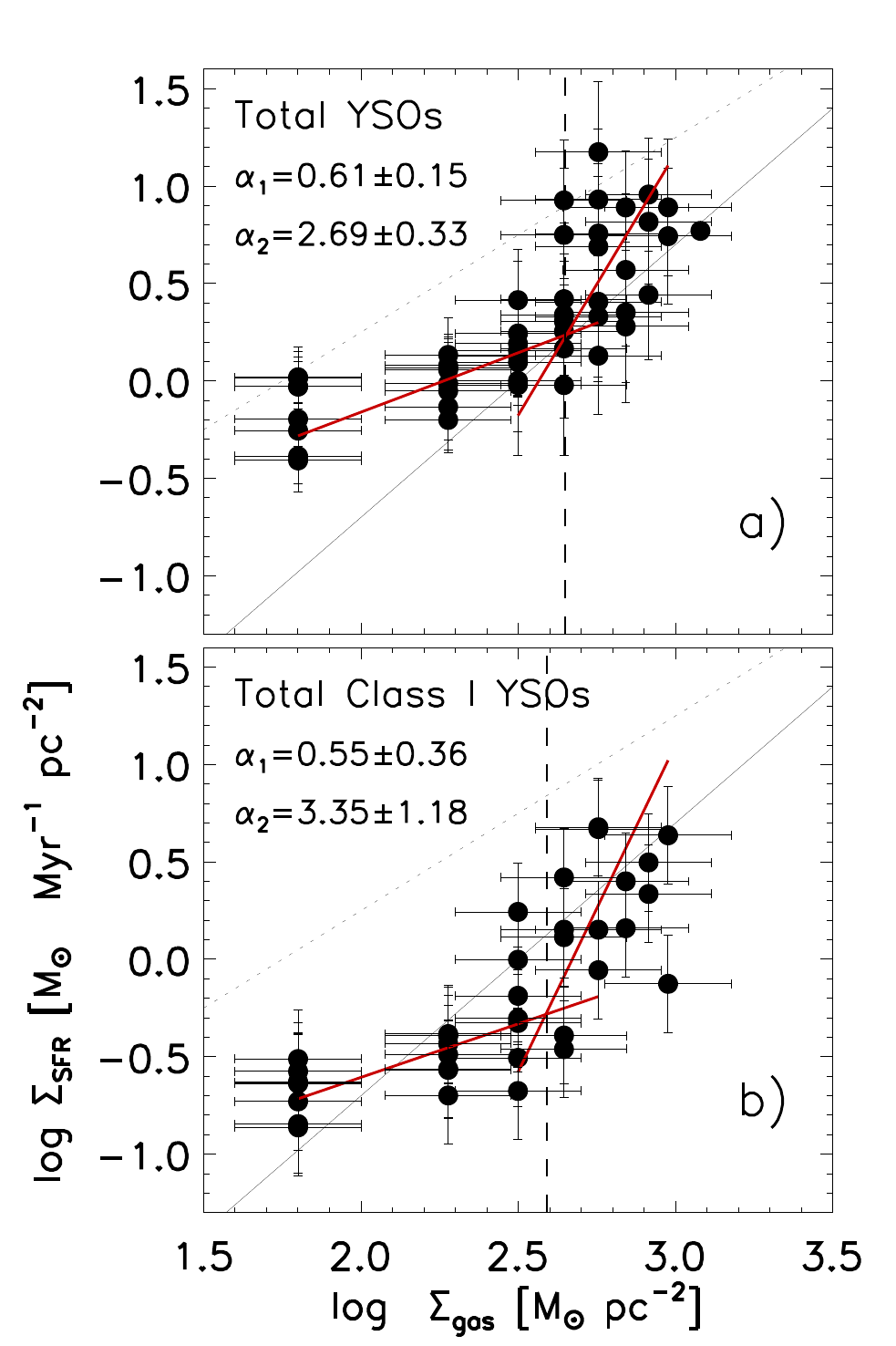}
    \end{center}
\caption{
\sigsfr-\siggas~relation for (a) the total all classes and (b) Class I YSOs 
in the clouds that presents the BPL trend.
The power-law indices for each gas density range ($\alpha_{1}$ and $\alpha_{2}$) 
are given in the top-left of the plot. The dashed vertical line indicates the density 
at the break in \sigsfr. The Kennicutt-Schmidt law (thin solid line) 
and the \citet{wu+05} relation (dotted line) are shown for reference.
}
   \label{figure11}
\end{figure}

In order to explore if the resulting trend for a BPL fit to the data is 
statistically significant, we used a total YSOs sample for the clouds that 
show the trend and several fits were tested. This was made for two cases, 
for all Classes and Class I YSOs. First, in each cloud it was obtained the 
y-value (\sigsfr) (column~9, Table~7) related to the threshold value in 
\siggas\ (column~8, Table~7). Then each data point of the \sigsfr-\siggas 
plot is divided by this y-value and the resulting data distribution is added 
in a total normalized sample.  In this sample, the best error weighted fit 
is given by a broken power law relation with a slightly slope for lower 
densities ($\alpha$=0.6,~0.6) and steeper slope for higher densities 
($\alpha$=2.7,~3.4), similarly as previously was found in the individual 
fits for clouds. Therefore, we suggested that the BPL fit 
is adequate to describe the data point distribution for each cloud with the 
BPL trend.  The fitted parameters are shown in the Table~7 and BPL fit 
of the normalized samples (all Classes and Class I) is plotted in the 
Figure~\ref{figure11}.

\subsection{Global Star formation law}

\begin{figure}[!ht]
     \begin{center}
            \includegraphics[width=9.5cm]{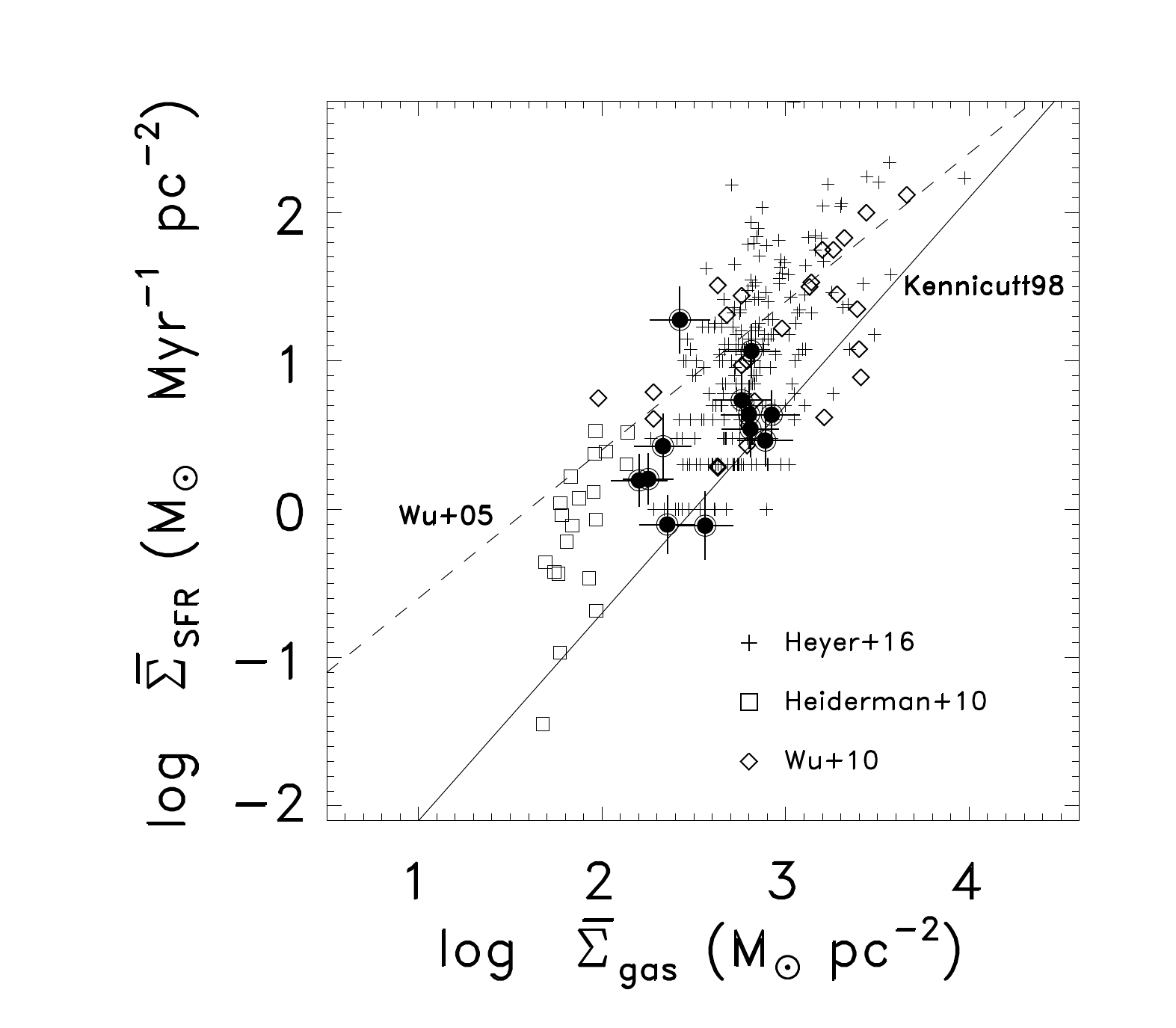}
    \end{center}
\caption{
Globally averaged Schmidt law for the clouds. The Kennicutt (1998) relation 
is plotted with solid line, while the linear relation of Wu et al. (2005) is 
plotted with dashed line. The data of previous studies for galactic \starf~
regions are taken from c2d+GB surveys (square symbol; 
\cite{heiderman+10}) and dense massive clumps (diamond symbol, 
\cite{wu+10}; cross symbol from \cite{heyer+16}). The filled symbols shown 
the values for the clouds.
}
\label{figure12}
\end{figure}

We now investigate the position of each cloud as a whole in the global SF law.
For doing this, we compare the globally averaged values of \sigsfrav\ and \siggasav\ 
for our clouds (given in Table~6), with the 
global values for Galactic clouds and clumps obtained in other studies
in Figure~\ref{figure12}. Our \sigsfrav--\siggasav\ 
values are shown by solid circles, while the data from the study of 
\citet{heiderman+10} are plotted with open squares, and  the massive dense 
clumps from \citet{wu+10} and \citet{heyer+16} are plotted with diamonds and cross 
symbols, respectively. 
The linear relation from \citet{wu+05} and Kennicutt relation \citep{kennicutt98} 
are shown by dashed and solid lines, respectively. \citet{heiderman+10} had found 
that their global results for low mass \starf~regions lie above the KS relation 
by a factor of up to 17. They found the factors to be as high as 54 for 
regions containing the youngest sources (Class I and Class Flat). This latter
value overlaps with values for the high-density massive clumps of \citet{wu+10} 
and \citet{heyer+16}.   
Our clouds lie between the linear relation of \citet{wu+05} and the canonical 
Kennicutt relation, and mostly related to the lower SFR dense clumps from 
\citet{heyer+16} and \citet{wu+10}.

In recent work by \citet{heiderman+10}, it has been found that the use of 
\co\ in the calculation of mass of a molecular cloud results in an underestimation 
of the mass compared to if visual extinction is used, which is considered better 
tracer of the mass of the cloud. According to their results, the mass of the 
cloud is underestimated by a factor 4-5. If this correction factor is applied in 
calculating the mass of our sample of clouds plotted in Figure 12, the result 
is that the values of \siggas\ are increased by a value of 0.6 (in log scale) 
and our distribution follows well to distributions for dense clumps reported by 
\citet{heyer+16,wu+10}. The corrected values of our clouds lie on the KS 
relationship.

It can be noted in the figure that the average gas densities of several MCs from 
our sample  are clustered around $\log$(\siggasav)$=2.6\pm0.3$, compared to 
clouds from \citet{heiderman+10} clustered around $\log$(\siggas)$=2.0$. 
On the other hand, \siggasav\ for the dense clumps 
are distributed over a much wider dynamic range. \citet{lada+13} also
found almost constant average \siggasav\ for the MCs they analyzed, which 
they used to argue against the existence of a Schmidt law between MCs. 
They further argued that the clustering of global \siggasav\ is due to 
the well-known scaling law of \citet{larson+81} between cloud size and mass, 
where \siggasav\ is a constant for a given value of the limiting column density 
used to define the MCs. \citet{heyer+09} found \siggasav$\sim40$~\msunpc\ 
for MCs with a defining boundary of $A_v=1$~mag. 
Values obtained in our study are around a factor of 4--9 higher, 
and our lower values compare well with the \siggasav$\sim170$~\msunpc\  obtained 
by \citet{solomon+87}.

\begin{figure}[ht!]
     \begin{center}
            \includegraphics[width=8cm]{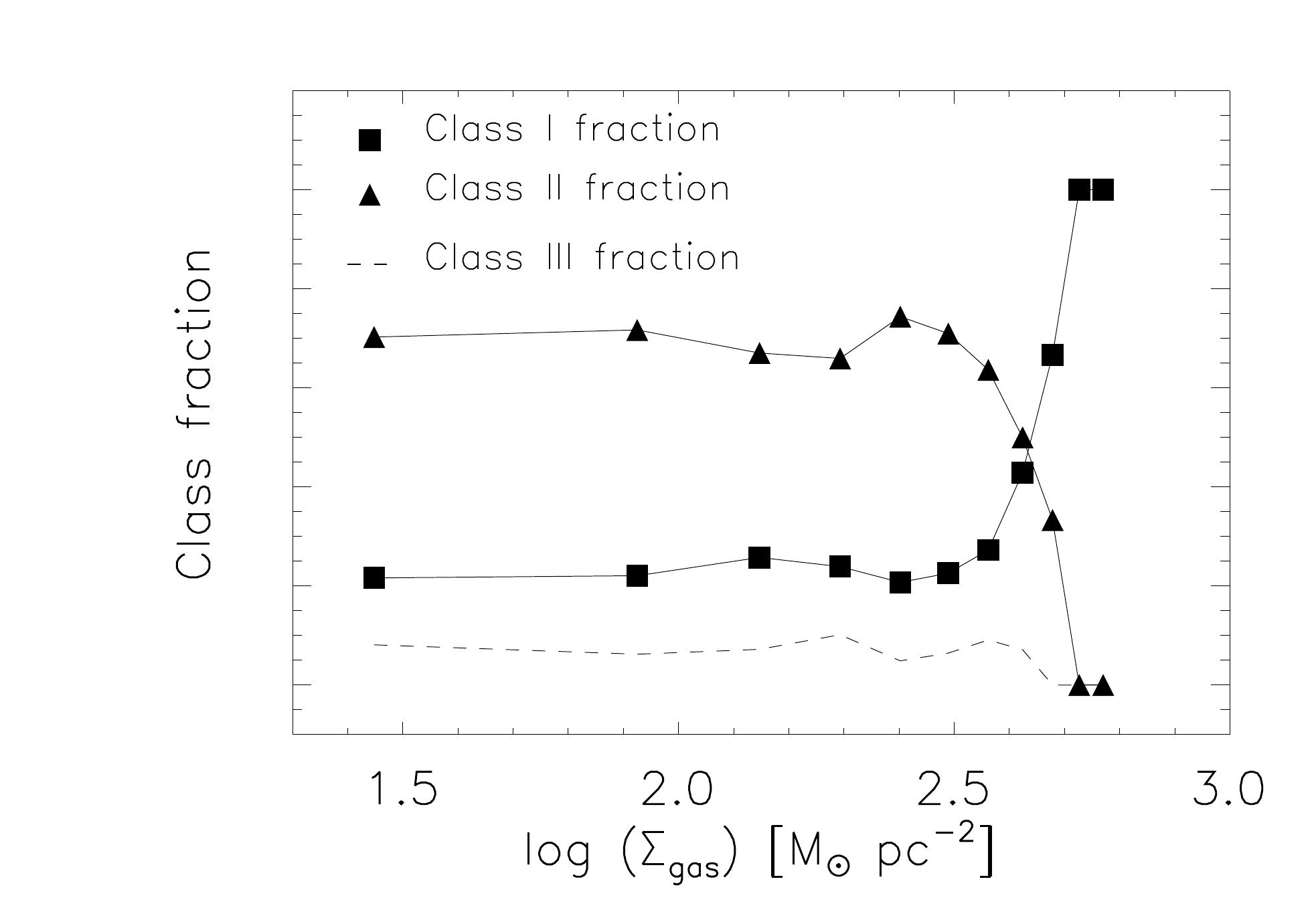} \\
            \includegraphics[width=8cm]{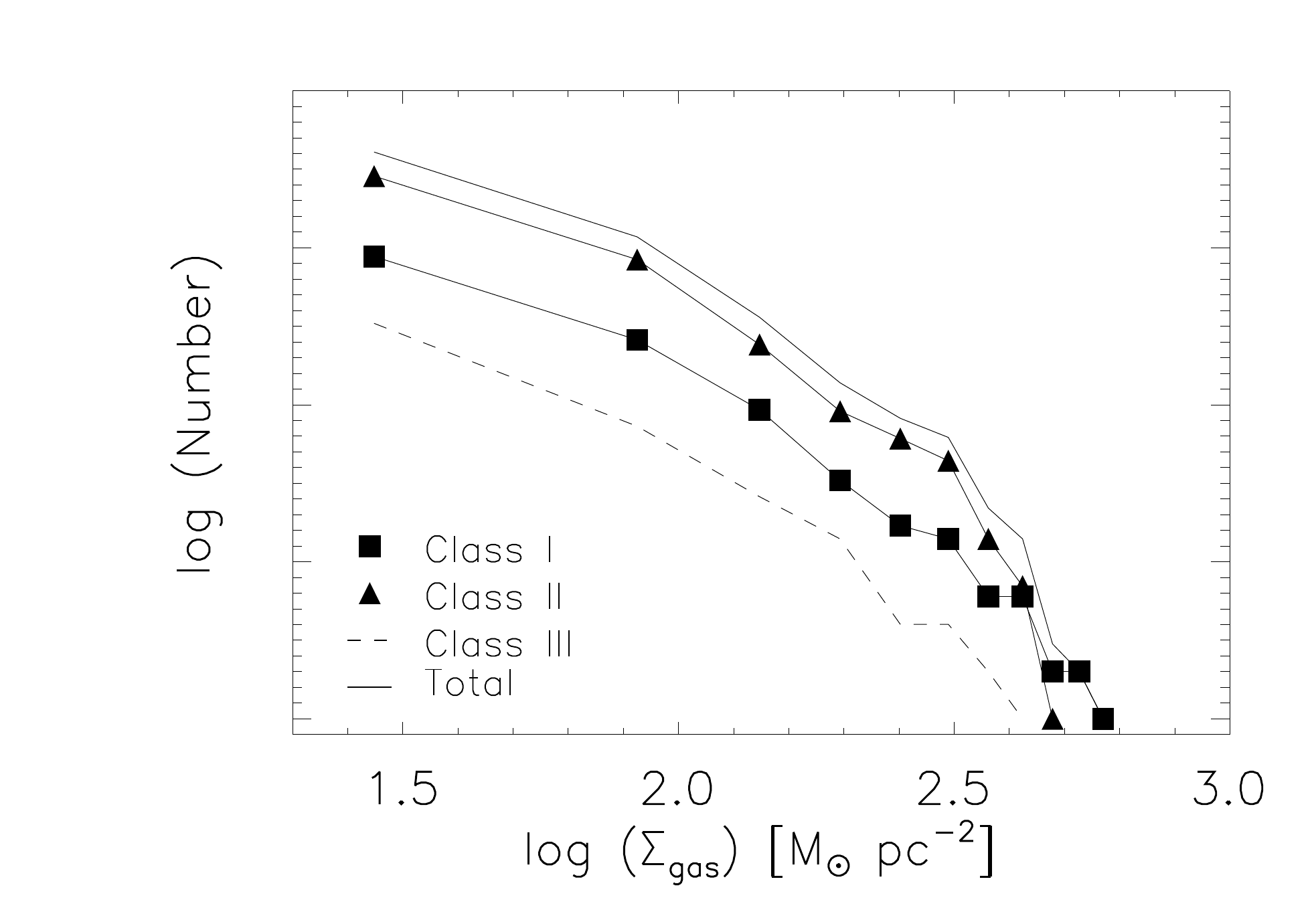}
    \end{center}
\caption{
The number (top) and fraction of total (bottom) YSOs separated
by Class as a function of surface density of the gas. 
The range of threshold densities in our sample of clouds is
shown by the dotted vertical lines. It can be seen that the
relative fractions are almost constant below the threshold density,
with the fraction of Class I YSOs increasing steeply at higher
densities.
}
\label{figure13}
\end{figure}

\subsection{Star formation below the threshold density}

The threshold densities, \sigth\ for our clouds are comparable to the values 
obtained in the studies of \cite{heiderman+10} and \cite{lada+12}. However, 
there is a clear difference in the star formation relation below \sigth\ in 
our clouds as compared to the clouds studied by \citet{heiderman+10}. 
In the latter study, SF decreases abruptly below \sigth,
whereas in our case SF almost remains constant or decreases less steeply with index values
being sub-linear below \sigth. 
This trend below \sigth\ is also found in the recent study by \citet{willis+15}
(see their Figure~9). However, they suspected an increasing contamination from giant 
stars and background galaxies at decreasing densities  as a possible reason for the
apparent excess of YSOs below \sigth, and analyzed the star formation law only for the 
data points above \siggas$>$200~\msunpc.

Is the apparent SF below \sigth\ an artifact of the method used or is it real?
The adopted filtering method (see \S3) has effectively cleaned our sample of
contaminants such as bright foreground/background stars or background galaxies,
and hence the presence of YSOs at low densities is real. While calculating
the \sigsfr, we have included all YSOs, including those of Class II and III. 
\citet{heiderman+10} have used only Class I sources in obtaining the steeply
decreasing \sigsfr\ below \sigth. We now investigate whether the apparent
constant \sigsfr\ below \sigth\ is due to the inclusion of Class II and III
YSOs in our study. In Figure~\ref{figure13}, we plot the number (upper panel) 
and fraction (lower panel) of YSOs separated by Class in each gas density bin. 
YSOs from all MCs are added at the corresponding gas densities to obtain 
this plot. It can be seen that Class II sources dominate the observed number 
of YSOs below \sigth. The Class~III fraction remains almost constant at 
$\sim$10\% at all gas densities. Thus, evolved Class III objects do not 
contribute significantly to \sigsfr\ below \sigth.

The presence of Class I/II YSOs associated to gas below \sigth, could imply 
either (1) stellar migration or (2) effect of beam dilution or both.
We discuss these two possibilities below: \\
1. Stellar migration from their birth sites: star-forming cores can migrate from
their formation site in a dense environment to the current location. For typical 
velocities of dense cores of 0.33 to 0.55~\kms \citep{muench+07,kirk+10}, 
Class I and II objects can be as far as 1--2 parsecs away from their formation 
sites. Our clouds measure
typically $>$5~pc, implying the Class I and II YSOs in our clouds had no time
to migrate from dense regions to low surface gas density regions. \\
2. Beam dilution: The \siggas\ in our study is obtained using \co\ beams 
(0.3--1.2~pc) that are in general larger than the typical core size (0.1~pc). 
The cores are expected to be surrounded by dense clump gas whose typical
size (1~pc) is larger than our beam size. Under such a situation, the observed
\co\ column densities are expected to be that of typical clump densities, which
would be higher than \sigth. The observed low column densities would
then imply that the star-forming cores at densities below \sigth\ 
are isolated and are not be surrounded by the dense clump gas.

\section{Conclusions}

We here analyzed the local and global star formation law in a sample of 
12 Galactic molecular clouds with signposts of ongoing high-mass star formation.
The study is different from similar previous studies in that 
(1) we investigate the SF relation in the whole cloud, and not just in the
dense clumps, and (2) Class-dependent masses are obtained for each MIPS source. 
The number of high luminosity sources ($L_{\rm bol}>$10~\msun) is found to be 
clearly in excess of that expected for a Chabrier mass function, suggesting
a combined Chabrier-Salpeter mass function for the YSOs in our clouds.
YSOs with $L_{\rm bol}>$10~\msun\ contribute more than 30\% of the total mass
in each of our clouds, with a mean value of 55\%.
The SFR is obtained by the standard technique of counting the YSOs 
enclosed within contours of gas
density as traced by the \co\ emission above $A_v=1$~mag (\siggas=20~\msunpc). 
The relation between \sigsfr\ and \siggas\ has a break at $\sim$150--360~\msunpc 
for almost of the clouds,
with power-law forms on either side of the break, which is also found using the 
total YSO population of the clouds with the trend. The power-law index above
the break lies between 1.3---3.0 in different clouds, which is
in general higher than the Kennicutt value of 1.4 for extra-galactic regions, but
is consistent with the values of Galactic regions found in recent studies. 
At densities below the break, we find \sigsfr\ almost independent of density.
The density at the break is consistent with the threshold density \sigth\ 
found in Galactic and extra-galactic \starf\ regions. 
Globally averaged \siggasav\ for our sample of molecular clouds are clustered around 
a value of 350~\msunpc within a factor of 2, a range too small to explore
a relation with these data alone. The global \sigsfrav\ lies above the
Kennicutt values for the observed \siggasav\ by factors between 1 to 60,
but agrees within a factor of 2 with the linear relation for massive clumps.

\acknowledgments
This work was supported by the CONACyT (Mexico) research fellowship 199495 granted
to RR, and research grants CB-2010-01-155142-G3 (PI:YDM) and 182841 (PI:AL).

\end{document}